\documentclass[twocolumn,secnumarabic,amssymb, nobibnotes, aps, prb]{revtex4-2}
\usepackage{amsmath}
\usepackage{graphicx}
\newcommand{\ul}[1]{\underline{#1}}
\newcommand{\dul}[1]{\underline{\underline{#1}}}
\newcommand{\mc}[1]{\mathcal{#1}}

\begin{document}

\title{Theory of perturbatively nonlinear quantum transport II: Hilbert space truncation, gauge invariance, and second order transport in a spatially uniform, time-varying electric field}

\author{Varga Bonbien$^1$}%
\email[]{bonbien.varga@kaust.edu.sa}
\author{Aur\'elien Manchon$^{2}$}
\affiliation{$^1$Physical Science and Engineering Division (PSE), King Abdullah University of Science and Technology (KAUST), Thuwal 23955-6900, Saudi Arabia\\
$^2$Aix-Marseille Univ, CNRS, CINaM, Marseille, France}

\begin{abstract}
This article is the second of a trilogy that addresses the perturbative response of general quantum systems, with possibly nontrivial ground state geometry, beyond linear order. Here, we establish concise, general formulae for second order response to a spatially uniform, time-varying electric field in the velocity gauge that are \textit{manifestly} free of static limit spurious divergences. We first discuss general quantum evolution in a curved space, then detail how such a situation is a natural byproduct of Hilbert space truncation, and point out crucial subtleties associated with the resulting finite curvatures. We then present a geometric perspective of the two popular gauges often used in quantum transport theories, the velocity gauge and the length gauge, and discuss how they, taking truncation-induced curvature effects into account, naturally lead to the same results in spite of the truncation. We highlight subtle formal discrepancies in the literature. Finally, we provide a general scheme for removing static limit spurious divergences in the velocity gauge \textit{without} frequency expansions and present concise and comprehensive Green's function formulae for responses up to second order. As an application of specific aspects of our theory, second order charge current responses in selected cases are analyzed in Refs. \cite{Bonbien2021a} and \cite{Bonbien2021c}. 
\end{abstract}

\maketitle
\tableofcontents

\section{Introduction}

The study of transport phenomena in an electric field has a long history. The seminal work of Kubo \cite{Kubo1956,Kubo1957} in the 1950s opened up a path towards laying bare the quantum mechanical intricacies of such phenomena. This is the second paper in our series on nonlinear quantum transport and, after the generalities of the first paper \cite{Bonbien2021a}, we narrow our focus to the study of nonlinear responses to a spatially uniform, possibly time-dependent electric field.\\

The lifting of the shroud over the quantized Hall effect’s topological underbelly \cite{TKNN} was followed by a flurry of revelations regarding the subtle geometric structures, Berry’s phase being a prime example, at the root of many potentially interesting quantum phenomena \cite{SimonBerry,Berry1984,WilczekZee1984,AvronBerry}. Over the years, this newfound insight was simplified \cite{SundaramNiu}, led to a deeper understanding of Hall effects \cite{NagaosaAHErev,SinovaSHE}, and heralded the topological approach to materials \cite{HasanKane,TQC}. Keeping transport phenomena in sight, despite the focus on static transport effects first order in the driving field, geometric concepts were spilling over to nonlinear optics \cite{SipeShkrebtii2000} and have gradually become more widespread \cite{MooreOrenstein,SodemannFu2015,MorimotoNagaosa2016,QCPGE2017,Ahn2020,Watanabe2021} culminating in their pervasive presence \cite{Orenstein2021,Ma2021}. In this paper we provide a very general geometric framework for computing second order and, by extension, higher order time-dependent and static electric field responses.\\

\subsection{The velocity gauge, the length gauge, and spurious divergences}

There are two recipes for coupling a uniform electric field to a quantum system. The first is via the minimal coupling prescription $\textbf{p}\to \textbf{p}-e\textbf{A}(t)$, where $\textbf{E}(t) = -\partial \textbf{A}(t)/\partial t$ with $\textbf{E}(t)$ being the electric field and $\textbf{A}(t)$ the vector potential, and can be considered as being the dipole approximation of the more general Coloumb gauge \cite{KobeGauge}. This prescription is commonly referred to as the \textit{velocity gauge}. The second option is to couple the electric field directly to the position operator $\textbf{x}\cdot \textbf{E}(t)$, which corresponds to the dipole approximation of the multipolar gauge \cite{KobeGauge}, and is dubbed the \textit{length gauge}. The initial approaches to the computation of nonlinear optical responses, second harmonic generation for instance, made use of the velocity gauge and it was shown that spurious divergences arise in the static limit that could nevertheless be cancelled via the use of delicate sum rules \cite{Aspnes1972}. Similarly, the early theories describing the bulk photovoltaic effect, a direct current response of semiconductors to an optical field, also used the velocity gauge \cite{KrautBaltz1979,KrautBaltz1981}. However, the methods developed for handling the spurious divergences assumed the ideal circumstance of an infinite number of bands \cite{Sipe1993} and, in the case of sum rules, required their identification on a case-by-case basis \cite{AversaSipeSumRules}. Eventually, problems with this spurious divergence and disagreements between calculations performed in the two gauges resulted in the velocity gauge losing prominence and the length gauge garnering favour \cite{SipeShkrebtii2000,AversaSipeSumRules}. More than a decade passed before the issue was revisited \cite{Ventura2017,Taghizadeh2017,Passos2018} and the intimate connection between the divergences and truncation of the infinite number of bands to a finite subset was emphasized. The earlier sum rules requiring an infinite number of bands were reformulated for the finite band case \cite{Ventura2017} and it was shown, both formally and numerically, that the velocity gauge yields results identical to the length gauge even for finite band models \cite{Passos2018} moreover, the sum rules were given an interpretation as the vehicles for moving between the two gauges. The velocity gauge has since made a resurgence and the diagrammatic techniques based on Green’s function expressions of the retarded correlators, already enjoying widespread use for first order responses, are starting to gain traction \cite{Parker2019,Joao2020,Holder2020}.\\

The issues with the velocity gauge are, however, yet to be fully resolved. While the general sum rules guarantee the non-existence of spurious divergences and equivalence to the length gauge \cite{Ventura2017,Passos2018}, the fact that the sum rules appear only when moving to the length gauge---in other words, that the non-existence of spurious divergences is not manifest \textit{within} the velocity gauge---, is a significant drawback. The reason for this is that even though the `default' \textit{combination} of retarded correlators giving the second (and higher) order response function in the velocity gauge (see Eq. \eqref{eq:KAAomega}) is, by virtue of the sum rules, divergence-free, crucially, the `default' combination in question is not \textit{manifestly} divergence-free. Rather, the divergence-free property of the `default' combination is only unmasked when moving to the length gauge and realizing that the part that would otherwise provide the divergence in the velocity gauge is precisely the part that vanishes by virtue of the sum rules. This devoids us of insights regarding the role of the rather different velocity gauge correlators in the `default' combination: in general, only their totality has physical relevance. Then, a natural question to ask is whether there exists an equivalent combination of correlators that is \textit{manifestly} free of spurious divergences? Should the answer be affirmative, it would provide a rather robust handle on the response functions since all concerns about the divergences could be laid to rest \textit{without having to move to the length gauge}. An important result of this paper is that such a combination indeed exists and can be arrived at in a rather simple manner. We show that the standard technique of shifting the vector potential \cite{Rammer1986,RammerQT}, now by a constant, $\textbf{A}(t)\to\textbf{A}(t)+\textbf{c}$, and requiring the resulting physical response to be independent of this constant, leads to constraints that we refer to as `gauge conditions' and a specific conflation of these conditions (Eq. \eqref{eq:secondOrderCondK}) allows us to express the default combination of retarded correlators using a different combination (Eq. \eqref{eq:secondOrderCondAA}). This permits us to rewrite the second order, finite-frequency response function as an expression (Eq. \eqref{eq:secondOrderFinal}) that is \textit{manifestly} free of spurious divergences which can then be expressed using Green’s functions, and ‘collapsed’ into a relatively simple formula (Eq. \eqref{eq:secondOrderGreenFinal}). To the best of our knowledge, the subtleties concerning the spurious divergence alluded to above have not been recognized and the form of the response function we present is absent from the literature. Crucially, should we be interested in taking the static limit, we would not be required to perform a Taylor expansion in the frequency but could simply set the latter to zero in the `collapsed’ formula. As a technical note, we point out that the explicit form of the correlators is only required for the proof of the gauge conditions, with the special combination for manifest divergence-cancellation being a very general, formal expression between the retarded correlators, in a way providing the essence of why the divergences are, in fact, spurious. Furthermore, this special combination is not arbitrary but admits a pattern that can be generalized to higher orders and was carried out by us also for third order \cite{Bonbien3rdOrder}.\\

\subsection{Band truncations and geometry}

\begin{figure}
\begin{center}
\includegraphics[width=8cm]{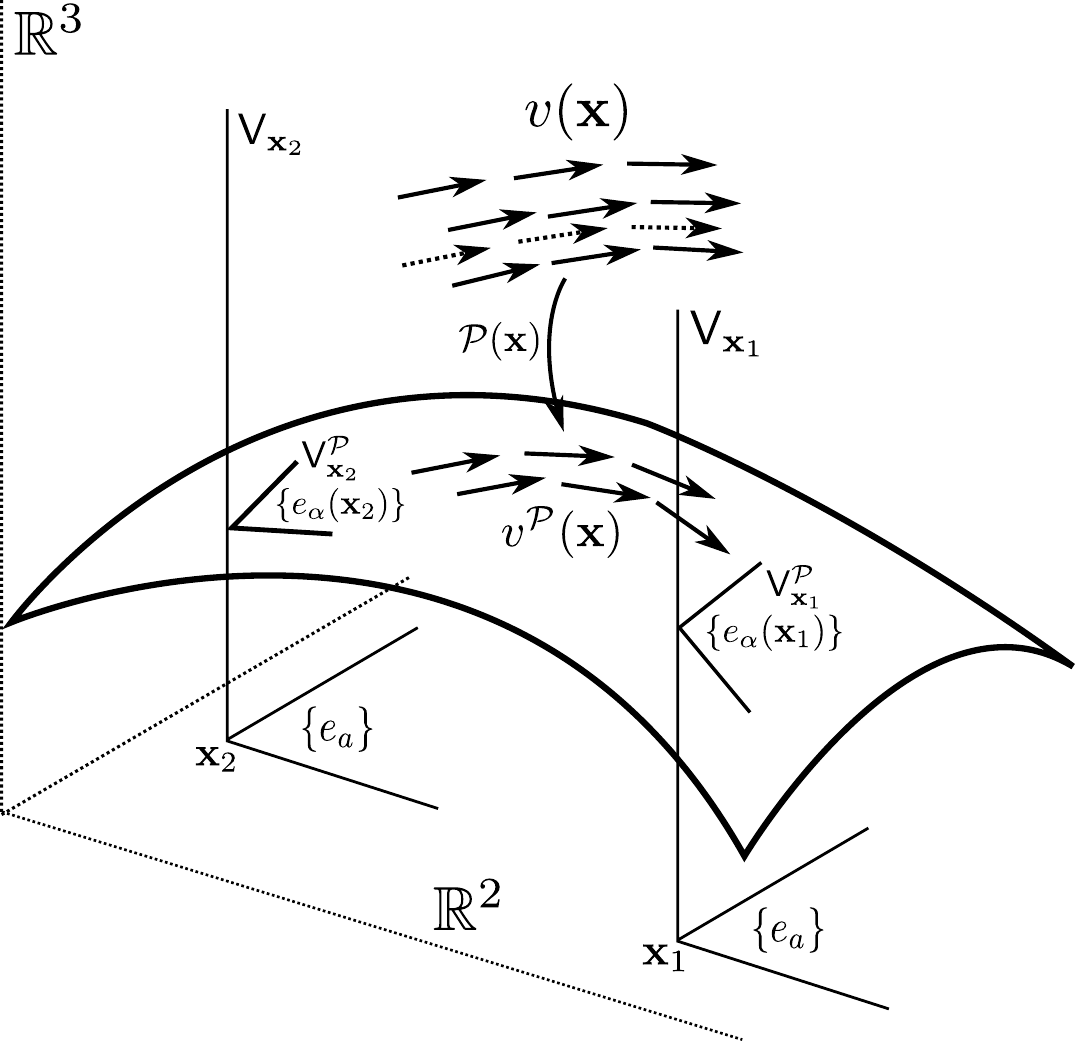}
\caption{Projection onto a subspace. We have $\textbf{x}\in\mathbb{R}^2$ as points of the base space and vector spaces $\mathsf{V}_{\textbf{x}}\cong\mathbb{R}^3$ as 3-dimensional fibres. The standard bases $\{e_a\}$ of each $\mathsf{V}_{\textbf{x}}$ can be trivially identified for all $\textbf{x}$ and $v(\textbf{x})\in \mathsf{V}_{\textbf{x}}$ is a smooth vector field. $\mathsf{V}^{\mc{P}}_{\textbf{x}}$ are 2-dimensional vector spaces obtained via an $\textbf{x}$-dependent projection and $\{e_{\alpha}(\textbf{x})\}$ are their bases which can no longer be trivially identified for different $\textbf{x}$. This means that we have to use different bases at different $\textbf{x}$ to expand the projected vector field $v^{\mc{P}}(\textbf{x})$ within the projected subspace and need a way to transfer the resulting components between the bases in order to facilitate a consistent comparison and differentiation.\label{fig:projectionIntro}}
\end{center}
\end{figure}

While the detrimental effects of band truncations on calculations within the velocity gauge have been studied broadly, linked to the failing of sum rules and thereby electromagnetic gauge-invariance \cite{Taghizadeh2017,Passos2018}, the underlying interplay with geometry has, we believe, yet to be fully appreciated. It is well-known that a Berry connection with non-vanishing curvature arises upon projection to a well-separated set of bands \cite{AvronBerry,XiaoBerryRev}. This is a special case of a more general phenomenon rooted in the behaviour of derivatives and closely linked to vector bundles \cite{AtiyahYangMills}. Roughly speaking, if we have a vector field $v(\textbf{x})$, where $\textbf{x}\in\mathbb{R}^n$, with values in $N$-dimensional vector spaces $\mathsf{V}_{\textbf{x}}$ at each $\textbf{x}$ such that $\mathsf{V}_{\textbf{x}}$ are copies of the same vector space $\mathsf{V}$, i.e., $\mathsf{V}_{\textbf{x}}$ are considered as fibres of a vector bundle over $\mathbb{R}^n$ with the vector field $v(\textbf{x})$ being a section of the bundle, and we can find a basis $\{e_a\}$ with $a\in\{1,\dots,N\}$---the standard basis---that is the \textit{same} for each $\mathsf{V}_{\textbf{x}}$ independent of $\textbf{x}$, then the vector field $v(\textbf{x})$ can be identified with its components $\underline{v}(\textbf{x})=(v_1(\textbf{x}),\dots,v_a(\textbf{x}),\dots,v_N(\textbf{x}))^T$ in this basis and differentiation of the vector field $\mathcal{D}_iv(\textbf{x})$ is equivalent to partial differentiation $\partial_i\underline{v}(\textbf{x})$ of its components. On the other hand, should we perform an \textbf{x}-dependent smooth projection $\mathcal{P}(\textbf{x})$ to subspaces $\mathsf{V}_{\textbf{x}}^{\mathcal{P}}$ at each \textbf{x}, such a `constant' basis no longer exists within the collection of subspaces and we have to define different bases $\{e_{\alpha}(\textbf{x})\}$ with $\alpha\in\{1,\dots,\text{dim}\mathsf{V}_{\textbf{x}}^{\mathcal{P}}\}$ (note that \textit{all} $\mathsf{V}_{\textbf{x}}^{\mathcal{P}}$ have the same dimension independent of $\textbf{x}$) for $\mathsf{V}_{\textbf{x}}^{\mathcal{P}}$ at \textit{each} \textbf{x}. This means that differentiation of the projected vector field within the projected subspace is now a rather delicate matter since we would have to compare components attached to different bases, the latter being $\textbf{x}$-dependent, and therefore need a way to connect the bases at different $\textbf{x}$ (see FIG. \ref{fig:projectionIntro}). This is provided by a connection on the vector bundle with fibres $\mathsf{V}_{\textbf{x}}^{\mathcal{P}}$ and we are lead to a covariant derivative $\mathcal{D}^{\mathcal{P}}_iv^{\mc{P}}(\textbf{x})$ \cite{AtiyahYangMills}. A key difference between the two cases is that even though in the former case we could perform an \textbf{x}-dependent change of basis in each $\mathsf{V}_{\textbf{x}}$ resulting in the fact that the components of the vector field derivative $\mathcal{D}_iv(\textbf{x})$ would naturally not be equivalent to the partial derivative of the vector field components in this basis, but, rather, they would have the structure of a `covariant derivative', we could always move back to the constant basis and obtain the partial derivative. This cannot be done in the latter case: no matter what we do, we will not be able to arrive at the constant basis and the derivative will always be `covariant'.  The corresponding connections are said to be `flat' in the former case and `curved' in the latter case. We can see this concretely by writing the covariant derivative in terms of (local) connection components $\dul{\omega_i}$, i.e., $\omega^b_{ia}(\textbf{x})$, by defining its action on the bases at each $\textbf{x}$ as

\begin{equation}
\nonumber
\mc{D}_ie_a(\textbf{x})=\sum_b \omega^b_{ia}(\textbf{x})e_b(\textbf{x}),
\end{equation}

and on component functions $v^a(\textbf{x})\in\ul{v}(\textbf{x})$ of a vector field as $\mc{D}_iv^a(\textbf{x})=\partial v^a(\textbf{x})/\partial x^i$. Then, we have for its action on a vector field

\begin{equation}
\nonumber
\mathcal{D}_iv(\textbf{x})=\sum_a\left(\frac{\partial v^a(\textbf{x})}{\partial x^i}+\sum_b\omega^b_{ia}(\textbf{x})v^b(\textbf{x})\right)e_a(\textbf{x}).
\end{equation}

Suppose there exists a basis, the standard basis, in which $\dul{\omega_i}(\textbf{x})=0$. Then, the covariant derivative of a vector field is simply the partial derivative of its components \textit{in this basis}. Moving to another \textbf{x}-dependent basis the connection components $\dul{\omega_{i}}(\textbf{x})$ will not remain zero (they transform non-covariantly, see Appendix \ref{diffgeo}.\ref{diffgeo1}). On the other hand, should we not have a standard basis available, $\dul{\omega_{i}}(\textbf{x})$ will never vanish, regardless of what basis we choose. We can connect these notions to flatness via the curvature components $\dul{\Omega_{ij}}(\textbf{x})$ given as

\begin{equation}
\nonumber
[\mc{D}_i,\mc{D}_j]e_a(\textbf{x})=\sum_{b}\left(\dul{\Omega_{ij}}(\textbf{x})\right)^b_{a}e_b(\textbf{x}),
\end{equation}

where

\begin{equation}
\nonumber
\dul{\Omega_{ij}}=\frac{\partial}{\partial x^i}\dul{\omega_{j}}-\frac{\partial}{x^j}\dul{\omega_{i}}+\left[\dul{\omega_{i}},\dul{\omega_{j}}\right].
\end{equation}

Clearly, if $\dul{\omega_i}(\textbf{x})=0$ then $\dul{\Omega_{ij}}(\textbf{x})=0$. What about in another $\textbf{x}$-dependent basis? It turns out that $\dul{\Omega_{ij}}(\textbf{x})$ remains zero (it transforms covariantly under a change of basis) and such a connection is rightfully called flat. Conversely, should no basis exist in which $\dul{\omega_i}(\textbf{x})=0$, the curvature cannot vanish and the connection is said to be curved.\\

This discussion above is particularly relevant for our investigation of band or Hilbert space truncations. Formally, we have Hilbert spaces $\mathsf{H}_{(t,\textbf{k})}$ for each time $t$ and crystal momentum \textbf{k}; spanned by periodic Bloch states $\{|u_a(\textbf{k})\rangle\}$, where $a$ is the band index; considered as fibres of a Hilbert bundle over $\mathbb{R}\times\mathcal{B}$, with $\mathbb{R}$ referring to time and $\mathcal{B}$ the first Brillouin zone. Just as a vector field was a section of a vector bundle, a state $|\psi(t,\textbf{k})\rangle$ is a section of this Hilbert bundle, and to evolve it in time we need a way to move it between different fibres which requires a connection on the Hilbert bundle. We define the connection through its components in the standard basis $\{|e^{\text{F}}_a\rangle\}$ of $\mathsf{H}_{(t,\textbf{k})}$, which is \textit{not} the basis consisting of the periodic Bloch states, rather, it is the standard, `$\textbf{k}$-independent' basis of the total space of fibres $\mathsf{H}_{(t,\textbf{k})}$, each of which are \textit{spanned} by $\{|u_a(\textbf{k})\rangle\}$ (cf. $\{e_a\}$ for the standard basis of the vector bundle considered earlier). In this basis, let the time-direction components of the connection be given by the Bloch Hamiltonian $\dul{\mathcal{H}}(\textbf{k})$ and in the $\textbf{k}$ direction be vanishing $\dul{\mathcal{A}^{\text{F}}_i}(\textbf{k})=0$ (see Section \ref{curvedSpace} and Appendix \ref{crystal}.\ref{dynamics} for the general justification). This corresponds to a flat connection in the $\textbf{k}$ direction.  We can also project via $\mathcal{P}(\textbf{k})$ along \textbf{k} to a subspace and obtain states $|\psi^{\mathcal{P}}(t,\textbf{k})\rangle$ within the projected subspaces corresponding to the fibres $\mathsf{H}^{\mathcal{P}}_{(t,\textbf{k})}$ (with respect to a local trivialization of the projected bundle) that are connected by the projected connection, now curved even over \textbf{k}, with components $\dul{\mathcal{H}}^{\mathcal{P}}(\textbf{k}),\,\dul{\mathcal{A}_i^{\mathcal{P}}}(\textbf{k})$. We thus clearly have to distinguish between evolution in the `total space' and the `projected subspace'. In the former case the connection is flat and we can always choose a basis (locally), the standard basis, in which its components vanish $\dul{\mathcal{A}_i^{\text{F}}}(\textbf{k}) = 0$, meaning that we can simply consider the Bloch Hamiltonian $\dul{\mathcal{H}}(\textbf{k})$ as describing the total evolution in this basis, whereas in the latter case the projection renders the connection curved and we have to look at \textit{both} $\dul{\mathcal{H}}^{\mathcal{P}} (\textbf{k})$
and $\dul{\mathcal{A}_i^{\mathcal{P}}}(\textbf{k})$ in order to properly account for the evolution. In other words, we have to beware of the fact
that the standard quantum evolution in the `total space' has to be carried over to a modified quantum evolution within the `projected subspace'. To put this rather abstract description into perspective, we give two practical examples of total spaces and projected subspaces: the `total space' consists of an infinite number of bands and the `projected subspace' contains only a finite number of bands; the `total space' is made up of a finite number of bands and the `projected subspace' also consists of a finite, albeit not as numerous number of bands. We assume that the bands in the projected subspaces are well-separated from the rest. Should we obtain a finite band model by truncating an infinite band model and continue our calculations within the finite band space while not taking into account the fact that the truncation (projection) gave rise to a \textit{curved} connection, we shall naturally get fictitious, gauge invariance breaking results. Similarly, should we start out with a finite band model, truncate it to a lesser number of bands and continue working in the truncated band space, we would have to work with a curved connection. The sum rules are only valid if we are working in the \textit{total} spaces, regardless of them consisting of an infinite or finite number of bands since what matters is whether the corresponding connection is flat or not and we always take the connection on the Hilbert bundle corresponding to the total spaces to be flat. As an example, we have the trivial relation $(\partial_i\partial_j-\partial_j \partial_i)\psi(t,\textbf{k})\propto [x_i,x_j]\psi(t,\textbf{k}) =0$, where $x_i = i\partial_i = i\partial/\partial k^i$, which turns into a sum rule upon choosing the basis to be one in which the Bloch Hamiltonian is diagonal at each \textbf{k} and is responsible for the cancellation of spurious divergences in the velocity gauge at second order \cite{AversaSipeSumRules}. It is clear that this sum rule implicitly assumes a \textit{flat} connection since it is defined in a basis in which the components of the connection vanish. Thus, the validity of the sum rules rests not on the number of bands, rather, on the curvature of the connection.\\

\subsection{Quantum evolution and parallel transport}

We have referred to quantum evolution as some kind of `movement' of states between fibres $\mathsf{H}_{(t,\textbf{k})}$ of the Hilbert bundle, but have not specified precisely what it is. In fact, we have a connection on the Hilbert bundle and can thus go for the most straightforward option: parallel transport. While this sounds rather unfounded, it is precisely what is happening. In the case without an electric field, evolution is only in the time-direction and parallel transport simply means that the state evolves in such a way that it remains covariantly constant along time; $\mathcal{D}_t|\psi(t,\textbf{k})\rangle = 0$. Here $\mathcal{D}_t$ is the covariant derivative in the time-direction and its components in a local basis correspond to the Hamiltonian $\dul{\mathcal{H}}(\textbf{k})$, meaning that writing $\mathcal{D}_t|\psi(t,\textbf{k})\rangle = 0$ in a basis results in the standard Schrödinger equation (see Section \ref{curvedSpace} for details)

\begin{equation}
\label{eq:SchIntro}
i\hbar\frac{d}{dt}\ul{\psi}(t,\textbf{k})=\dul{\mathcal{H}}(\textbf{k})\ul{\psi}(t,\textbf{k}).
\end{equation}

Upon the application of an electric field, in the velocity gauge, \textbf{k} gets modified to $\textbf{k}(t)=\textbf{k}-e\textbf{A}(t)$ and we also have evolution in the \textbf{k} direction. The total evolution then happens along a path $(t,\textbf{k}(t))$ with tangent $(1, d\textbf{k}/dt)$, where $d\textbf{k}/dt = e\textbf{E}(t)$, meaning that the parallel transport is $\mathcal{D}_t|\psi(t, \textbf{k}(t))\rangle + \frac{dk^i}{dt} \mathcal{D}_i|\psi(t, \textbf{k}(t))\rangle = 0$, where we assumed automatic summation on $i$ and $\mathcal{D}_i$ is the covariant derivative along $\textbf{k}$  in the $i$ direction with components $\dul{\mathcal{A}_i}$ in a (local) basis. Writing out this parallel transport equation in an arbitrary (local) basis, we obtain the `extended' Schrödinger equation

\begin{equation}
\label{eq:SchParIntro}
i\hbar\frac{d}{dt}\ul{\psi}(t,\textbf{k}(t))=\dul{\mathcal{H}}(\textbf{k}(t))\ul{\psi}(t,\textbf{k}(t))+eE^i\dul{\mathcal{A}_i}(\textbf{k}(t))\ul{\psi}(t,\textbf{k}(t)).
\end{equation}

Should the connection over \textbf{k} be flat, i.e., the above equation describes evolution in the `total space', we can always choose a local basis along each point of the path, labelled as `F', in which $\dul{A^{\text{F}}_i }(\textbf{k}(t)) = 0$ meaning that, \textit{in this basis}, the extended Schrödinger equation reduces to the standard Schrödinger equation. Similarly, we can choose a basis in which the Hamiltonian is diagonal, or any other basis for that matter. However, if the connection over \textbf{k} is \textit{not} flat, in other words, curved, i.e., we are working in a `projected subspace', then there is no local basis in which the components $\dul{\mathcal{A}^{\mathcal{P}}_i} (\textbf{k}(t))$ vanish and \textit{we have to describe quantum evolution with the extended Schrödinger equation}! From the perspective of perturbative response, we have to expand the connection components $\dul{\mathcal{A}_i}(\textbf{k}(t))$ with respect to $\textbf{A}(t)$ \textit{in addition} to the Hamiltonian. At first order, the response function will be formally similar to the flat case, however, at second order, path-dependence in \textbf{k} starts to matter and the response function is qualitatively different when compared with the flat case: a term proportional to the curvature appears (see Eq. \eqref{eq:secondOrderFinal}). This is the fundamental reason for why such an integrated geometric framework is necessary and is an important result of this paper. Indeed, should we want to accurately account for Hilbert space truncation effects and work within the truncated subspace, our description \textit{requires} a curved connection on a Hilbert bundle. To the best of our knowledge, no such consistent framework has been developed for nonlinear responses.\\

While our integration of nonlinear perturbative response theory into a fully geometric framework is, to the best of our knowledge, novel, the application of Hilbert bundle techniques to quantum evolution is an old story. The interpretation of the standard Schrödinger equation as always being a parallel transport on a Hilbert bundle with the Hamiltonian being components of a connection on this bundle was first considered by \citet{AsoreyParallelTransport}. These authors also discuss several mathematical subtleties related to the construction of the connection (see also ref. \cite{SardParallelTransport}). Later on, in a series of papers \cite{Iliev1,Iliev2,Iliev3}, Iliev used similar ideas for a formulation of quantum mechanics in fibre bundle language, but did not discuss the case of the Hamiltonian depending on time-evolving parameters, such as the crystal momentum $\textbf{k}(t)$ in the velocity gauge, a key and central theme of our paper. This was done in a rather sophisticated manner by Sardanashvily \cite{SardTimeDependentPar} via a connection on a composite bundle, and the author arrived at the extended Schrödinger equation \eqref{eq:SchParIntro}, however, no explicit construction of the connection was provided and the author did not proceed beyond a discussion of quantum evolution. This was partially remedied by Viennot \cite{ViennotParameter} who generalized the composite bundle approach and arrived at a concrete connection, albeit, yet again, failed to progress beyond a detailed analysis of quantum evolution. In comparison, our approach, though detailed, does not make use of advanced techniques and is rather elementary, since the primary focus of this paper is application, and our aim is to get up to speed with the geometric background as quickly as possible, so that we can go ahead and delve into the subtleties of perturbative response calculations.\\

\subsection{A reader's guide}

Establishing a rigorous theory of nonlinear responses to an electric field in a curved space involves multiple concepts from differential geometry, and requires clarifying a number of fundamental aspects pertaining to quantum evolution and the choice of gauge. We have thus organized the paper into five main sections. We develop our geometric framework in section \ref{curvedSpace} by starting out with a formulation of quantum evolution in a Hilbert bundle with a curved connection. We then discuss parallel transport; show how to perform perturbative expansions in a proper manner, all the while highlighting some fundamental inconsistencies present in the literature; and wrap up via a discussion of density matrix evolution and expectation values. We move on to Section \ref{velLength}, in which we present a detailed study of the velocity and length gauges from a geometric perspective and also present a consistent definition of the velocity operator in a curved space. Following a brief review of some results from our paper I \cite{Bonbien2021a}, in Section \ref{pertresponse} we present the calculation of the perturbative response upto second order---including the gauge conditions and manifest cancellation of spurious divergences---, and finally arrive at a finite frequency, second order response formula valid for a Hilbert bundle with a curved connection and provide an expression for it in terms of Green’s functions. We finish with a discussion in section \ref{discussion} and draw our conclusions in section \ref{conclusion}. Among the appendices, the first two contain more than just technical details of calculations, but we thought of them best suited as supplements to the main text. In Appendix \ref{diffgeo}.\ref{diffgeo1} we provide some basic notions of vector bundles and connections, and in \ref{diffgeo}.\ref{covDerAbuse} discuss some major abuses of notation widespread in the condensed matter and nonlinear optics literature. In Appendix \ref{crystal}.\ref{dynamics} we discuss electron dynamics within a periodic crystal and provide a derivation, within our framework, of the semiconductor Bloch equations, before moving on to a discussion of Blount’s position operator in \ref{crystal}.\ref{Blount}.\\

\section{Quantum evolution in a curved space}
\label{curvedSpace}

Let $\textbf{E}(t)$ be a spatially uniform, time-varying electric field. In the velocity gauge, it can be represented using the vector potential as $\textbf{E}=-\partial\textbf{A}(t)/\partial t$. The momentum-dependent equilibrium Hamiltonian $\mathcal{H}_0(\textbf{p})$ changes according to the minimal-coupling prescription to $\mathcal{H}_0(\textbf{p}-e\textbf{A}(t))\equiv\mathcal{H}_0(\textbf{p}(t))$, where we defined $\textbf{p}(t)=\textbf{p}-e\textbf{A}(t)$. It is thus clear, that the Hamiltonian can be considered as dependent on time only \textit{implicitly}, through the change of momentum. In light of this, we require a discussion of quantum evolution with a clear distinction between Hamiltonians containing implicit time-dependence through a parameter and/or explicit time-dependence.\\

For reasons that will become clear, in this section, we shall treat an abstract state vector $|\psi\rangle$ of a Hilbert space and the collection of its components $\ul{\psi}$ in a basis as differing objects in an explicit manner. For example, in a countable basis $\{|e_a\rangle\}$, we have

\begin{equation}
|\psi\rangle = \sum_a\psi_a|e_a\rangle \to \ul{\psi} = (\psi_1,\psi_2,\dots)^T.
\end{equation}

Of course, we could have also chosen a continuous basis. Similarly, the collection of components of an operator $\mathcal{O}$ in a basis shall be labeled as $\dul{\mathcal{O}}$.

\subsection{The extended Schrödinger equation}

Suppose we are given a Hamiltonian $\mc{H}(t)$ dependent on time explicitly, that is, not through a parameter. Let $\{|e_a(t)\rangle\}$ be a basis at each instant of time and define

\begin{equation}
\label{eq:DtDef}
\mathcal{D}_t|e_a(t)\rangle = \frac{i}{\hbar}\sum_b\mathcal{H}_{ba}(t)|e_b(t)\rangle.
\end{equation}

In other words, $\mathcal{D}_t$ is an operation that maps a state vector at $t$ to another state vector at $t$ such that its action on a basis state is determined by the Hamiltonian. Next, let the action of $\mathcal{D}_t$ on a component $\psi_a(t)$ of $|\psi(t)\rangle$ be the standard derivative, i.e., $\mathcal{D}_t\psi_a(t)\equiv \partial_t\psi_a(t)$, where we defined $\partial_t \equiv \partial/\partial t$ and let it satisfy the Leibniz rule. Then, we can consider the action of $\mathcal{D}_t$ on $|\psi(t)\rangle$ which yields

\begin{equation}
\label{eq:DtPsi}
\mathcal{D}_t|\psi(t)\rangle = \sum_a\left(\partial_t\psi_a(t)+\frac{i}{\hbar}\sum_b\mathcal{H}_{ab}(t)\psi_b(t)\right)|e_a(t)\rangle.
\end{equation}

We want $|\psi(t)\rangle$ to be determined by the condition

\begin{equation}
\label{eq:SchNoFrame}
\mathcal{D}_t|\psi(t)\rangle=0,
\end{equation}

which, by virtue of \eqref{eq:DtPsi}, means 

\begin{equation}
\label{eq:SchFrame}
i\hbar\frac{d}{dt}\ul{\psi}(t)=\dul{\mathcal{H}}(t)\ul{\psi}(t),
\end{equation}

and we have arrived at the standard Schrödinger equation. The Schrödinger equation should retain its form in a different basis. In order to check this, let us consider an \textit{explicitly} time-dependent change of basis using a unitary operator

\begin{equation}
|e_a(t)\rangle = \sum_b U_{ba}(t)|e_b^U(t)\rangle.
\end{equation}

The components of $|\psi(t)\rangle$ become $\ul{\psi}^U(t)=\dul{U}(t)\ul{\psi}(t)$ in the new basis and, from definition \eqref{eq:DtDef}, $\dul{\mathcal{H}}(t)$ changes to 

\begin{equation}
\label{eq:Htransform}
\dul{\mathcal{H}}^U(t)=\dul{U}(t)\dul{\mathcal{H}}(t)\dul{U}^{\dagger}(t)+i\hbar(\partial_t \dul{U}(t))\dul{U}^{\dagger}(t).
\end{equation}

As can be checked straightforwardly, the Schrödinger equation \eqref{eq:SchFrame} retains its form in the new basis. Reading off the components of $\mathcal{D}_t|\psi(t)\rangle$ from \eqref{eq:DtPsi}, we can see via the transformation property \eqref{eq:Htransform}, that they transform just as $\ul{\psi}(t)$, i.e.,

\begin{equation}
(\mathcal{D}_t|\psi(t)\rangle)^U_a=\sum_b U_{ab}(t)(\mathcal{D}_t|\psi(t)\rangle)_b.
\end{equation}

We can look at $\dul{U}(t)$ as a `passive' gauge-transformation, and $\mathcal{D}_t$ thus behaves as a covariant derivative along $t$ acting on elements of Hilbert spaces at each $t$. Indeed, we can consider a trivial Hilbert bundle $\pi_{\mathsf{H}}: \mathbb{R}\times\mathsf{H}\to\mathbb{R}$, where $\mathsf{H}$ is a Hilbert space and $\mathbb{R}$ represents time, such that its fibres $\pi^{-1}(t)=\mathsf{H}_t$ are copies of the same Hilbert space attached to each $t$ and the Hamiltonian corresponds to the local form of a connection on this Hilbert bundle, as detailed in \cite{AsoreyParallelTransport}. (see Appendix \ref{diffgeo}.\ref{diffgeo1} for a brief overview of Hilbert bundles).\\

\begin{figure}
\begin{center}
\includegraphics[width=8cm]{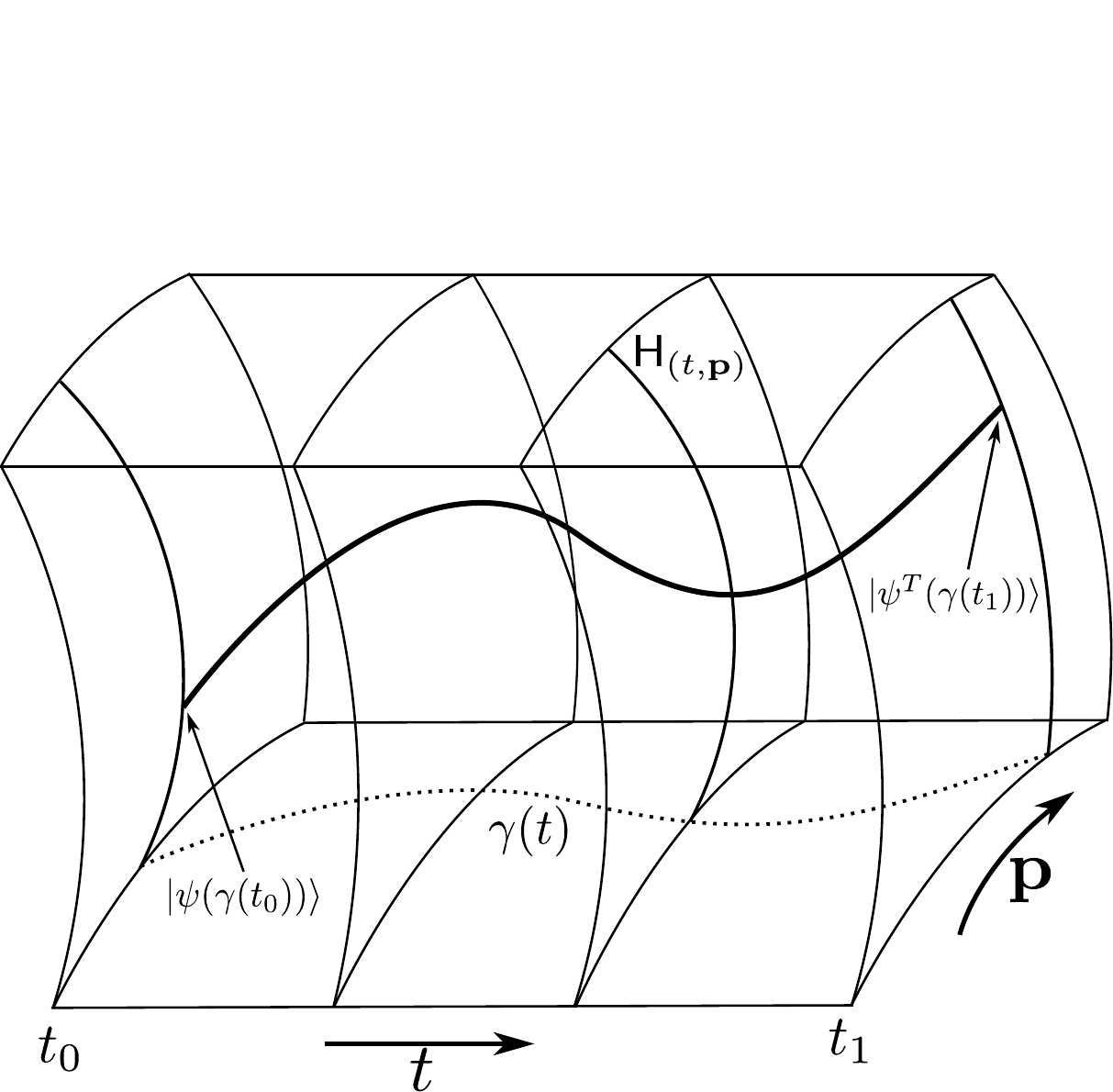}
\caption{Quantum evolution in a curved space. We have a Hilbert bundle over $(t,\textbf{p})$ with fibres $\mathsf{H}(t,\textbf{p})$ being Hilbert spaces. The state $|\psi(t_0,\textbf{p})\rangle = |\psi(\gamma(t_0))\rangle$ is parallel trans- ported via a connection with components $(\dul{\mathcal{H}}(t,\textbf{p}), \dul{\mathcal{A}_i}(t, \textbf{p}(t))$
and traces out a path $|\psi(\gamma(t))\rangle$ in the bundle, where $\gamma(t)$ is the corresponding path in the base space $(t, \textbf{p})$ \label{fig:curvedTp}}
\end{center}
\end{figure}

Suppose now that the system also depends on a set of parameters that we label as \textbf{p} (we assume that the parameter space is a smooth manifold). Then, as the system evolves in time, the latter's evolution traces out a path $(t,\textbf{p}(t))$ in time-parameter space (see FIG \ref{fig:curvedTp}). The question now, is how do we extend the Schrödinger equation \eqref{eq:SchFrame} to this case? From \eqref{eq:DtDef} and \eqref{eq:SchNoFrame} we see that the Hamiltonian is responsible for evolution in the time direction, meaning that we need another quantity that generates evolution in the parameter direction. In order to obtain this, let $\{|e_a(t,\textbf{p})\rangle\}$ be a local basis, or, in other words, a local frame of the Hilbert space attached to the point $(t,\textbf{p})$. Define

\begin{equation}
\label{eq:DiDef}
\mathcal{D}_i|e_a(t,\textbf{p})\rangle = \frac{i}{\hbar}\sum_b\mathcal{A}_{iba}(t,\textbf{p})|e_b(t,\textbf{p})\rangle,
\end{equation}

where $i$ ranges along the dimension of the parameter space and $\mathcal{A}_{iba}(t,\textbf{p})$ is a collection of components determining the operation $\mathcal{D}_i$ in a local frame. Furthermore, we demand that $\mathcal{D}_i$ act on component functions as $\mathcal{D}_i \psi_a(t,\textbf{p}) = \partial \psi_a(t, \textbf{p})/\partial p^i\equiv \partial_i\partial\psi_a(t, \textbf{p})$ and satisfy the Leibniz rule. Then, its action on a state vector becomes

\begin{equation}
\label{eq:DiPsi}
\begin{split}
\mathcal{D}_i|\psi(t,\textbf{p})\rangle = \sum_a\bigg(&\partial_i\psi_a(t,\textbf{p})\\
&+\frac{i}{\hbar}\sum_b\mathcal{A}_{iab}(t,\textbf{p})\psi_b(t,\textbf{p})\bigg)|e_a(t,\textbf{p})\rangle.
\end{split}
\end{equation}

Let $(t,\textbf{p}(t))$ be a path in time-parameter space with the tangent in the $p^i$ direction being $dp^i/dt$. We demand that the complete evolution of $|\psi(t,\textbf{p})\rangle$ along this path be determined by the condition

\begin{equation}
\label{eq:SchParNoFrame}
\mathcal{D}_t|\psi(t,\textbf{p}(t))\rangle+\frac{dp^i}{dt}\mathcal{D}_i|\psi(t,\textbf{p}(t))\rangle=0,
\end{equation}

where automatic summation is implied on $i$. Using \eqref{eq:DtPsi} and \eqref{eq:DiPsi} we can write this condition in a local frame as

\begin{equation}
\label{eq:SchParFrame}
i\hbar\frac{d}{dt}\ul{\psi}(t,\textbf{p}(t))=\left(\dul{\mathcal{H}}(t,\textbf{p}(t))+\frac{dp^i}{dt}\dul{\mathcal{A}_i}(t,\textbf{p}(t))\right)\ul{\psi}(t,\textbf{p}(t)),
\end{equation}

where $d/dt \equiv \partial_t + dp^i/dt \partial_i$ is the total derivative. We refer to this equation as the `extended' Schrödinger equation and shall provide a more elegant interpretation for it after the next few paragraphs below. This equation was also derived by Sardanashvily \cite{SardTimeDependentPar} using sophisticated techniques involving connections on composite bundles (see also ref. \cite{SardMechBook} for a more detailed exposition by the same author). In comparison, our approach is rather elementary, since our goal here is just to set up a consistent framework for perturbative response calculations and not delve into the depths of mathematical beauty. Note that when applying $\mathcal{D}_t$ to a component function $\psi_a(t,\textbf{p})$ with both time and parameter dependence, it is defined to act \textit{only} on the time argument, hence it acts as a \textit{partial} derivative with respect to time. Confusion might arise from the fact that the path along which the system evolves in time-parameter space is parameterized by time itself. We clarify this point in more detail below.\\
In order to see how \eqref{eq:SchParFrame} transforms under a local, unitary change of frame, or gauge-transformation, we first determine the transformation properties of $\dul{\mathcal{A}_i}(t,\textbf{p})$. The frame changes as

\begin{equation}
\label{eq:eParTransform}
|e_a(t,\textbf{p})\rangle = \sum_b U_{ba}(t,\textbf{p})|e_b^U(t,\textbf{p})\rangle,
\end{equation}

leading to the component functions transforming as $\ul{\psi}^U(t,\textbf{p})=\dul{U}(t,\textbf{p})\ul{\psi}(t,\textbf{p})$, which, by definition \eqref{eq:DiDef}, results in 

\begin{equation}
\label{eq:Aitransform}
\dul{\mathcal{A}^U_i}(t,\textbf{p})=\dul{U}(t,\textbf{p})\dul{\mathcal{A}_i}(t,\textbf{p})\dul{U}^{\dagger}(t,\textbf{p})+i\hbar(\partial_i \dul{U}(t,\textbf{p}))\dul{U}^{\dagger}(t,\textbf{p}).
\end{equation}

Performing the transformation along every point of the path we find that \eqref{eq:SchParFrame} retains its form in the new frame. Note that when extending the Hamiltonian's transformation rule \eqref{eq:Htransform} to the parameter-dependent case and performing it along every point of a path to see how the evolution equation \eqref{eq:SchParFrame} transforms, the Hamiltonian should be

\begin{equation}
\label{eq:HPartransform}
\begin{split}
\dul{\mathcal{H}^U}(t,\textbf{p}(t))=&\dul{U}(t,\textbf{p}(t))\dul{\mathcal{H}}(t,\textbf{p}(t))\dul{U}^{\dagger}(t,\textbf{p}(t))
\\
&+i\hbar(\partial_i \dul{U}(t,\textbf{p}(t)))\dul{U}^{\dagger}(t,\textbf{p}(t)),
\end{split}
\end{equation}

where we took the \textit{partial} derivative of the second term, since $\mathcal{D}_t$ acts only on the time argument of a component.\\

We have to provide an interpretation for $\dul{\mathcal{A}_i}(t,\textbf{p})$. Since they are components of the $\mathcal{D}_i$ operation, i.e., connection, in a local frame (see the definition \eqref{eq:DiDef}), suppose that there exists a local frame $\{|e^{\text{F}}_a (t, \textbf{p})\rangle\}$ in which they all vanish identically, i.e., $\dul{\mathcal{A}^{\text{F}}_i}(t, \textbf{p}) = 0$. By \eqref{eq:DiDef}, this means that, in this particular frame, $\mathcal{D}_i |e^{\text{F}}_a (t, \textbf{p})\rangle = 0$ for all $a$, so \eqref{eq:DiPsi} becomes

\begin{equation}
\label{eq:DiPsiF}
\begin{split}
\mathcal{D}_i|\psi(t,\textbf{p})\rangle = \sum_a\partial_i\psi^{\text{F}}_a(t,\textbf{p})|e^{\text{F}}_a(t,\textbf{p})\rangle,
\end{split}
\end{equation}

and $\mathcal{D}_i$ acts as a partial derivative on the component functions. The extended Schrödinger equation \eqref{eq:SchParFrame} reduces to

\begin{equation}
\label{eq:SchParFrameF}
i\hbar\frac{d}{dt}\ul{\psi}^{\text{F}}(t,\textbf{p}(t))=\dul{\mathcal{H}}^{\text{F}}(t,\textbf{p}(t))\ul{\psi}^{\text{F}}(t,\textbf{p}(t)),
\end{equation}

which is the standard Schrödinger equation for a Hamiltonian with both explicit time-dependence and a time-dependent parameter. It is clear that the validity of this equation rests on the existence of the special frame $\{|e^{\text{F}}_a(t, \textbf{p})\rangle\}$ in which all the components $\dul{\mathcal{A}^{\text{F}}_i} (t, \textbf{p})$ vanish and is thus a special case of the more general extended one \eqref{eq:SchParFrame}. We show in Appendix \ref{crystal}.\ref{dynamics} that a realization of a special case of equation \eqref{eq:SchParFrameF} with a Hamiltonian lacking explicit time-dependence is provided by the evolution under a spatially uniform electric field of periodic Bloch states in a crystal, with $\textbf{p} \to \hbar\textbf{k}$, where \textbf{k} is the crystal momentum, and the Hamiltonian corresponding to the Bloch Hamiltonian $\dul{\mc{H}}(t,\textbf{p}(t))\to\dul{\mc{H}}(\textbf{k}(t))$, where $\textbf{k}(t)\equiv\textbf{k}-\frac{ie}{\hbar}\textbf{A}(t)$ with $\textbf{A}(t)$ being the vector potential.\\

Now let us change the frame by a unitary transformation according to \eqref{eq:eParTransform}. Taking a glance at \eqref{eq:Aitransform}, we find that $\dul{\mathcal{A}}^{\text{F}}_i(t, \textbf{p})=0$ transforms to

\begin{equation}
\label{eq:AiFU}
\dul{\mathcal{A}_i}^{\text{F},U}(t,\textbf{p})=i\hbar(\partial_i \dul{U}(t,\textbf{p}))\dul{U}^{\dagger}(t,\textbf{p}),
\end{equation}

which is clearly non-vanishing. As expected, this results in the fact that equation \eqref{eq:SchParFrameF} does not retain its form under a change of frame. On the other hand, $\dul{\mathcal{A}_i}^{\text{F},U}(t, p)$ is not an undetermined quantity, rather, it is a definite expression given in terms of the unitary transformation used to change the frame. In the special case of a crystal, we can consider a change of frame $\dul{U}^{\dagger}(\textbf{k})$, that is not \textit{explicitly} time-dependent, such that the Bloch Hamiltonian becomes diagonal. In this case, $\dul{U}^{\dagger}(\textbf{k})$ simply contains the components of the periodic Bloch states $|u_a(\textbf{k})\rangle$ written in the ‘F-frame’ (see Appendix \ref{crystal}.\ref{dynamics}), and consequently $\dul{\mathcal{A}_i}^{\text{F},U}(\textbf{k})$ correspond to components of the flat Berry connection.\\

It would be conducive to provide a quantity, dependent on the components $\dul{\mc{A}_i}(t,\textbf{p})$, with the following property: should it vanish in one frame, it should go on to vanish in all frames, i.e., it should transform covariantly. Indeed, looking at \eqref{eq:DiPsiF} and \eqref{eq:DiPsi}, the only difference between them is that in the former we only find the partial derivative whereas in the latter we also stumble into the components $\dul{\mathcal{A}_i}(t,\textbf{p})$. Since the partial derivatives in different directions commute, i.e., $\partial_i \partial_j = \partial_j \partial_i$, we can quantify the importance of $\dul{\mathcal{A}_i}(t,\textbf{p})$ by checking the commutation of $\mathcal{D}_i$ in different directions. Using \eqref{eq:DiDef} we obtain the standard result

\begin{equation}
\label{eq:DiDk}
[\mathcal{D}_i,\mathcal{D}_k]|e_a(t,\textbf{p})\rangle = \frac{i}{\hbar}\sum_b\left(\dul{\mathcal{F}_{ik}}(t,\textbf{p})\right)_{ba}|e_b(t,\textbf{p})\rangle,
\end{equation}

where

\begin{equation}
\label{eq:curvature}
\dul{\mathcal{F}_{ik}}=\partial_{i}\dul{\mathcal{A}_k}-\partial_{k}\dul{\mathcal{A}_i}+\frac{i}{\hbar}\left[\dul{\mathcal{A}_i},\dul{\mathcal{A}_k}\right],
\end{equation}

and we dropped the explicit arguments for clarity. These are the components of the local form of the curvature of a connection on a vector bundle \cite{Nakahara,FrankelGOPh,TuDiffGeo} and they transform covariantly under a change of frame: $\dul{\mathcal{F}^U_{ik}}=\dul{U}\dul{\mathcal{F}_{ik}}\dul{U}^{\dagger}$. Thus, the special local frame $\{|e^{\text{F}}_a(t,\textbf{p})\rangle\}$ in which the local components of the connection vanish $\dul{\mathcal{A}^{\text{F}}_i}(t,\textbf{p})=0$ and evolution is described by \eqref{eq:SchParFrameF}, only exists if the curvature components vanish $\dul{\mathcal{F}_{ik}}=0$; in other words, if the connection is flat. However, note that, as can be checked by utilizing \eqref{eq:curvature}, \eqref{eq:AiFU} also has vanishing curvature. Indeed, not all bundles admit flat connections (a notable example being the tangent bundle of the 2-sphere $S^2$ \cite{FrankelGOPh}), but even if we look at a bundle that does, such as the bundle of periodic Bloch states over the Brillouin torus, we will not necessarily be able to choose a continuous frame defined over the entire bundle in which the connection components vanish, rather, we might be able to construct such a frame only locally, and use compatibility relations to find that the components will take the form \eqref{eq:AiFU} over other parts \cite{FrankelGOPh} (see also Appendix \ref{diffgeo}.\ref{diffgeo1}).\\

The extended Schrödinger equation \eqref{eq:SchParFrame} can thus describe quantum evolution in a Hilbert bundle with a non-flat connection in the parameter direction and is the equation we shall use for our calculation of responses. Before we see how such a curved connection can actually arise, we provide an interpretation for the frame-independent form \eqref{eq:SchParNoFrame} of the extended Schrödinger equation.\\

The standard Schrödinger equation \eqref{eq:SchFrame} is \eqref{eq:SchNoFrame} written in a frame. The latter equation simply expresses that the state $|\psi(t)\rangle$ does not change with respect to the covariant derivative in the time direction $\mathcal{D}_t$; it is covariantly constant. We can also provide a similar interpretation for the extended equation. Indeed, taking a glance at \eqref{eq:SchParNoFrame}, we are prompted to define $\mathcal{D}_0 \equiv \mathcal{D}_t$ and $\mathcal{D}_{\mu} \equiv (\mathcal{D}_0,\,\mathcal{D}_i)$. Similarly, we write $P^{\mu}(t)\equiv (t, p^i(t))$ for points of a path in $(t,\textbf{p})$ space parameterized by $t$ and $dP^{\mu}/dt \equiv (1,dp^i/dt)$ for the tangent to this path. Using this notation, \eqref{eq:SchParNoFrame} reduces to

\begin{equation}
\label{eq:SchParParallel}
\frac{dP^{\mu}}{dt}\mathcal{D}_{\mu}|\psi(P(t))\rangle=0.
\end{equation}

In other words, the extended Schrödinger equation states that $|\psi(t,\textbf{p})\rangle$ should evolve along a path in such a way that it remain covariantly constant along this path. In the case that will be of interest to us, the parameter will correspond to a momentum and the path will be determined by an externally applied electric field. Hence, there is no need for concern regarding the path-dependence of this formula.\\
The combined connection components $\dul{\mathcal{A}_{\mu}}(P) \equiv (\dul{\mathcal{H}}(P),\dul{\mathcal{A}_i}(P))$ lead us to the following remark. When
the Hamiltonian $\dul{\mc{H}}$ is looked at in isolation and considered as the `matrix' of components of a connection on a Hilbert bundle over time, it necessarily corresponds to a connection that is flat (time is 1-dimensional so we only have a covariant derivative in the $t$ direction which commutes with itself, cf. \eqref{eq:DiDk}). However, when thought of as a vector component of $\dul{\mc{A}_{\mu}}(P)$, then even if the curvature \eqref{eq:curvature} corresponding to the `latin index' part $\dul{\mc{A}_i}(P)$ vanishes, in other words, the connection corresponding to the covariant derivative in the parameter direction is flat, the total connection with components $\dul{{A}_{\mu}}(P)$ is not necessarily flat, since the curvature components

\begin{equation}
\label{eq:curvature0i}
\dul{\mathcal{F}_{0i}}=\partial_{t}\dul{\mathcal{A}_i}-\partial_{i}\dul{\mathcal{H}}+\frac{i}{\hbar}\left[\dul{\mathcal{H}},\dul{\mathcal{A}_i}\right],
\end{equation}

corresponding to $[\mc{D}_0,\mc{D}_i]$ do not necessarily vanish. The Hamiltonian, in this sense, is not part of a flat connection. In fact, the components $\dul{\mc{F}_{0i}}$ are proportional to those of the velocity operator. We expand on this in section \ref{velocity}.\\
By construction, Equation \eqref{eq:SchParParallel} is frame-independent, however, time is still handled in an awkward manner and not treated on an equal footing with the other parameters. This is, of course, due to the choice of parameterization of the path in $(t,\textbf{p})$. We can simply parameterize the path with another parameter, say, $\tau$, leading to $P^{\mu}(\tau) = (t(\tau), p^i(\tau))$ with tangent vector $dP^{\mu}/d\tau = (dt/d\tau,dp^i/d\tau)$. The extended Schrödinger equation \eqref{eq:SchParParallel} written out along this path is then

\begin{equation}
\label{eq:SchParParallelExpand}
\frac{dP^{\mu}}{d\tau}\mathcal{D}_{\mu}|\psi(P(\tau))\rangle=\left(\frac{dt}{d\tau}\mathcal{D}_{0}+\frac{dp^i}{d\tau}\mathcal{D}_{i}\right)|\psi(P(\tau))\rangle=0.
\end{equation}

It is thus clear that the extended Schrödinger equation \eqref{eq:SchParNoFrame} corresponds to the special case of $dt/d\tau = 1$.\\
We can generalize Eq. \eqref{eq:SchParParallelExpand} further. Suppose we choose the parameter to be space proper, i.e., $P^{\mu} = (t, p^i) \to x^{\mu} = (t, x^i)$ is space-time (we take $c=1$). Then, we have a Hilbert space of states containing $|\psi(x)\rangle$ defined at each point $x$ of spacetime and the parallel transport equation

\begin{equation}
\label{eq:SchParParallelST}
\frac{dx^{\mu}}{d\tau}\mathcal{D}_{\mu}|\psi(x(\tau))\rangle=0,
\end{equation}

over a path in spacetime parameterized by $\tau$ lifted to the tangent bundle with tangents $dx^{\mu}/d\tau$, is manifestly (general) relativistic. Note that if we let $g_{\mu\nu}(x)$ be the components of a metric on the tangent bundle of the spacetime manifold and use it to fix the tangent frames at each spacetime point to be orthonormal, then we can expand the product in \eqref{eq:SchParParallelST} with connection components labelled $\dul{\mathcal{P}_{\mu}}(x) \equiv (\dul{\mathcal{H}}(x),\dul{\mathcal{P}_i}(x))$ in a manner similar to \eqref{eq:SchParParallelExpand}  ($g_{\mu\nu}(x)$ can be written locally as the Minkowski metric). Equation \eqref{eq:SchParParallelST} simply expounds that the quantum state should evolve over a spacetime path in such a way that it remain covariantly constant. Following an appropriate interpretation of the paths and connection components, this turns out to be a fruitful approach to describing relativistic quantum evolution \cite{BonbienRelat}.\\

\begin{figure}[h]
\begin{center}
        \includegraphics[width=7cm]{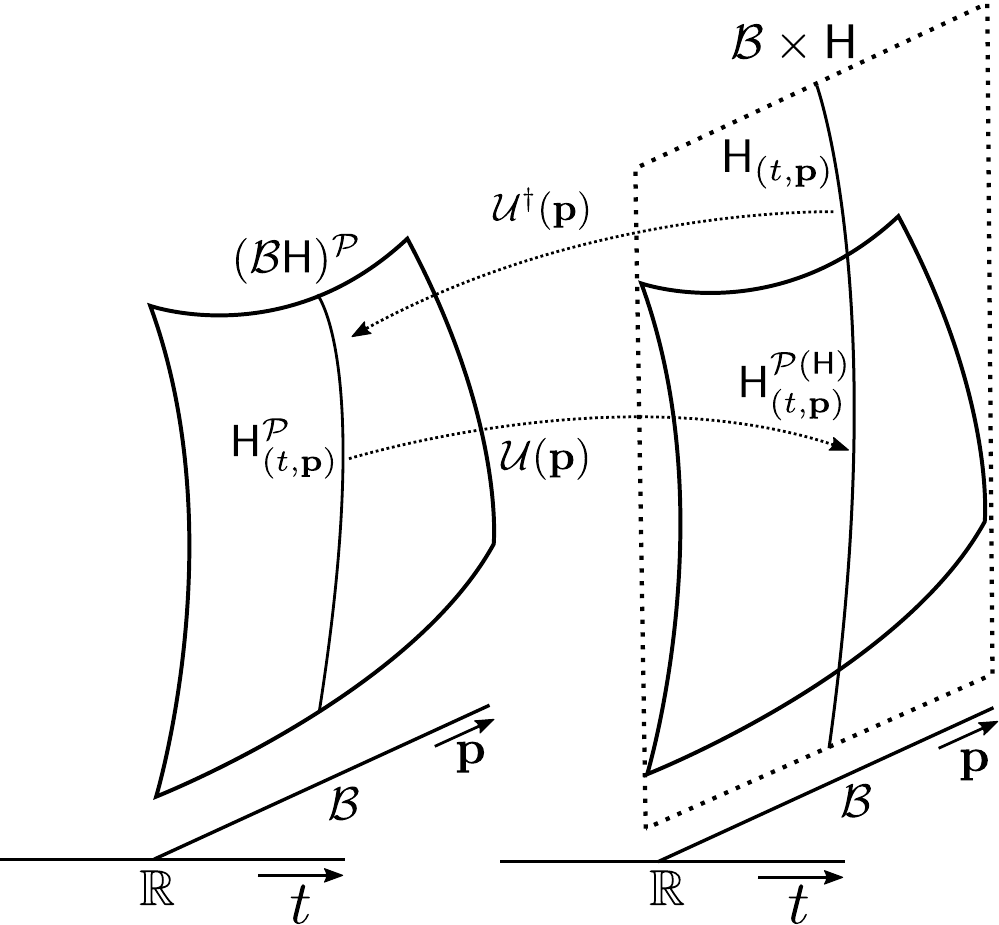}
      \caption{\label{fig:projectionmaps}}
\end{center}
\end{figure}

A natural question to ask is: how could a non-flat connection on the trivial Hilbert bundle over parameter space with components $\dul{\mc{A}_i}$ arise? For the purposes of this paper, we look at a standard construction that we adapt from \cite{AtiyahYangMills}, also discussed in \cite{BonbienRev2021} for the special case of periodic Bloch states. Suppose we have a state $|\psi(t,\textbf{p})\rangle$ as a local section of a trivial Hilbert bundle $\pi:\mathbb{R}\times\mathcal{B}\times\mathsf{H}\to \mathbb{R}\times\mathcal{B}$, where $\mc{B}$ is the parameter space and $\mathbb{R}$ refers to time. The fibres of this bundle at points $(t, \textbf{p}) \in \mathbb{R}\times \mc{B}$ are labeled as the Hilbert spaces $\mathsf{H}_{(t,\textbf{p})} \equiv \pi^{-1}(t,\textbf{p})$, moreover, we consider a local frame in which the connection components in the time direction are given by the Hamiltonian and in the parameter direction are vanishing. Note that this does not necessarily imply that the connection in the part of the bundle over parameter space $\mc{B}$ is trivial---unless $\mc{B}$ is contractible; we could very well have a flat connection that is nevertheless non-trivial, but we are looking at a local patch over parameter space in which the connection components vanish (see Appendix \ref{diffgeo}.\ref{diffgeo1} for a discussion of this). Now consider a sub-bundle whose fibres in a local trivialization are $\mathsf{H}^{\mc{P}}_{(t,\textbf{p})}$ obtained via a parameter-dependent projection $\mc{P}(\textbf{p})$ to a subspace of $\mathsf{H}_{(t,\textbf{p})}$. Let $\mathsf{H}^{\mc{P}(\mathsf{H})}_{(t,\textbf{p})} \subset \mathsf{H}(t,\textbf{p})$ be the projected subspace $\mathsf{H}^{\mc{P}}_{(t,\textbf{p})}$ considered as embedded within $\mathsf{H}_{(t,\textbf{p})}$ (see FIG. \ref{fig:projectionmaps}) and define the map $\mc{U}(\textbf{p}): \mathsf{H}_{(t,\textbf{p})}\to \mathsf{H}^{\mc{P}(\mathsf{H})}_{(t,\textbf{p})} \subset \mathsf{H}_{(t,\textbf{p})}$ together with $\mc{U}^{\dagger}(\textbf{p}): \mathsf{H}_{(t,\textbf{p})}\to \mathsf{H}^{\mc{P}}_{(t,\textbf{p})}$ such that $\mc{U}^{\dagger}(\textbf{p})\mc{U}(\textbf{p}) = \text{id}^{\mc{P}}_{\textbf{p}}$, the identity map on the projected space and $\mc{P}(\textbf{p}) \equiv \mc{U}(\textbf{p})\mc{U}^{\dagger}(\textbf{p}): \mathsf{H}_{(t,\textbf{p})}\to \mathsf{H}^{\mc{P}(\mathsf{H})}_{(t,\textbf{p})}$ an orthogonal projection. Note that we do not consider the projection $\mc{P}(\textbf{p})$ to be dependent on time explicitly. This results in the fact that the projected sub-bundle remains trivial with respect to time $\pi^{\mc{P}}:\mathbb{R}\times(\mathcal{B}\mathsf{H})^{\mc{P}}\to\mathbb{R}\times\mathcal{B}$, where $(\mc{B}\mathsf{H})^{\mc{P}}$ is the total space of the non-trivial bundle $\pi_t^{\mc{P}}: (\mc{B}\mathsf{H})^{\mc{P}} \to \mc{B}$, but only trivializes locally with respect to the parameter space---unless the parameter space $\mc{B}$ is contractible, in which case it is trivial too. As we show below, this will lead to a purely parameter-dependent non-flat connection on the $\pi_t^{\mc{P}}$ bundle, whose components in a local frame cannot be made to vanish any longer.\\

We can now use the maps $\mc{U},\,\mc{U}^{\dagger}$ to construct a derivative on the projected subspace $\mathsf{H}^{\mc{P}}_{(t,\textbf{p})}$. In order to do this,
consider a frame $\{|e^{\mc{P}}_{\alpha}(t,\textbf{p})\rangle\}$ of $\mathsf{H}^{\mc{P}}_{(t,\textbf{p})}$ defined locally in a neighbourhood of $(t,\textbf{p})$. We can use $\mc{U}(\textbf{p})$ to map this into $\mathsf{H}_{(t,\textbf{p})}$ with local frame $\{|e^{\text{F}}_a (t,\textbf{p})\rangle\}$ with vanishing connection components $\mc{D}_i|e^{\text{F}}_a (t, \textbf{p})\rangle = 0$, since the latter is a local frame in a trivial bundle $\pi_t: \mc{B}\times\mathsf{H}\to \mc{B}$ with a flat connection:

\begin{equation}
\label{eq:UprojF}
\mc{U}(\textbf{p})|e^{\mc{P}}_{\alpha}(t,\textbf{p})\rangle=\sum_a \mc{U}_{a\alpha}(\textbf{p})|e^{\text{F}}_a (t,\textbf{p})\rangle.
\end{equation}

Similarly, we have

\begin{equation}
\label{eq:UFproj}
\mc{U}^{\dagger}(\textbf{p})|e^{\text{F}}_a (t,\textbf{p})\rangle=\sum_{\alpha} \mc{U}^{*}_{a\alpha}(\textbf{p})|e^{\mc{P}}_{\alpha}(t,\textbf{p})\rangle.
\end{equation}

Since the result of \eqref{eq:UprojF} is in $\mathsf{H}^{\mc{P}(\mathsf{H})}_{(t,\textbf{p})}\subset \mathsf{H}_{(t,\textbf{p})}$ we can apply the operation $\mc{D}_i$ to it and use $\mc{P}(\textbf{p}) = \mc{U}(\textbf{p})\mc{U}^{\dagger}(\textbf{p})$ to project the result onto $\mathsf{H}^{\mc{P}(\mathsf{H})}_{(t,\textbf{p})}$ allowing us to read-off the connection components within the projected subspace $\mathsf{H}^{\mc{P}}_{(t,\textbf{p})}$. Indeed, we have

\begin{equation}
\begin{split}
\mc{U}(\textbf{p})&\mc{U}^{\dagger}(\textbf{p})\mc{D}_i\mc{U}(\textbf{p})|e^{\mc{P}}_{\alpha}(t,\textbf{p})\rangle
\\
&=\mc{U}(\textbf{p})\mc{U}^{\dagger}(\textbf{p})\sum_a \partial_i\mc{U}_{a\alpha}(\textbf{p})|e^{\text{F}}_a (t,\textbf{p})\rangle
\\
&=\mc{U}(\textbf{p})\sum_{\beta}\left(\sum_a\mc{U}^{*}_{a\beta}(\textbf{p})\partial_i\mc{U}_{a\alpha}(\textbf{p})\right)|e^{\mc{P}}_{\beta}(t,\textbf{p})\rangle,
\end{split}
\end{equation}

where we used the fact that $\mc{D}_i|e^{\text{F}}_a (t,\textbf{p})\rangle=0$ and \eqref{eq:UFproj}. The derivative in the projected subspace thus becomes

\begin{equation}
\label{eq:DiProj}
\mathcal{D}_i|e^{\mc{P}}_{\alpha}(t,\textbf{p})\rangle = \frac{i}{\hbar}\sum_{\beta}\mathcal{A}^{\mc{P}}_{i\beta\alpha}(\textbf{p})|e^{\mc{P}}_{\beta}(t,\textbf{p})\rangle,
\end{equation}

where the connection components are

\begin{equation}
\label{eq:AiProj}
\mathcal{A}^{\mc{P}}_{i\beta\alpha}(\textbf{p})=-i\hbar\sum_a\mc{U}^{*}_{a\alpha}(\textbf{p})\partial_i\mc{U}_{a\beta}(\textbf{p}).
\end{equation}

Note that in order to get a non-flat connection, $\dul{\mc{F}^{\mc{P}}_{\alpha\beta}}(\textbf{p})\neq 0$ (see Eq. \eqref{eq:curvature} with $\mc{A}_{iab}\to\mc{A}^{\mc{P}}_{i\alpha\beta}$), it is crucial for the frame labeled by greek indices to span a \textit{sub}space of the space spanned by the frame with latin indices, i.e., that we consider a `proper' projection, otherwise, we get the degenerate case of $\mc{U}$ being a unitary transformation, thereby relating the two frames according to \eqref{eq:eParTransform} (with $\mc{U}\to U^{\dagger}$). Given that the connection components vanish in the latin index frame (we took the latin indices to label elements of the `F-frame'), they would become \eqref{eq:AiFU} (with $\mc{U}\to \mc{U}^{\dagger}$) in the greek index frame and would continue to be the components of a flat connection which can be simply made to locally vanish by moving to a different local frame.\\

We can also see the curved connection arising directly, by looking at how the extended Schrödinger equation \eqref{eq:SchParFrameF} in the F-frame changes form when restricted to the projected subspace. Suppose we have a state $|\psi^{\mc{P}}(t,\textbf{p})\rangle\in \mathsf{H}^{\mc{P}}(t,\textbf{p})$ in the projected subspace and let us apply the map $\mc{U}(\textbf{p})$ to move it to the embedded projected subspace $\mathsf{H}^{\mc{P}(\mathsf{H})}_{(t,\textbf{p})} \subset \mathsf{H}(t,\textbf{p})$ as $|\psi(t,\textbf{p})\rangle=\mc{U}(\textbf{p})|\psi^{\mc{P}}(t,\textbf{p})\rangle$. The state $|\psi(t,\textbf{p})\rangle\in \mathsf{H}^{\mc{P}(\mathsf{H})}_{(t,\textbf{p})}$  obtained in this way clearly satisfies $|\psi(t,\textbf{p})\rangle=\mc{P}(\textbf{p})|\psi(t,\textbf{p})\rangle$, since $\mc{P}(\textbf{p})=\mc{U}(\textbf{p})\mc{U}^{\dagger}(\textbf{p})$. In the embedded space $\mathsf{H}^{\mc{P}(\mathsf{H})}_{(t,\textbf{p})}$ we have access to the F-frame and, using \eqref{eq:UprojF} to find the corresponding frame in the projected subspace, we can also relate the corresponding components of the states via
\begin{equation}
\psi_a^{\text{F}}(t,\textbf{p})=\sum_{\alpha}\mc{U}_{a\alpha}(\textbf{p})\psi^{\mc{P}}_{\alpha}(t,\textbf{p}).
\end{equation}

Writing this into \eqref{eq:SchParFrameF}, defining the projected subspace Hamiltonian

\begin{equation}
\mc{H}^{\mc{P}}_{\alpha\beta}(t,\textbf{p})=\sum_{a,b}\mc{U}^*_{a\alpha}(\textbf{p})\mc{H}^{\text{F}}_{ab}(t,\textbf{p})\mc{U}_{b\beta}(\textbf{p}),
\end{equation}

recognizing the curved connection components \eqref{eq:AiProj}, and rearranging, we obtain

\begin{equation}
\begin{split}
&i\hbar\frac{d}{dt}\ul{\psi}^{\mc{P}}(t,\textbf{p}(t))
\\
&=\left(\dul{\mc{H}}^{\mc{P}}(t,\textbf{p}(t))+\frac{dp^i}{dt}\dul{\mc{A}_i^{\mc{P}}}(\textbf{p}(t))\right)\ul{\psi}^{\mc{P}}(t,\textbf{p}(t)),
\end{split}
\end{equation}

which is the extended Schrödinger equation \eqref{eq:SchParFrame} in a local frame $\{|e^{\mc{P}}_{\alpha}(t,\textbf{p})\rangle\}$ of the projected subspace.\\

Throughout this construction we have only considered the projection to a generic subspace of $\mathsf{H}_{(t,\textbf{p})}$. However, there are certain distinguished subspaces, such as the eigensubspaces of the Hamiltonian. Should the projection be onto an isolated 1-dimensional eigenspace of the Hamiltonian, \eqref{eq:AiProj} becomes the Abelian Berry connection, on the other hand, should it be onto a well-separated multi-dimensional eigenspace—--possibly containing isolated 1-dimensional eigenspaces—--we get the more general non-Abelian Berry connection \cite{AvronBerry}.\\

In general, we can consider a truncation of the Hilbert space $\mathsf{H}_{(t,\textbf{p})}$ via a projection to a subspace and this is the perspective we would like to emphasize. The connection components can be looked at as representing the data contained within the truncated degrees of freedom on the subspace under investigation. Thus, we can choose to work within the subspace by sacrificing the flatness of the connection in the paramater direction. Should we tread this path, we clearly have to use the extended Schrödinger equation \eqref{eq:SchParFrame} to describe quantum evolution, since the standard one \eqref{eq:SchParFrameF} assumes a flat connection. However, the effort is worthwhile since we shall find that going beyond first order in perturbative response calculations, path-dependence begins to matter when integrating the extended Schrödinger equation \eqref{eq:SchParFrame} and, compared to the flat case, this changes the structure of the response functions (see section \ref{secondorderresponse} below).

\subsection{The parallel transporter and infinitesimal expansions}
\label{Taylor}

The extended Schrödinger equation \eqref{eq:SchParParallel} asserts that the quantum state $|\psi(P)\rangle$ evolves in such a way that it remains covariantly constant along the path $\gamma(t): P^{\mu}(t)=(t,\textbf{p}(t))$ with tangent $\dot{\gamma}(t): dP^{\mu}/dt = (1,dp^i/dt)$. We can thus think of the operator that evolves the state as a parallel transporter. Let the initial time be $t_0$ and we define the parallel transporter $T(\gamma_{t_0\to t})$ that evolves the quantum state from $P(t_0)$ to $P(t)$ along the path $\gamma$ as

\begin{equation}
\label{eq:TransporterDef}
|\psi^T_{\gamma}(P(t))\rangle = T(\gamma_{t_0\to t})|\psi(P(t_0))\rangle.
\end{equation}

The components of the parallel transporter in a local frame can be defined as follows. Suppose $\{|e_a(P(t_0))\rangle\}$ is a frame at $P(t_0)$. Since $T(\gamma_{t_0\to t})$ maps this frame to the Hilbert space at $P(t)$ we should be able to expand the result in a frame at $P(t)$:

\begin{equation}
T(\gamma_{t_0\to t})|e_a(P(t_0))\rangle = \sum_b T_{ba}(\gamma_{t_0\to t})|e_b(P(t))\rangle,
\end{equation}

which means that \eqref{eq:TransporterDef} can be written in component form as

\begin{equation}
\label{eq:TransporterFrame}
\ul{\psi}^T_{\gamma}(P(t))=\dul{T}(\gamma_{t_0\to t})\ul{\psi}(P(t_0)).
\end{equation}

We can obtain an expression for these components of the parallel transporter by writing the extended Schrödinger equation \eqref{eq:SchParParallel} in a local frame, and, the result being an ordinary differential equation in time, integrating it along $\gamma(t)$. Indeed, we have for \eqref{eq:SchParParallel} in a local frame

\begin{equation}
\frac{d}{dt}\ul{\psi}(P(t)+\frac{i}{\hbar}\frac{dP^{\mu}}{dt}\dul{\mathcal{A}_{\mu}}(P(t))\ul{\psi}(P(t))=0,
\end{equation}

where we defined $\dul{\mathcal{A}_{\mu}}\equiv (\dul{\mathcal{H}},\dul{\mc{A}_i})$, and, using the standard iterative procedure, obtain

\begin{equation}
\label{eq:TransporterExp}
\dul{T}(\gamma_{t_0\to t})=\mathtt{T}\,\exp\left(-\frac{i}{\hbar}\int_{t_0}^t dt'\,\dul{\mathcal{A}_{\mu}}(P(t'))\frac{dP^{\mu}}{dt'}\right),
\end{equation}

where $\mathtt{T}$ is the time-ordering symbol. Furthermore since $\dul{\mc{A}_{\mu}}$ is Hermitian, the parallel transporter is unitary, meaning that it preserves the modulus of the states it transports. Note that, fundamentally, this is a path-ordered expression, but due to choosing time as our parameter—--under the assumption that time only goes forward--—we are left with a time-ordered expression. In its form, the parallel transporter is similar to Dyson’s time-ordered evolution operator, however it is fundamentally different: our state is being transported not only over time, but also in the non-trivial Hilbert bundle over parameter space, meaning that the parallel transporter is heavily path-dependent. For the specific case of interest to us, as alluded to earlier, the choice of path is something we will not have to be concerned with since we shall interpret the tangent $dp^i/dt$ in the parameter direction as a known external classical field, the electric field, and a path is thereby handed to us.\\

It is important to know how the components of the parallel transporter transform under a local change of frame, or `passive' gauge transformation $\dul{U}(P(t))$. This can be deduced in a straightforward manner from \eqref{eq:TransporterFrame}. We want the transformed components to be connected by the transformed parallel transporter which means

\begin{equation}
\dul{U}(P(t))\dul{T}(\gamma_{t_0\to t})\ul{\psi}(P(t_0))=\dul{T}^U(\gamma_{t_0\to t})\dul{U}(P(t_0))\ul{\psi}(P(t_0)),
\end{equation}

and we thus have

\begin{equation}
\label{eq:TransporterTransform}
\dul{T}^U(\gamma_{t_0\to t})=\dul{U}(P(t))\dul{T}(\gamma_{t_0\to t})\dul{U}^{\dagger}(P(t_0)),
\end{equation}

a relation that will be of crucial importance. Using the parallel transporter we can now gain some intuition for the covariant derivative defined earlier. Consider the infinitesimal transporter that transports from point $P(t_0)\equiv P + dP$ to another point $P(t) \equiv P$. That is, we take \eqref{eq:TransporterExp} along an infinitesimal path

\begin{equation}
\label{eq:TransporterInf}
\dul{T}(P,P+dP)=\dul{1}+\frac{i}{\hbar}\dul{\mc{A}_{\mu}}dP^{\mu},
\end{equation}

where $\dul{T}(P,P+dP)$ refers to components of the parallel transporter that moves states along an infinitesimal path from $P +dP$ to $P$. We wish to compare states at $P$ and $P + dP$, however, the states at these two points live in different Hilbert spaces. Hence, the comparison is only possible if we transport the state from the space at the latter point to the one at the former point. Doing this we can compare their components in a local frame

\begin{equation}
\label{eq:DTransportFrame}
\begin{split}
&\ul{\psi}^{T}_{dP}(P)-\ul{\psi}(P)
\\
&\quad=\dul{T}(P,P+dP)\ul{\psi}(P+dP)-\ul{\psi}(P)
\\
&\quad\approx dP^{\mu}\left(\partial_{\mu}\ul{\psi}(P)+\frac{i}{\hbar}\dul{\mathcal{A}_{\mu}}\ul{\psi}(P)\right),
\end{split}
\end{equation}

where the $dP$ subscript refers to the fact that we are transporting along an infinitesimal path; we also expanded the \textit{component functions} to first order in $dP$. This corresponds to the frame-independent expression

\begin{equation}
\label{eq:DTransportNoFrame}
\begin{split}
&|\psi^T_{dP}(P)\rangle-|\psi(P)\rangle=dP^{\mu}\mathcal{D}_{\mu}|\psi(P)\rangle,
\end{split}
\end{equation}

which is another interpretation of the covariant derivative $\mc{D}_{\mu}$.

At this point we are compelled to say a few words about operators, such as certain observables, acting on states. Let $\mc{O}(P)$ be an operator acting on the Hilbert space at $P$ (more precisely, we could define it as a local section of the bundle of operators acting on the Hilbert bundle, but we do not venture further into mathematical details, as it is not the purpose of our paper). As usual, we can define a covariant derivative on an operator based on the Leibniz rule

\begin{equation}
\label{eq:DopDef}
\begin{split}
(\mc{D}_{\mu}\mc{O}(P))|\psi(P)\rangle &= \mc{D}_{\mu}(\mc{O}(P)|\psi(P)\rangle)-\mc{O}(P)\mc{D}_{\mu}|\psi(P)\rangle
\\
&\equiv [\mc{D}_{\mu},\mc{O}(P)]|\psi(P)\rangle.
\end{split}
\end{equation}

The operator at $P$ acts on states at $P$. Hence it maps an element of a local frame at $P$ to a linear combination of local frame elements at $P$. This means that we can give the operator’s components in a local frame as

\begin{equation}
\label{eq:opFrame}
\mc{O}(P)|e_a(P)\rangle=\sum_b\mc{O}_{ba}(P)|e_b(P)\rangle,
\end{equation}

and, after a straightforward calculation, find the components of the operator's covariant derivative \eqref{eq:DopDef} to be

\begin{equation}
\label{eq:DopFrame}
\dul{\mc{D}_{\mu}\mc{O}}=\partial_{\mu}\dul{\mc{O}}+\frac{i}{\hbar}\left[\dul{\mc{A}_{\mu}},\dul{\mc{O}}\right],
\end{equation}

where we dropped the $P$ arguments for clarity. Now let us change frames. Using \eqref{eq:eParTransform} and \eqref{eq:opFrame} we find that, as expected, the components of the operator become $\dul{\mc{O}}^U(P) = \dul{U}(P)\dul{\mc{O}}(P)\dul{U}^{\dagger}(P)$ in the new frame, or, in other words, they transform covariantly. This can be combined with the transformation rule of the connection components and it follows that the components of the covariant derivative \eqref{eq:DopFrame} also transform covariantly.\\

Just as we have a parallel transporter along a path for states we can also define one for operators. We want the transported operator to act on a transported state in such a way that the result is equivalent to transporting a state being acted on by the operator:

\begin{equation}
\mc{O}^T_{\gamma}(P(t))|\psi^T_{\gamma}(P(t))\rangle = T(\gamma_{t_0\to t})\mc{O}(P(t_0))|\psi(P(t_0))\rangle.
\end{equation}

Using \eqref{eq:TransporterDef} and the fact that the transporter is unitary,
we arrive at

\begin{equation}
\label{eq:opTransporter}
\mc{O}^T_{\gamma}(P(t))=T(\gamma_{t_0\to t})\mc{O}(P(t_0))T^{\dagger}(\gamma_{t_0\to t}).
\end{equation}

This expression provides a straightforward way to show that, in analogy to the case of the state \eqref{eq:DTransportNoFrame}, the covariant derivative of the operator can likewise be interpreted as the difference between the operator at $P$ and the operator transported from $P + dP$ to $P$ :

\begin{equation}
\label{eq:DopTransporterNoFrame}
\mc{O}^T_{\gamma}(P)-\mc{O}(P)=dP^{\mu}\mc{D}_{\mu}\mc{O}(P).
\end{equation}

We have finally arrived at the first justification for the enormous emphasis we have placed on separating `abstract' states and operators from their components in a local frame. Components are simply scalar-valued functions, perhaps collected into arrays, and can thereby be expanded in a standard Taylor series around a point, i.e.,

\begin{equation}
\begin{split}
&\ul{\psi}(P+dP)=\ul{\psi}(P)+dP^{\mu}\partial_{\mu}\ul{\psi}(P)+O(dP^2),
\\
&\dul{\mc{O}}(P+dP)=\dul{\mc{O}}(P)+dP^{\mu}\partial_{\mu}\dul{\mc{O}}(P)+O(dP^2),
\end{split}
\end{equation}

for the components of a state, and the components of an operator. On the other hand, `abstract' states (operators) are elements of (act on) different Hilbert spaces---different copies of the same Hilbert space to be precise---at different points, meaning that we simply cannot expand them as

\begin{equation}
\label{eq:PsiBadExpansion1}
|\psi(P+dP)\rangle \neq |\psi(P)\rangle+dP^{\mu}\partial_{\mu}|\psi(P)\rangle+\dots,
\end{equation}

neither as

\begin{equation}
\label{eq:PsiBadExpansion2}
|\psi(P+dP)\rangle \neq |\psi(P)\rangle+dP^{\mu}\mc{D}_{\mu}|\psi(P)\rangle+\dots,
\end{equation}

etc., since, we emphasize again, $|\psi(P)\rangle$ and $|\psi(P+dP)\rangle$ live in \textit{different} spaces! There is, however, a special case in which an expansion like \eqref{eq:PsiBadExpansion1} can be made sense of. This possibility hinges on the ability to choose the same frame or basis for the Hilbert spaces at $P + dP$ and $P$, i.e., a frame that is (covariantly) constant with respect to $P$. Then, we can identify the abstract states with their components at each $P$ and compare them. Of course, this is only possible if the connection is flat, since otherwise we cannot choose a local frame that is covariantly constant. Indeed, a covariantly constant local frame requires $\mc{D}_{\mu}|e_a^{\text{F}}(P)\rangle=0$, which means that in this local frame $\dul{\mc{A}^{\text{F}}_{\mu}}=0$ and the curvature components vanish $\dul{\mc{F}^{\text{F}}_{\mu\nu}}=0$. Since the curvature components transform covariantly, their components will vanish in any other frame and the connection is flat. Should we have such a local frame at our disposal, we do not need to worry about the components in this frame transforming as we move between points since we are expanding with respect to the \textit{same} basis at each point. This can be seen clearly for, say, quantum states from the component form of their parallel transport \eqref{eq:TransporterFrame}. If $\dul{\mc{A}^{\text{F}}_{\mu}}=0$ then, by \eqref{eq:TransporterExp}, $\dul{T}=\dul{1}$ which is the identity, meaning that \eqref{eq:TransporterFrame} simply asserts that the transported \textit{components} do not need to be transformed by the parallel transporter (see also section \ref{velLength} for a related discussion). Taking a closer look at only the right hand side of \eqref{eq:PsiBadExpansion2}, we can note that the expression is, in fact, well-defined. It only contains states living in the Hilbert space at point $P$, since the covariant derivative maps between states living in the same space. We can further imagine adding symmetrized combinations of higher-order covariant derivatives to this expression and arrive at the concept of a jet \cite{SaundersJet}. Jets can be thought of as abstract Taylor polynomials, and provide the natural language for generalizing partial differential equations to general spaces in a frame-independent manner, however, they only map states within the same space and do not approximate an object living in a different space at a different point.\\

 We could also think of expanding the component functions, e.g. of an operator $\mc{O}(P)$, in the following, rather unnatural way

\begin{equation}
\label{eq:opDbadExpansion}
\dul{\mc{O}}(P+dP)\neq\dul{\mc{O}}(P)+dP^{\mu}\dul{\mc{D}_{\mu}\mc{O}}(P)+O(dP^2),
\end{equation}

where $\mc{D}_{\mu}\mc{O}(P)$ are the components \eqref{eq:DopFrame} of the
covariant derivative of the operator. While this expansion makes sense, insofar as we only consider the comparability of the expressions on the two sides---they are just component functions, it is in conflict with \eqref{eq:DopTransporterNoFrame}, which establishes that the components of the operator’s covariant derivative measure the difference between the components of operators at the \textit{same} point one of them being components of an operator \textit{transported} to the point. The components $\dul{\mc{O}}(P +dP)$ are those of an operator at $P +dP$, not those of one that was transported to $P$.\\

We now look at the special case of a flat connection over the parameter space. We choose the local frame $\{|e^{\text{F}}_a(P)\rangle\}$ in which the connection components over a part of parameter space vanish $\dul{\mc{A}}^{\text{F}}_i(P) = 0$. This means
that, locally, Hilbert spaces at different points of parameter space can be trivially identified, i.e., the local frame is covariantly constant in the parameter direction, and the parallel transporter \eqref{eq:TransporterExp} only needs to transport in the time direction. What happens if we formally move from $(t_0, \textbf{p} + d\textbf{p})$ to $(t_0, \textbf{p})$; a path purely within parameter space for simplicity? The components of the parallel transporter \eqref{eq:TransporterExp} become $\dul{T}^{\text{F}}((t_0,\textbf{p}),(t_0,\textbf{p})+(t_0,d\textbf{p})) = \dul{1}$. Now let us change to a different frame via a local unitary transformation $U(P)$. By \eqref{eq:TransporterTransform}, the components of the parallel transporter become

\begin{equation}
\label{eq:TransporterFU}
\dul{T}^{\text{F},U}((t_0,\textbf{p}),(t_0,\textbf{p})+(t_0,d\textbf{p}))=\dul{U}(t_0,\textbf{p})\dul{U^{\dagger}}(t_0,\textbf{p}+d\textbf{p}).
\end{equation}

Note that a crucial difference from the curved case is that this parallel transporter is not given by a path-ordered product, in fact, instead of an infinitesimal one we could have chosen an arbitrary path and the result would still just be the product of unitary matrices at the two ends of the path reflecting the fact that these are just the components of a `flat' parallel transporter. Now consider the components $\ul{\psi}^{\text{F}}(t_0, \textbf{p}+d\textbf{p})$ and $\ul{\psi}^{\text{F},U}(t_0, \textbf{p} + d\textbf{p}) = \dul{U}(t_0, \textbf{p} + d\textbf{p})\ul{\psi}^{\text{F}}(t_0, \textbf{p} + d\textbf{p})$ in the two frames. Since these are just collections of scalar-valued functions, we can perform standard Taylor expansions around $\textbf{p}$. Let us expand the components in the non-transformed frame

\begin{equation}
\ul{\psi}^{\text{F}}(t_0,\textbf{p} + d\textbf{p})=\ul{\psi}^{\text{F}}(t_0,\textbf{p})+dp^{i}\partial_{i}\ul{\psi}^{\text{F}}(t_0,\textbf{p})+O(d\textbf{p}^2),
\end{equation}

and rewrite them with the components in the `U-frame'

\begin{equation}
\begin{split}
\dul{U}^{\dagger}(t_0,\textbf{p} + &d\textbf{p})\ul{\psi}^{\text{F},U}(t_0,\textbf{p} + d\textbf{p})=\dul{U}^{\dagger}(t_0,\textbf{p})(\ul{\psi}^{\text{F},U}(t_0,\textbf{p})
\\
&+dp^{i}(\partial_{i}+\dul{U}\partial_i\dul{U}^{\dagger}\ul{\psi}^{\text{F},U}(t_0,\textbf{p})+O(d\textbf{p}^2)).
\end{split}
\end{equation}

Multiplying both sides by $\dul{U}(t_0, \textbf{p})$; recognizing the flat, transformed parallel transporter \eqref{eq:TransporterFU}  on the left; and using the flat connection expression \eqref{eq:AiFU} we find

\begin{equation}
\label{eq:PsiExpansionFU}
\begin{split}
&\dul{T}^{\text{F},U}((t_0,\textbf{p}),(t_0,\textbf{p})+(t_0,d\textbf{p}))\ul{\psi}^{\text{F},U}(t_0,\textbf{p} + d\textbf{p})
\\
&=\ul{\psi}^{\text{F},U}(t_0,\textbf{p})+dp^{i}\left(\partial_{i}+\frac{i}{\hbar}\dul{\mc{A}_i}^{\text{F},U}\right)\ul{\psi}^{\text{F},U}(t_0,\textbf{p})
\\
&\quad+O(d\textbf{p}^2).
\end{split}
\end{equation}

We can recognize a Taylor expansion with the derivative being replaced by a `covariant derivative' on the right hand side, however, most importantly, the left hand side is not simply $\ul{\psi}^{\text{F},U}(t_0,\textbf{p} + d\textbf{p})$, rather, it is multiplied by the parallel transporter in the ‘$U$-frame’, thereby giving the `transported' components in complete consistency with the curved case (cf. Eq. \eqref{eq:DTransportFrame}). The same analysis applies for an operator, verbatim.\\

The reason we are expending so much effort with expansions and the role of curved and flat connections in different frames is because these are precisely what are required for perturbative response calculations and ill-defined, inconsistent expansions, several of which were presented above, are abundant in the literature (see for example refs. \cite{Passos2018,Parker2019}). As an example, \citet{Parker2019} consider a special case of the expansion \eqref{eq:PsiExpansionFU} for the Hamiltonian without the parallel transporter appearing on the left hand side. Indeed, suppose the Hamiltonian $\dul{\mc{H}}(\textbf{p}(t))=\dul{\mc{H}}(\textbf{p}_0-e\textbf{A}(t))$ is not dependent on time \textit{explicitly} only implicitly through the change of the parameter according to the minimal-coupling prescription. Since we only have component functions, we can expand them around $\textbf{p}_0$ in powers of $\textbf{A}(t)$ in the standard way regardless of the connection on the Hilbert bundle over parameter space being flat or not:

\begin{equation}
\label{eq:Hexpansion}
\dul{\mc{H}}(\textbf{p}(t))=\dul{\mc{H}}(\textbf{p}_0)-eA^i(t)\partial_{i}\dul{\mc{H}}(\textbf{p}_0)+O(A^2).
\end{equation}

Curvature effects would arise from the fact that in the non-flat case we would have to use the extended Schrödinger equation \eqref{eq:SchParFrame} and the connection components in the parameter direction also demand expansion. On the other hand, we can also add difficulty to our lives by rewriting the expansion in terms of components in a different frame---see the steps leading to \eqref{eq:PsiExpansionFU}---, for instance in one wherein $\dul{\mc{H}}(\textbf{p}_0)$ is diagonal. The Hamiltonian transforms covariantly under a change of frame by a purely parameter-dependent unitary transformation $\dul{\mc{H}}^U (\textbf{p}(t))=\dul{U}(\textbf{p}(t))\dul{\mc{H}}(\textbf{p}(t))\dul{U}^{\dagger}(\textbf{p}(t))$ at each point of the path in parameter space. Note that this transformation is \textit{not} dependent on time explicitly, only implicitly, since, by \eqref{eq:HPartransform}, the non-covariant term only appears if the change of frame is dependent on time \textit{explicitly}. Taking the flat case for simplicity, expansion \eqref{eq:Hexpansion} for the transformed components yields

\begin{equation}
\label{eq:HexpansionU}
\begin{split}
&\dul{U}(\textbf{p}_0)\dul{U}^{\dagger}(\textbf{p}(t))\dul{\mc{H}}^U(\textbf{p}(t))\dul{U}(\textbf{p}(t))\dul{U}^{\dagger}(\textbf{p}_0)
\\
&\qquad=\dul{\mc{H}}^U(\textbf{p}_0)-eA^i(t)\dul{\mc{D}_i\mc{H}^U}(\textbf{p}_0)+O(A^2),
\end{split}
\end{equation}

where $\dul{\mc{D}_{i}\mc{H}}^U=\partial_{i}\dul{\mc{H}}^U+\frac{i}{\hbar}\left[\dul{\mc{A}_{i}}^{\text{F},U},\dul{\mc{H}}^U\right]$ with $\dul{\mc{A}_{i}}^{\text{F},U}$ being the flat connection \eqref{eq:AiFU}. Note that, as it should, the `flat' parallel transporter within parameter space from point $\textbf{p}(t)$ to $\textbf{p}_0$ has appeared on the left hand side. This flat parallel transporter is absent in \citet{Parker2019}, wherein, referring to Blount’s prescription for the position operator (see Appendix \ref{crystal}.\ref{Blount} for our discussion of this), the authors simply expand the Hamiltonian in a series of covariant derivatives, which, as discussed at length in the previous paragraphs, is completely inconsistent. The saving grace is the following. Suppose we take the `flat' parallel transporters on the left hand side into account, and expand the Hamiltonian in the consistent way. We get the correct expression on the right hand side. Now let us disregard the parallel transporters on the left and expand the Hamiltonian in the wrong way, i.e., using the covariant derivative akin to \eqref{eq:opDbadExpansion}. We get the same correct expression on the right hand side. \textit{It is clear that, miraculously, the two errors cancelled each other out}. In the flat case, this error-correction works for all orders---the `flat' parallel transporter is not path-dependent, but it is pointless to move away from the local frame---the `F-frame'---in which the connection components vanish, rather, we could help ourselves by only changing to, say, the frame in which the Hamiltonian is diagonal, after we have gone as far as we could with the calculation. In the curved case, there is no local frame in which the connection components vanish and, in contrast to the flat case, we do not have to change frame for them to appear. In this case the error resulting from neglecting the parallel transporter and performing an expansion with the covariant derivative such as \eqref{eq:opDbadExpansion} would only be compensated for first order, since, beyond that, path-dependence in the parallel transporter begins to matter (cf. Eq. \eqref{eq:TransporterExp}). We shall get back to this briefly in the next section when discussing expectation values.

\subsection{Density matrix evolution and expectation values}
\label{densmat}

In order to compute general expectation values we are in need of another tool: the density matrix, or statistical operator $\rho$. Under the assumption that any explicit or implicit time-dependence of the Hamiltonian is the result of external perturbations that leave the probability of microstate occupation within the statistical ensemble of interest unaltered, i.e., we remain close to equilibrium, the standard density matrix satisfies the von Neumann equation. Just as for the Schrödinger equation, if we have implicit time-dependence through a parameter and want to properly account for curvature effects in a possibly non-trivial bundle over the parameter space, we have to consider an extended von Neumann equation. Since the density matrix $\rho(P)$ is a weighted sum of projections to pure states and the components of the latter transform as vectors under a change of local frame, the components of $\rho(P)$ transform covariantly $\dul{\rho}\to\dul{U}\dul{\rho} \dul{U}^{\dagger}$ and we can use \eqref{eq:DopFrame} to obtain the covariantly transforming components of the density matrix’s covariant derivative.\\
The interpretation of the extended Schrödinger equation as a condition that the state remain covariantly constant \eqref{eq:SchParParallel} can be rolled-over to the density matrix

\begin{equation}
\label{eq:NeumannParNoFrame}
\frac{dP^{\mu}}{dt}\mc{D}_{\mu}\rho(P)=0,
\end{equation}

where the action of $\mc{D}_{\mu}$ on an operator is defined in \eqref{eq:DopDef}. In a local frame, \eqref{eq:NeumannParNoFrame} becomes

\begin{equation}
\label{eq:NeumannParFrame}
i\hbar\frac{d}{dt}\dul{\rho}(P(t))=\left[\dul{\mc{A}_{\mu}}(P(t))\frac{dP^{\mu}}{dt},\dul{\rho}(P(t))\right].
\end{equation}

While the extended von Neumann equation is formally covariant under a local change of frame or `passive' gauge transformation $U(P) = U (t, \textbf{p})$, we make the following remark. Suppose the equilibrium density matrix is the Boltzmann weight $\dul{\rho_0} = e^{-\beta\dul{\mc{H}_0}}/Z$, where $Z$ is the partition function. The objects on the two sides of this equality are fundamentally different since the components $\dul{\mc{H}_0}$ transform as those of a connection and thus non-covariantly under an explicitly time-dependent change of frame, whereas the components $\dul{\rho_0}$ transform
covariantly under such a transformation. This points to a fundamental discrepancy within quantum-statistical mechanics most likely related to subtleties involving the concept of temperature and `energy', however, such considerations are beyond the scope of this paper and will be addressed elsewhere.\\

We finally have all the necessary ingredients for computing general expectation values. Consider a physical observable operator $\mc{O}(\textbf{p})$ that depends on the parameters of the system. The local frame components of an observable transform covariantly, i.e., $\dul{\mc{O}}(\textbf{p}) \to \dul{U}(p)\dul{\mc{O}}(\textbf{p})\dul{U}^{\dagger}(\textbf{p})$. In the Schrödinger picture, the density matrix always has explicit time-dependence $\rho(t,\textbf{p}(t))$, and it can also have implicit time-dependence through the evolution of parameters. On the other hand, still in the Schrödinger picture, the observable does not have explicit time-dependence, but can depend on time \textit{implicitly} $\mc{O}(\textbf{p}(t))$. The observable's expectation value is the weighted trace

\begin{equation}
\langle\mathcal{O}(t)\rangle=\text{tr}\bigg(\rho(t,\textbf{p}(t))\mathcal{O}(\textbf{p}(t))\bigg),
\end{equation}

where the trace is taken over both, the Hilbert spaces at each $(t,\textbf{p})$ and over the parameter space, leading to the result being dependent only on time (we assume that all the operators are well-behaved such that the trace exists). Note that some observables, such as the velocity $(v_x,v_y,v_z)$, are vectors of operators $v_i$ whose vector indices transform as components of a vector in the tangent bundle of the parameter space. Integrating such objects over parameter space can be problematic (in order to add up vectors from different tangent spaces we have to transport them all to one point requiring a choice of connection) unless, 1: the latter’s tangent bundle is trivial, in which case all tangent spaces can be trivially identified via the trivial connection and there is no need to be concerned about adding up tangent vectors from different tangent spaces, an example being the torus, or 2: the integrand is actually a differential $d$-form on the tangent bundle, $d$ being the real dimension of the parameter space, in disguise and integration of $d$-forms can be defined for a $d$-dimensional (orientable) smooth manifold \cite{TuManifolds}. Taking these points into consideration, in order to keep the presentation simple, henceforth, we shall assume that the parameter space has a trivial tangent bundle. Although the non-trivial case might also be interesting, we do not delve into it any further here.\\

The trace of products of operators can be naturally written as the trace of the products of their component matrices in a local frame. Due to both the density matrix and the observable transforming covariantly under a unitary change of local frame $U(\textbf{p})$, the cyclic property of the trace ensures that the expectation value will be independent of the chosen frame, or, in other words, will be gauge-invariant. Indeed, suppose that the components of the density matrix have evolved according to the extended von Neumann equation \eqref{eq:NeumannParFrame}, meaning that they have been parallel-transported along the path $\gamma(t)$ from $t_0$ to $t$. Then

\begin{equation}
\begin{split}
&\text{tr}\bigg(\dul{\rho}(t,\textbf{p}(t))\dul{\mc{O}}(\textbf{p}(t))\bigg)
\\
&=\text{tr}\bigg(\dul{T}(\gamma_{t_0\to t})\dul{\rho}(t_0,\textbf{p}(t_0))\dul{T}^{\dagger}(\gamma_{t_0\to t})\dul{\mc{O}}(\textbf{p}(t))\bigg).
\end{split}
\end{equation}

Now let us change the frame at each point of the path, and use the transformation property \eqref{eq:TransporterTransform} of the parallel transporter and the one of other operators to obtain

\begin{widetext}
\begin{equation}
\begin{split}
&\text{tr}\bigg(\dul{U}(\textbf{p}(t))\dul{T}(\gamma_{t_0\to t})\dul{U}^{\dagger}(\textbf{p}(t_0))\dul{U}(\textbf{p}(t_0))\dul{\rho}(t_0,\textbf{p}(t_0))\dul{U}^{\dagger}(\textbf{p}(t_0))\dul{U}(\textbf{p}(t_0))\dul{T}^{\dagger}(\gamma_{t_0\to t})\dul{U}^{\dagger}(\textbf{p}(t))\dul{U}(\textbf{p}(t))\dul{\mathcal{O}}(\textbf{p}(t))\dul{U}^{\dagger}(\textbf{p}(t))\bigg)
\\
&=\text{tr}\bigg(\dul{T}(\gamma_{t_0\to t})\dul{\rho}(t_0,\textbf{p}(t_0))\dul{T}^{\dagger}(\gamma_{t_0\to t})\dul{\mathcal{O}}(\textbf{p}(t))\bigg),
\end{split}
\end{equation}
\end{widetext}

confirming the local frame-independence or gauge-invariance. Note that since we are changing the frames in the Hilbert spaces at each point along the path, it is crucial for a parameter-dependent observable to be dependent on time implicitly through the parameter for this result to hold.\\ 

Finally, not letting our interest in perturbative response theory dissipate, we settle our discussion of Taylor expansions, this time in the context of expectation values. We restrict ourselves to the parameter space, since our goal here is to further illustrate the subtle inconsistency within the literature already presented. Consider the infinitesimal transport of the density matrix $\rho_0(\textbf{p}_0)$ from point $\textbf{p}_0$ to point $\textbf{p}_0 + d\textbf{p}$. Using \eqref{eq:TransporterInf} we have

\begin{equation}
\begin{split}
&\text{tr}\bigg(\dul{T}(\textbf{p}_0+d\textbf{p},\textbf{p}_0)\dul{\rho_0}(\textbf{p}_0)\dul{T}^{\dagger}(\textbf{p}_0+d\textbf{p},\textbf{p}_0)\dul{\mc{O}}(\textbf{p}_0+d\textbf{p})\bigg)
\\
&\approx\text{tr}\bigg(\left(\dul{1}-\frac{i}{\hbar}\dul{\mathcal{A}_i}dp^i\right)\dul{\rho_0}(\textbf{p}_0)\left(\dul{1}+\frac{i}{\hbar}\dul{\mathcal{A}_i}dp^i\right)\dul{\mathcal{O}}(\textbf{p}_0+d\textbf{p})\bigg)
\\
&\approx\text{tr}\bigg(\left(\dul{\rho_0}+\frac{i}{\hbar}\left[\dul{\rho_0},\dul{\mathcal{A}_i}\right]dp^i\right)(\dul{\mathcal{O}}+\partial_{j}\dul{\mathcal{O}}dp^j)\bigg)
\\
&=\text{tr}(\dul{\rho_0}\dul{\mathcal{O}})+\text{tr}(\dul{\rho_0}\dul{\mathcal{D}_{i}\mathcal{O}})dp^i+O(dp^2),
\end{split}
\end{equation}

where we dropped the $\textbf{p}_0$ arguments starting from the second line and used the cyclicity of the trace. The components of the covariant derivative have appeared, and, naturally, we have a frame-independent or gauge-invariant expression, since all objects within the trace transform covariantly under a change of frame. It should be noted that should the connection not be flat, we would not be allowed to expand in such a simple manner beyond first order since path-dependence starts to matter in the parallel transporter and we have to use the path-ordered product \eqref{eq:TransporterExp}.\\

Now let us neglect the parallel transporter and perform the \textit{wrong} expansion using the covariant derivative. We get

\begin{equation}
\begin{split}
&\text{tr}\bigg(\dul{\rho_0}(\textbf{p}_0)\dul{\mc{O}}(\textbf{p}_0+d\textbf{p})\bigg)
\\
&\qquad\neq\text{tr}(\dul{\rho_0}\dul{\mathcal{O}})+\text{tr}(\dul{\rho_0}\dul{\mathcal{D}_{i}\mathcal{O}})dp^i+O(dp^2),
\end{split}
\end{equation}

which is the correct result. However, in general, the error-correction does not work beyond first order due to the heralded path-dependence of the parallel transporter.\\

Let us now consider the case of a flat connection, and, as before, work in a particular local frame $\{|e^{\text{F}}_a(\textbf{p})\rangle\}$ in which the connection's components vanish: $\dul{\mc{A}^{\text{F}}_i}(\textbf{p}) = 0$. Then, yet
again, we can make our lives more difficult by changing the local frame via $U(\textbf{p})$ and rewriting the expansion performed in the `F-frame' in terms of components in the `U-frame'. The parallel transporter from $\textbf{p}_0$ to $\textbf{p}_0 + d\textbf{p}$ in the `F-frame' becomes the identity and we have for the expectation value

\begin{equation}
\begin{split}
&\text{tr}\bigg(\dul{\rho_0}^{\text{F}}(\textbf{p}_0)\dul{\mc{O}}^{\text{F}}(\textbf{p}_0+d\textbf{p})\bigg)
\\
&=\text{tr}(\dul{\rho_0}^{\text{F}}(\textbf{p}_0)\dul{\mathcal{O}}^{\text{F}}(\textbf{p}_0))+\text{tr}(\dul{\rho_0}^{\text{F}}(\textbf{p}_0)\dul\partial_{i}\dul{\mathcal{O}}^{\text{F}}(\textbf{p}_0))dp^i+O(dp^2),
\end{split}
\end{equation}

which, upon writing the components in the `U-frame', becomes

\begin{equation}
\begin{split}
&\text{tr}\bigg(\dul{U}(\textbf{p}_0+d\textbf{p})\dul{U}^{\dagger}(\textbf{p}_0)\dul{\rho_0}^{\text{F},U}(\textbf{p}_0)\dul{U}(\textbf{p}_0)\dul{U}^{\dagger}(\textbf{p}_0+d\textbf{p})
\\
&\qquad\qquad\qquad\times\dul{\mc{O}}^{\text{F}}(\textbf{p}_0+d\textbf{p})\bigg)
\\
&=\text{tr}(\dul{\rho_0}^{\text{F},U}\dul{\mathcal{O}}^{\text{F},U})+\text{tr}(\dul{\rho_0}^{\text{F},U}\dul{\mc{D}_{i}\mathcal{O}}^{\text{F},U})dp^i+O(dp^2),
\end{split}
\end{equation}

where, in the last line, we defined

\begin{equation}
\dul{\mc{D}_{\mu}\mc{O}}^{\text{F},U}=\partial_{\mu}\dul{\mc{O}}^{\text{F},U}+\frac{i}{\hbar}\left[\dul{\mc{A}_{\mu}}^{\text{F},U},\dul{\mc{O}}^{\text{F},U}\right],
\end{equation}

with $\dul{\mc{A}_{\mu}}^{\text{F},U}$ being the flat, transformed connection
components \eqref{eq:AiFU}. Notice that the `flat' parallel transporter in the `U-frame' has appeared on the left. Just as in the previous section (expansion \eqref{eq:HexpansionU} and the surrounding discussion), simply neglecting the parallel transporter and expanding the operator using the covariant derivative yields the correct result via the expounded error-correcting; something we shall not repeat here.\\

The conclusion of this subsection is that when expanding implicitly time-dependent objects for perturbative calculations, we have to perform the \textit{standard} Taylor expansion. In the case of a flat connection on the Hilbert bundle over parameter space, we can simply perform all calculations in a local frame with vanishing connection components and quantum evolution is described by the standard Schrödinger and von Neumann equations. \textit{In the curved case, on the other hand, we cannot choose a local frame in which the connection components vanish and we have to use the extended Schrödinger and the corresponding extended von Neumann equations for the quantum evolution}.\\

\section{The velocity gauge, the length gauge, and the velocity operator}
\label{velLength}

\subsection{The velocity gauge}
Having developed the formulation of quantum evolution in a curved space, we now move back to electric field responses. As already noted in the very beginning of section \ref{curvedSpace}, a spatially uniform, time-dependent electric field $\textbf{E}(t)$ can be represented in the Coulomb gauge as $\textbf{E}(t)=-\partial\textbf{A}(t)/\partial t$, with $\textbf{A}(t)$ being the vector potential. Consider the momentum-dependent equilibrium Hamiltonian $\dul{\mathcal{H}_0}(\textbf{p})$. According to the minimal coupling prescription $\dul{\mathcal{H}_0}(\textbf{p})$ changes to $\dul{\mathcal{H}_0}(\textbf{p}-e\textbf{A}(t))$ upon the application of an electric field represented as described and is commonly referred to as the velocity gauge in the literature. Let us consider this Hamiltonian in slightly more detail. Suppose that the applied field was turned on at a time $t_0$, and before this instant the momenta took on reference values denoted $\textbf{p}_0$. Upon application of the field, the momenta start varying in time as $\boldsymbol{\Pi}(\textbf{p}_0,t)\equiv\textbf{p}(t)=\textbf{p}_0-e\textbf{A}(t)$, and we assume that $\textbf{A}(t)=0$ if $t\leq t_0$. What we have then is a Hamiltonian $\dul{\mathcal{H}_0}(\boldsymbol{\Pi}(\textbf{p}_0,t))$ with implicit time-dependence through the variation of a parameter.\\
 The process just described can also be formulated as follows. Suppose momentum space is $\mathcal{B}$ and define the map $\boldsymbol{\Pi}(\cdot,t):\mathcal{B}\to \mathcal{B}$ acting as $\textbf{p}_0\mapsto \boldsymbol{\Pi}(\textbf{p}_0,t)$. Thus, $\boldsymbol{\Pi}(\textbf{p}_0,t)$ describes paths in $(t,\textbf{p})$ space with starting reference points $\textbf{p}_0\in\mathcal{B}$. This formulation of motion, that follows a reference point as it evolves, is known as the Lagrangian description and is widely used in fluid and continuum mechanics \cite{Batchelor2000}. It is clear that, as already recognized by the authors of \cite{SipeShkrebtii2000}, the minimum coupling prescription, and thereby the velocity gauge, corresponds to a Lagrangian description.\\

Should the connection on the Hilbert bundle over momentum space be curved, as in the case of a Hilbert bundle with fibres being truncated Hilbert spaces (see the discussion leading to Eq. \eqref{eq:AiProj}), we have to use the formalism developed in section \ref{curvedSpace} and describe the evolution using the extended Schrödinger equation \eqref{eq:SchParFrame}. Indeed, we see that $dp^i/dt\equiv d\Pi^i(\textbf{p}_0,t)/dt=d(p_0^i-eA^i(t))/dt=eE^i(t)$ so the latter equation becomes

\begin{equation}
\label{eq:SchParFrameE}
\begin{split}
&i\hbar\frac{d}{dt}\ul{\psi}(t,\boldsymbol{\Pi}(\textbf{p}_0,t))
\\
&= \left(\dul{\mathcal{H}_0}(\boldsymbol{\Pi}(\textbf{p}_0,t))+e\dul{\mathcal{A}_i}(\boldsymbol{\Pi}(\textbf{p}_0,t))E^i(t)\right)\ul{\psi}(t,\boldsymbol{\Pi}(\textbf{p}_0,t)).
\end{split}
\end{equation}

We can recognize that the Hamiltonian accounting for the complete time evolution in the velocity gauge or Lagrangian description is

\begin{equation}
\label{eq:velHamiltonian}
\dul{\mc{H}}^{\text{vel}}(\boldsymbol{\Pi}(\textbf{p}_0,t))=\dul{\mathcal{H}_0}(\boldsymbol{\Pi}(\textbf{p}_0,t))+e\dul{\mathcal{A}_i}(\boldsymbol{\Pi}(\textbf{p}_0,t))E^i(t),
\end{equation}

with both the equilibrium Hamiltonian $\dul{\mc{H}_0}$ and connection components $\dul{\mc{A}_i}$ requiring expansion when calculating the perturbative response (see section \ref{pertresponse}. below).\\

\subsection{The length gauge}

Another way to prescribe the action of an electric field on a quantum system is through the so-called length gauge, which is the dipole approximation of the multipolar gauge \cite{KobeGauge}. In the length gauge, the electric field couples directly to the position operator and the coupling term can be formulated as $\textbf{x}\cdot\textbf{E}(t)$, where $\textbf{x}=i\hbar\partial_{\textbf{p}}$; the partial derivative. Recall that the partial derivative acts on components of a state in a local frame and the result corresponds to the components of the covariant derivative of the state in a local frame with vanishing connection components (see Eq. \eqref{eq:DiPsiF}). Thus, we can write $x_i\ul{\psi}^{\text{F}}\to i\hbar(\dul{\mc{D}_i|\psi\rangle})^{\text{F}}$, where, in order to avoid a notational conundrum, we indicated that we are looking at the \textit{components} of the covariant derivative’s action on an `abstract' state, since, while the covariant derivative can act on the abstract states, the partial derivative can act only on the components of the state in a local frame. We can, of course, move to a different frame and the position operator transforms to $x^U_i=x_i-\dul{\mc{A}_i}^{\text{F},U}$, where $\dul{\mc{A}_i}^{\text{F},U}= i\hbar(\partial_i\dul{U})\dul{U}^{\dagger}$ are the components of a flat connection in the `$U$-frame' (see Eq. \eqref{eq:AiFU}). It is clear that the emergence of these components is merely a consequence of our choice of local frame, and can always be made to vanish by simply reverting back to the F-frame. On the other hand, should the connection be curved, an F-frame in which the components of the connection locally vanish cannot be chosen and we have to define the position operator as a ‘covariant derivative’ $r_i\ul{\psi}= (x_i-\dul{\mc{A}_i})\ul{\psi}\to i\hbar\dul{\mc{D}_i|\psi\rangle}$. In this case, the coupling to the electric field becomes $\textbf{r}\cdot\textbf{E}(t)$. Note that this is the covariant derivative in a local frame and it acts on the components of `abstract' states. Indeed, just like the Hamiltonian is a the collection of a connection's components in the time-direction with respect to a local frame (see Eq. \eqref{eq:DtDef}), the position ‘operator’ in its form as $r_i$ is always referenced to a particular frame. Thus, even though the latter is an ‘operator’, as in a ‘matrix’ of components, it cannot be looked at as an `abstract' operator in the sense of having an existence without reference to a frame (see Appendix \ref{diffgeo}.\ref{covDerAbuse} for further elaboration of this point), however, in contrast to the Hamiltonian, it, in fact, has an abstract counterpart; the covariant derivative itself. Naturally, the length and velocity gauges are intimately related and we should be able to obtain the length gauge prescription from the velocity gauge prescription. There are two approaches pursued in the literature: working in the position basis of the total Hilbert space and introducing a unitary transformation of the form $\propto \exp(ie \textbf{A}\cdot \dul{\textbf{x}})$ where $\dul{\textbf{x}}$ are the components of the position operator in the position basis, i.e., just position (examples of this approach are in refs. \cite{SipeShkrebtii2000,Ventura2017,Lamb1987}), or, working in the momentum basis and introducing the same unitary transformation but with the position operator in this latter basis. In the context of crystals, this second choice amounts to working in the energy eigenbasis with the two labels $\textbf{k}$ for crystal momentum and $a$ for band index (see ref. \cite{Passos2018} for a recent example of this). We shall comment on these general approaches momentarily, but, before we do so, we present a different perspective that delves into the heart of the matter: the Lagrangian and Eulerian descriptions of motion.\\
We already showed above that the velocity gauge corresponds to the Lagrangian description. Now let us expand the total derivative with respect to time on the left hand side of the extended Schrödinger equation \eqref{eq:SchParFrameE}. By the chain rule, we have

\begin{equation}
\label{eq:totalDer}
\begin{split}
&i\hbar\frac{d}{dt}\ul{\psi}(t,\boldsymbol{\Pi}(\textbf{p}_0,t))
\\
&=i\hbar\frac{\partial}{\partial t}\ul{\psi}(t,\boldsymbol{\Pi}(\textbf{p}_0,t))+i\hbar\frac{d\Pi^j(\textbf{p}_0,t)}{dt}\partial_j\ul{\psi}(t,\boldsymbol{\Pi}(\textbf{p}_0,t))
\\
&=i\hbar\frac{\partial}{\partial t}\ul{\psi}(t,\boldsymbol{\Pi}(\textbf{p}_0,t))+i\hbar eE^j(t)\partial_j\ul{\psi}(t,\boldsymbol{\Pi}(\textbf{p}_0,t))
\\
&=i\hbar\frac{\partial}{\partial t}\ul{\psi}(t,\boldsymbol{\Pi}(\textbf{p}_0,t))+eE^j(t)x_j\ul{\psi}(t,\boldsymbol{\Pi}(\textbf{p}_0,t)).
\end{split}
\end{equation}

This is known as the material or convective derivative in fluid mechanics \cite{Batchelor2000}. Looking first at the curved case, we combine this with the right hand side of \eqref{eq:SchParFrameE} and get

\begin{equation}
\begin{split}
&i\hbar\frac{\partial}{\partial t}\ul{\psi}(t,\boldsymbol{\Pi}(\textbf{p}_0,t))
\\
&=(\dul{\mathcal{H}_0}(\boldsymbol{\Pi}(\textbf{p}_0,t))-eE^l(t)(x_l-\dul{\mathcal{A}_l}(\boldsymbol{\Pi}(\textbf{p}_0,t))))\ul{\psi}(t,\boldsymbol{\Pi}(\textbf{p}_0,t))
\\
&=(\dul{\mathcal{H}_0}(\boldsymbol{\Pi}(\textbf{p}_0,t))-eE^j(t)r_j(\boldsymbol{\Pi}(\textbf{p}_0,t)))\ul{\psi}(t,\boldsymbol{\Pi}(\textbf{p}_0,t)).
\end{split}
\end{equation}

Now, following the standard procedure, we define the Eulerian coordinate $\textbf{p} = \boldsymbol{\Pi}(\textbf{p}_0,t)$ and note that the Jacobian is the identity to get

\begin{equation}
\label{eq:SchParFrameLength}
\begin{split}
&i\hbar\frac{\partial}{\partial t}\ul{\psi}(t,\textbf{p})
=(\dul{\mathcal{H}_0}(\textbf{p})-eE^j(t)r_j(\textbf{p}))\ul{\psi}(t,\textbf{p}),
\end{split}
\end{equation}

which is the Eulerian description of motion and the standard expression for the evolution equation in the length gauge. We refer to standard fluid mechanics textbooks such as \cite{Batchelor2000} for a detailed overview of the Lagrangian and Eulerian descriptions.\\
In the Eulerian case, we are not following the evolution of a reference point, rather, we are looking at a fixed point and examining the change at this particular point. The Hamiltonian accounting for the complete time evolution is then

\begin{equation}
\label{eq:lengthHamiltonian}
\dul{\mc{H}}^{\text{length}}(\textbf{p})=\dul{\mathcal{H}_0}(\textbf{p})-eE^j(t)r_j(\textbf{p}).
\end{equation}

We note how the position operator as a covariant derivative and the length gauge arose naturally from the velocity gauge. Crucially, we did not perform a unitary transformation of the extended Schrödinger equation \eqref{eq:SchParFrameE} in the Lagrangian description to get its form \eqref{eq:SchParFrameLength} in the Eulerian description. Rather, we performed a careful analysis of the time-derivative on the left hand side and noted that in the case of the velocity gauge it is a total derivative along a path, whereas in the Eulerian case it is a partial derivative at a point. It is important to remark that, as we have demonstrated extensively, a unitary transformation of the components represents a change of local frame and in this new frame we can again consider the Lagrangian and Eulerian descriptions and move between them by expanding the total time-derivative. The velocity and length gauge thus do not correspond to different frames, but different descriptions in the same frame. This is far from obvious in the position basis approach.\\

Let us now look at the flat case. In the F-frame the connection components vanish and we have the standard Schrödinger equation

\begin{equation}
\label{eq:SchParFrameEF}
i\hbar\frac{d}{dt}\ul{\psi}^{\text{F}}(t,\boldsymbol{\Pi}(\textbf{p}_0,t))
=\dul{\mathcal{H}^{\text{F}}_0}(\boldsymbol{\Pi}(\textbf{p}_0,t))\ul{\psi}^{\text{F}}(t,\boldsymbol{\Pi}(\textbf{p}_0,t)).
\end{equation}

Expanding the derivative on the left as in \eqref{eq:totalDer}, rearranging, and moving to the Eulerian description, we find

\begin{equation}
\label{eq:SchParFrameLengthF}
i\hbar\frac{\partial}{\partial t}\ul{\psi}^{\text{F}}(t,\textbf{p})
=(\dul{\mathcal{H}^{\text{F}}_0}(\textbf{p})-eE^j(t)x_j(\textbf{p}))\ul{\psi}^{\text{F}}(t,\textbf{p}),
\end{equation}

which is the standard length gauge prescription. Moving to a different frame, the components of the state and Hamiltonian transform covariantly, but the position operator becomes $x^U_i = x_i-i\hbar(\partial_i\dul{U})\dul{U}^{\dagger}$ (see the beginning of this section). The common practice is to choose $U$ such that the Hamiltonian be diagonal in the $U$-frame and, in this case, we see the natural `emergence' of Blount’s position operator \cite{Blount}. In the curved scenario, the position operator $r_i = x_i-\dul{\mc{A}_i}$ can be considered as a `generalized' version of Blount’s position operator, albeit the two are fundamentally different. Indeed, in the ordinary case the connection components can be made to vanish via a change of frame, but this cannot be done in the curved case. Should the curved connection arise as the result of a projection to a subspace (see Eq. \eqref{eq:AiProj}), a formal similarity between the flat and curved cases exists, and we could talk of Blount’s `projected' position operator. We present a more thorough discussion of Blount’s position operator in Appendix \ref{crystal}.\ref{Blount}.\\

As highlighted earlier, the fact that the velocity gauge is an essentially Lagrangian and the length gauge an Eulerian description was already recognized by Sipe and Shkrebtii \cite{SipeShkrebtii2000}, however, they use the phase shift in the position basis approach which obscures the fact that we are really just changing descriptions and not frames or bases. On the other hand, the unitary transformation in the energy eigenbasis approach carries subtle problems with the way it is implemented, for example by \citet{Passos2018}, closely related to our discussion of Taylor expansions in section \ref{Taylor}. Following \citet{Passos2018} consider the time-dependent transformation $S^U(t) = e^{\frac{i}{\hbar}eA^j(t)x^U_j}$, where $x^U_i = x_i-\dul{\mc{A}_i}^{\text{F},U}= x_i-i\hbar(\partial_i\dul{U})\dul{U}^{\dagger}$, with $x_i=i\hbar\partial_i$ and $U$ chosen to be the operator that diagonalizes the Hamiltonian. As an example, in the case of a periodic crystal, the Hamiltonian is taken to be the Bloch Hamiltonian and the elements of $U$ are the components of cell-periodic Bloch states in a countable basis (see Appendix \ref{crystal}.\ref{dynamics}). The motivation for such a transformation is, of course, to get rid of the vector potential via a translation: $\ul{\psi}(t, \boldsymbol{\Pi}(\textbf{p}_0,t)) = \ul{\psi}(t, \textbf{p}_0-e\textbf{A}(t))\to \ul{\psi}(t, \textbf{p}_0)$. However, recall from section \ref{Taylor} that components are simply scalar-valued functions and should be expanded via a \textit{standard} Taylor expansion around a point regardless of the connection over parameter space being flat or not. For a finite displacement in momentum, we can formally write

\begin{equation}
\label{eq:PsiTranslation}
\ul{\psi}(t, \textbf{p}_0-e\textbf{A}(t))=e^{-eA^i(t)\partial_i}\ul{\psi}(t, \textbf{p}_0)=e^{\frac{i}{\hbar}eA^i(t)x_i}\ul{\psi}(t,\textbf{p}_0),
\end{equation}

and identify the translation operator $S(t) = e^{\frac{i}{\hbar}eA^i(t)xi}$. Note that this operator is fundamentally different from the unitary transformations we have used to move between local frames, since the latter were all performed at each momentum, whereas the former contains the derivative with respect to $\textbf{p}$ and thereby connects different momenta. Similarly, we can apply the translation operator to matrices of components as $\dul{\mc{H}_0}(\textbf{p}_0-e\textbf{A}(t))= S(t)\dul{\mc{H}_0}(\textbf{p}_0)S^{\dagger}(t)$. According to \citet{Passos2018}, we should be doing

\begin{equation}
\begin{split}
\ul{\psi}(t, \textbf{p}_0-e\textbf{A}(t))\neq& e^{-eA^i(t)(\partial_i+\frac{i}{\hbar}\dul{\mc{A}_i}^{\text{F},U})}\ul{\psi}(t, \textbf{p}_0)
\\
&=e^{\frac{i}{\hbar}eA^i(t)x^U_i}\ul{\psi}(t,\textbf{p}_0),
\end{split}
\end{equation}

which corresponds to a Taylor expansion in terms of the covariant derivative. While the right hand side can be formalized in a frame-independent manner as the action of an infinite jet \cite{SaundersJet} on the abstract state, this will be a map from $\textbf{p}_0$ to $\textbf{p}_0$ parameterized by $t$ and will \textit{not} provide a state at $\textbf{p}_0-e\textbf{A}(t)$. (See also section \ref{Taylor} and Appendix \ref{diffgeo}.\ref{covDerAbuse}).\\

Inserting \eqref{eq:PsiTranslation} into the extended Schrödinger equation \eqref{eq:SchParFrameE}, we obtain

\begin{equation}
\label{eq:SchParFrameLengthTrans}
\begin{split}
i\hbar\frac{\partial}{\partial t}\ul{\psi}(t,\textbf{p}_0)
=(&\dul{\mathcal{H}_0}(\textbf{p}_0)-i\hbar S^{\dagger}(t)\partial_t S(t)
\\
&+eE^i(t)\dul{\mc{A}_i}(\textbf{p}_0))\ul{\psi}(t,\textbf{p}_0).
\end{split}
\end{equation}

Since $[x_i,x_j] = 0$, we can simply differentiate $S(t)$ with respect to time and arrive at the standard length gauge evolution \eqref{eq:SchParFrameLength}. We note the difference between this and our approach of simply changing description; apart from conceptual clarity, when moving to the Eulerian description by expanding the total time derivative we do not need to define operators such as $S(t)$ which are mathematically difficult (`derivative' of infinite expansions with unbounded operators), moreover, the fact that we are not changing frames, rather, just changing descriptions is completely transparent.\\

\begin{figure}
\begin{center}
        \includegraphics[width=10cm]{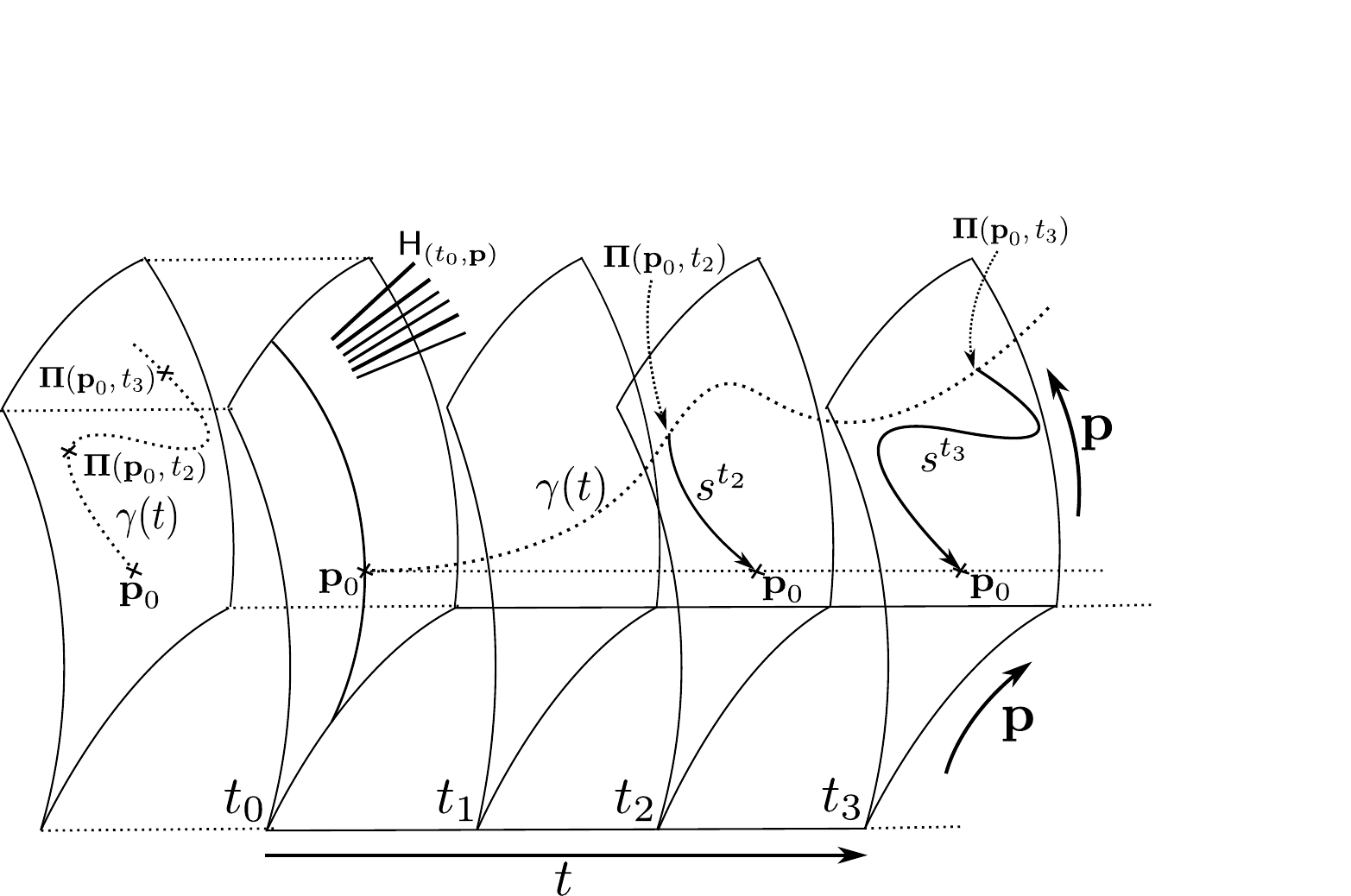}
      \caption{Parallel transport over a curve in momentum space. The transverse curved planes illustrate momentum space, and are not the fibres of the Hilbert bundle. A few fibres $\mathsf{H}_{(t_0,\textbf{p})}$ of the latter are depicted at $t_0$ in order to avoid confusion.  $\gamma(t)$ is the path in $(t,\textbf{p})$ over which the state evolves. $s^{t_i}$ are paths in momentum space at fixed times $t_i$ from $\boldsymbol{\Pi}(\textbf{p}_0,t_i)$ to $\textbf{p}_0$ parameterized by $\lambda$ with tangent $\propto\textbf{E}(\lambda)$ over which the states are parallel transported. The left most transverse plane shows the momentum space trajectory for all times.\label{fig:curvedlengthgauge}}
\end{center}
\end{figure}

Returning to the translation operator $S(t)$, it acts on the \textit{components} $\ul{\psi}(\textbf{p}_0)$ as functions and moves them from $\boldsymbol{\Pi}(\textbf{p}_0, t) = \textbf{p}_0-e\textbf{A}(t)$ to $\textbf{p}_0$. What about the `abstract' state $|\psi(\boldsymbol{\Pi}(\textbf{p}_0,t))\rangle$ itself? Could we, perhaps, move it to the fibre at $\textbf{p}_0$? Yes, we most definitely could, by recruiting the parallel transporter within the Hilbert bundle over momentum space. For each $t$ we choose a natural path $s^t : [0, 1] \to \mc{B}$ in momentum space with endpoints $s^t(\lambda = 0) = \boldsymbol{\Pi}(\textbf{p}_0,t),\, s^t(\lambda= 1) = \textbf{p}_0$ and tangent $eE^i(\lambda)$. Then, the components of the parallel transporter within the Hilbert bundle over momentum space are

\begin{equation}
\label{eq:momentumTransporter}
\dul{T}^{\textbf{p}}(s^t_{\boldsymbol{\Pi}(\textbf{p}_0,t)\to\textbf{p}_0})=\mathtt{P}\,\text{exp}\left(-\frac{ie}{\hbar}\int_{0}^{1}d\lambda\dul{\mc{A}_i}(\textbf{p}(\lambda))E^i(\lambda)\right),
\end{equation}

where $\text{P}$ is a path-ordering symbol. This moves states over a curve $s^t$ in momentum space at fixed times $t$ (see FIG. \ref{fig:curvedlengthgauge}). We are thus able to get the components of the transported state at $\textbf{p}_0$ as

\begin{equation}
\ul{\psi}^{T(t)}(t,\textbf{p}_0)=\dul{T}^{\textbf{p}}(s^t_{\boldsymbol{\Pi}(\textbf{p}_0,t)\to\textbf{p}_0})\ul{\psi}(t,\boldsymbol{\Pi}(\textbf{p}_0,t)),
\end{equation}

where the $^{T(t)}$ superscript on the left hand side labels that $\ul{\psi}^{T(t)}(t,\textbf{p}_0)$ are the components of a \textit{transported} state and that the parallel transporter is time-dependent, i.e., we are performing the transportation over momentum space for each time $t$. Let us use this to transform the extended Schrödinger equation \eqref{eq:SchParFrameE}. Noting that the parallel transporter is unitary and the fact that the components $\dul{T}^{\textbf{p}}$ of the parallel transporter themselves satisfy

\begin{equation}
\nonumber
i\hbar\frac{d}{dt}\dul{T}^{\textbf{p}}(s_{\boldsymbol{\Pi}(\textbf{p}_0,t)\to\textbf{p}_0})=\dul{\mathcal{A}_i}(\boldsymbol{\Pi}(\textbf{p}_0,t))E^i(t)\dul{T}^{\textbf{p}}(s_{\boldsymbol{\Pi}(\textbf{p}_0,t)\to\textbf{p}_0}),
\end{equation}

we have

\begin{equation}
\label{eq:SchParETrans}
\begin{split}
i\hbar\frac{d}{dt}\ul{\psi}^{T(t)}(t,\textbf{p}_0)
= \dul{\mathcal{H}_0^{T(t)}}(\textbf{p}_0)\ul{\psi}^{T(t)}(t,\textbf{p}_0)\rangle^T,
\end{split}
\end{equation}

where we defined the transported equilibrium Hamiltonian 

\begin{equation}
\begin{split}
&\dul{\mathcal{H}_0^{T(t)}}(\textbf{p}_0)
\\
&=\dul{T}^{\textbf{p}}(s_{\boldsymbol{\Pi}(\textbf{p}_0,t)\to\textbf{p}_0})\dul{\mathcal{H}_0}(\boldsymbol{\Pi}(\textbf{p}_0,t))(\dul{T}^{\textbf{p}}(s_{\boldsymbol{\Pi}(\textbf{p}_0,t)\to\textbf{p}_0}))^{\dagger}.
\end{split}
\end{equation}

What we have done is transferred the dynamics resulting from the curved space evolution over momentum space to the Hamiltonian and obtained an equation akin to the standard Schrödinger equation. This is analogous to the interaction picture, in which the dynamics due to the `simple part' is transferred onto the `difficult part' of the Hamiltonian. Indeed, below we shall see how a geometric interpretation can be given to the relationship between the evolution pictures. Despite offering conceptual clarity, actually working with \eqref{eq:SchParETrans} seems to be rather impractical since operator exponentials have appeared, and performing perturbation expansions with these are rather cumbersome due to the fact that, in general, $\dul{\mc{A}_i}$ does not commute with $\dul{\mc{A}_j}$ leading to the necessity of considering the `derivative of the exponential map' \cite{Suzuki1985}.\\

On a more abstract note, the construction just presented can be looked at as follows. Consider a diffeomorphism $f$ of momentum space that acts as $f:\mathcal{B}\to\mathcal{B}$, $\textbf{p}\mapsto \textbf{p}'$ and we would like to lift this to the states in the Hilbert bundle, i.e., $|\psi(\textbf{p})\rangle \to|\psi'(f(\textbf{p}))\rangle$. There is no canonical way to do this and a connection has to be chosen. Once this is done, we can use the parallel transporter to transport the state over some path $\gamma$ and get $|\psi'(f(\textbf{p}))=|\psi^T_{\gamma}(f(\textbf{p}))\rangle$. The diffeomorphism of the base space has induced a mapping of the bundle to itself, albeit this mapping is connection and path-dependent. Such constructions have been widely used in theoretical physics, specifically Yang-Mills theories, and we refer to \cite{GockSchuck} for a more thorough discussion. We also remark that there is a category of bundles, known as natural bundles, for which there is a canonical way to perform this lift. One example is the tangent bundle, with the lift being the push-forward (or differential) and over curved space-time these base space diffeomorphism induced transformations are known as general covariant transformations \cite{Sardanashvili}. As a final note, it is important to recognize that active gauge transformations which map a state in a fibre to `another' state in the same fibre and can thereby be considered `redundancies' or `internal symmetries' are fundamentally different from the latter transformations that move the fibres around and are thereby `external symmetries'. Such concepts are widely discussed in the context of real space gauge theories and general relativity and we refer to \cite{Grensing} for a relatively recent and exhaustive discussion.\\

In light of the above, we can now provide a geometric interpretation of the relationship between the Schrödinger and Heisenberg pictures. For simplicity, we do not consider parameters and suppose the Hamiltonian $\dul{\mc{H}}(t)$ is just time-dependent. Recall that the Hamiltonian can be considered as a collection of components of a connection on a Hilbert bundle over time (see section \ref{curvedSpace}). The standard Schrödinger equation is

\begin{equation}
\label{eq:SchFrameDiffeo}
i\hbar\frac{d}{dt}\ul{\psi}(t)=\dul{\mathcal{H}}(t)\ul{\psi}(t).
\end{equation}

Now let us consider a diffeomorphism of time given by $f : \mathbb{R}\to \mathbb{R}$ such that $t\mapsto t_0$. Using the parallel transporter over time with components given by

\begin{equation}
\label{eq:timeTransporter}
\dul{T}^{t}(t\to t_0)=\mathtt{T}\,\exp\left(-\frac{ie}{\hbar}\int_{t_0}^{t}dt'\,\dul{\mc{H}}(t')\right),
\end{equation}

where $\mathtt{T}(\dots)$ prescribes time-ordering, we can lift the base space diffeomorphism to the bundle and transport a state $|\psi(t)\rangle$ at $t$ to $t_0$ with the result being $|\psi^T(f(t))\rangle =|\psi^T(t_0)\rangle$ and in component form

\begin{equation}
\ul{\psi}^T(t_0)=\dul{T}^t(t\to t_0)\ul{\psi}(t),
\end{equation}

together with the standard Schrödinger equation \eqref{eq:SchFrameDiffeo} becoming

\begin{equation}
\frac{d}{dt}\ul{\psi}^T(t_0) = 0.
\end{equation}

This is the fixed state at $t_0$ in the Heisenberg picture. It is thus clear that the Schrödinger and Heisenberg pictures are related by an analogue of a general covariant transformation widely-studied within general relativity and their equivalence reflects an `external symmetry'.\\

It is important to emphasize that the \textit{translation} of components as functions $\ul{\psi}(\textbf{p}_0) = S(t)\ul{\psi}(\textbf{p}_0-e\textbf{A}(t))$, where $S(t) = e^{eA_i(t)\partial_i}$, and the parallel transport of `abstract' states $|\psi^{T(t)}(\textbf{p}_0)\rangle = T(s_{\lambda_0(t)\to\lambda_1)})|\psi(\textbf{p}_0-e\textbf{A}(t))\rangle$ over some path $s^t$ parameterized by $\lambda$ with $s^t(\lambda_0 (t)) = \textbf{p}_0-e\textbf{A}(t)$ and $s^t(\lambda_1) = \textbf{p}_0$ are fundamentally different concepts. From the component point of view, in the latter case, we are simply accounting for the fact that we need to take the components with respect to different bases at each $\textbf{p}$, i.e., we have to consider $\textbf{p}$-dependent local frames. The two notions of translation and parallel transport are often confused and even `hybridized'. For example, in a recent pa-
per, \citet{Wilhelm2021} define a ``boost operator" $\propto \mathtt{P}\,\exp(-\int_{\infty}^t dt'\, E^i(t')\partial_i )$, where $\mathtt{P}$ is a path-ordering symbol in momentum space, and use it to move from the velocity to the length gauge. Based on our discussion till now, it is clear why this `operator' does not make sense. Is it a parallel transporter in a frame with $\partial_i$ being the `connection components'? Or is it a translation operator acting on the component functions of a state in a frame? The authors use and interchange both of these interpretations. We remark that path-dependence within the multi-dimensional momentum space only arises if the connection on the Hilbert bundle over momentum space is not flat; a case the authors do not discuss.\\

In conclusion, we found that the length gauge can be derived from the velocity gauge by moving to the Eulerian description via the material derivative and there is no fundamental need to perform any `unitary' transformation when in the momentum basis, or energy eigenbasis in the case of a crystal. If we do want to take the transformation route we have to keep in mind that it is the standard Taylor expansion which is defined for the frame components of the relevant object and no expansion is defined for the abstract states since these live in different spaces. Furthermore, abstract states can, in general, only be moved between spaces over different momenta, i.e., between different fibres, via the Parallel transporter. Despite the difficulties with the position operator, the length gauge has still not withered away and, to the detriment of clarity, continues to justify its use within nonlinear transport calculations via the non-existence of apparent divergences in the static limit, lesser sensitivity to band truncations and supposed ‘easier’ comparison to semiclassical approximations \cite{Ventura2017,Parker2019}, however, our main goal is to show that all of these issues can be handled in a relatively simple manner within the velocity gauge: the divergence by the implementation of certain gauge conditions allowing the bypass of sum rules, the truncation by a combination of our curved space formalism and the gauge conditions (see the respective sections \ref{sumRules} for our discussion of sum rules and \ref{secondorderresponse} for the latter), and, in paper III of our series \cite{Bonbien2021c}, we use the decompositions presented in paper I \cite{Bonbien2021a} to derive exact, velocity gauge formulas in the spectral representation from which semiclassical limits can be found with little effort.\\

\subsection{The velocity operator}
\label{velocity}

We wrap up this section by taking a look at the velocity operator $\textbf{V}$. This is an important physical observable but its definition in a curved space, as we shall see below, is rather interesting. Velocity describes how position changes in time. Since we want it to be a physical observable operator its expectation value should be gauge invariant, meaning that it should transform covariantly under a gauge transformation, or, equivalently, its components in a local frame should transform covariantly under a change of frame. This means that we should have $\dul{V_i}(\textbf{p})\to \dul{U}(t,\textbf{p})\dul{V_i}(\textbf{p})\dul{U}^{\dagger}(t,\textbf{p})$ under a change of frame. It is possible to construct such an operator as the covariant derivative in the time direction of a covariantly transforming position operator. That is, we take the action of the curved position operator $r_i(\textbf{p})\ul{\psi}(t,\textbf{p})\to i\hbar\mathcal{D}_{i}|\psi(t,\textbf{p})\rangle$, with the resulting components transforming covariantly due to them being components of a covariant derivative, and consider the latter's covariant derivative in the time direction

\begin{equation}
\label{eq:velocityNoFrame}
V_i(\textbf{p})=\mathcal{D}_0(r_i(\textbf{p}))\equiv i\hbar [\mc{D}_0,\mc{D}_i].
\end{equation}

We can recognize the `$0i$' components of the curvature $\dul{\mc{F}_{\mu\nu}}$ of the connection with components $\dul{\mc{A}_{\mu}} = (\dul{\mc{H}_0}, \dul{\mc{A}_i})$.
The velocity operator can thus be interpreted as a contribution to a curvature of a connection on a bundle over time-momentum space and, going forward with the analogy to electromagnetism, we can think of it as a non-Abelian `electric field' over time-momentum space (the `$ij$' components of the curvature correspond to the non-Abelian `magnetic field'). We have to emphasize, though, that the combined time-momentum indices $\mu$ on the curvature $\dul{\mc{F}_{\mu\nu}}$, where $\mu,\nu \in {0,\dots,3}$ with $\mu=0$ corresponding to time and $\mu=i\,(i\in\{1,2,3\})$ to momentum components, are \textit{not} Lorentz indices and we should treat this non-Abelian `electromagnetic field' as a Euclidean one. The components in a local frame become

\begin{equation}
\label{eq:velocityFrame}
\begin{split}
\dul{V_i}(\textbf{p})&=i\hbar\left(\partial_t\dul{\mc{A}_i}(\textbf{p})-\partial_i\dul{\mc{H}_0}(\textbf{p})+\frac{i}{\hbar}\left[\dul{\mc{H}_0}(\textbf{p}),\dul{\mc{A}_i}(\textbf{p})\right]\right)
\\
&=-i\hbar\left(\partial_i\dul{\mc{H}_0}(\textbf{p})+\frac{i}{\hbar}\left[\dul{\mc{A}_i}(\textbf{p}),\dul{\mc{H}_0}(\textbf{p})\right]\right)
\\
&=-i\hbar\dul{\mc{D}_i\mc{H}_0},
\end{split}
\end{equation}

where we supposed that the components $\dul{\mc{A}_i}(\textbf{p})$ are not dependent on time explicitly. Note that should these components arise via an explicitly time-dependent projection to a subspace of the Hilbert space, the components would themselves be dependent on time (see the discussion in section \ref{curvedSpace}), and our efforts to reduce the velocity operator to the second equality would be stymied. Such a case might arise for Floquet systems in which the projection could be onto a quasi-band of the Floquet Hamiltonian. Under this circumstance we would have the components of the total curvature $\dul{\mc{F}_{\mu\nu}}(t, \textbf{p})$ with $\dul{\mc{F}_{0i}}(t,\textbf{p})= -\frac{i}{\hbar}\dul{V_i}(t,\textbf{p})$, the velocity operator and $\dul{\mc{F}_{ij}}(t, \textbf{p})$ the curvature components \eqref{eq:curvature} of $\dul{\mc{A}_i}(t,\textbf{p})$. Consider now the special case of a 3-dimensional periodic crystal with $\textbf{p} = \hbar\textbf{k}$, where $\textbf{k}$ is the crystal momentum. The time direction is periodic and is thus topologically a circle $S^1$, whereas the first Brillouin zone is a 3-torus $T^3=S^1\times S^1 \times S^1$. The base space for our Hilbert bundle is then the 4-torus $T^4 = S^1 \times S^1\times S^1 \times S^1$. We can use the curvature $\dul{\mc{F}_{\mu\nu}}(t,\textbf{p})$ to construct the second Chern character \cite{Nakahara,TuDiffGeo} of the bundle

\begin{equation}
\begin{split}
\varepsilon_{\mu\nu\rho\lambda}&\text{tr}_{\mathsf{H}}\left(\dul{\mc{F}_{\mu\nu}}(t,\textbf{p})\dul{\mc{F}_{\rho\lambda}}(t,\textbf{p})\right)
\\
&=-\frac{4i}{\hbar}\varepsilon_{ijk}\text{tr}_{\mathsf{H}}\left(\dul{V_i}(t,\textbf{p})\dul{\mc{F}_{jk}}(t,\textbf{p})\right),
\end{split}
\end{equation}

where $\varepsilon_{\mu\nu\rho\lambda}$ and $\varepsilon_{ijk}$ are the totally antisymmetric Levi-Civita symbols, moreover, the trace is taken only over the Hilbert spaces at each $(t,\textbf{p})$. Integrating this over the base 4-torus $T^4$, that is, over one period of time and the 3-dimensional first Brillouin zone, provides a topological invariant of the Hilbert bundle over $T^4$ proportional to the bundle's second Chern numbers. A question is, whether this can be realized as a physical quantity. Indeed it can, as a second order current response to an electric field. We shall deal with this issue in future work.\\

As an electric field is applied, in the velocity gauge, the momentum changes according to the minimal-coupling prescription and the velocity operator will get an implicit time-dependence through the momentum

\begin{equation}
V_i(\textbf{p})\to V^{\text{vel}}_i(\boldsymbol{\Pi}(\textbf{p}_0,t)).
\end{equation}

We remark an important subtlety. Should the connection components not vanish, we also have $\dul{\mc{A}_i}(\textbf{p})\to \dul{\mc{A}_i}(\boldsymbol{\Pi}(\textbf{p}_0,t))$, since we are taking the covariant derivative at each point of the path. The components $\dul{\mc{A}_i}(\textbf{p})$ do not vanish if the connection is curved or, in the flat case, by choice of local frame. This is crucial for perturbative current responses, because, in the non-vanishing case, we also have to expand the connection components in terms of the perturbing field. Indeed, to first order in the vector potential we have

\begin{equation}
\begin{split}
&\dul{\mc{D}_i\mc{H}_0}(\boldsymbol{\Pi}(\textbf{p}_0,t))\approx \dul{\mc{D}_i\mc{H}_0}(\textbf{p}_0)-eA^j(t)\partial_j\partial_i\dul{\mc{H}_0}(\textbf{p}_0)
\\
&\qquad -\frac{ie}{\hbar}A^j(t)\left(\left[\partial_j\dul{\mc{A}_i}(\textbf{p}_0),\dul{\mc{H}_0}(\textbf{p}_0)\right]+\left[\dul{\mc{A}_i}(\textbf{p}_0),\partial_j\dul{\mc{H}_0}(\textbf{p}_0)\right]\right)
\\
&=\dul{\mc{D}_i\mc{H}_0}(\textbf{p}_0)-eA^j(t)\partial_j\dul{\mc{D}_i\mc{H}_0}(\textbf{p}_0).
\end{split}
\end{equation}

This is, of course, expected, since, in general, we cannot have a covariant derivative at $\textbf{p}$ acting on an operator at $\textbf{p}'$. In particular, \citet{Parker2019} for example work with a flat connection in a frame with non-vanishing components, but do not consider the implicit time-dependence of the connection components, rather, the authors take the covariant derivative, or position operator $\textbf{r}$ as fixed at the reference momentum and expand using the covariant derivative, i.e.,

\begin{equation}
\dul{\mc{D}_i\mc{H}_0}(\boldsymbol{\Pi}(\textbf{p}_0,t))\neq\dul{\mc{D}_i\mc{H}_0}(\textbf{p}_0)-eA^j(t)\dul{\mc{D}_i\mc{D}_j\mc{H}_0}(\textbf{p}_0),
\end{equation}

and then use the fact that the connection is flat and covariant derivatives in different directions commute implying that $\mc{D}_i$ can be exchanged with $\mc{D}_j$, thereby providing a `covariant derivative expansion' of the left hand side. The reason why it works is the same error-correcting mechanism we discussed for Taylor expansions in section \ref{Taylor} for flat connections, i.e., moving to a local frame in which the flat connection’s components are nonvanishing, and then not taking into account the `flat' parallel transporter but at the same time expanding wrongly with the flat covariant derivative. However, in the curved case, this leads to serious issues with gauge invariance, so great care has to be taken in order to perform the expansions consistently.\\

How does the above velocity operator transfer to the length gauge? As discussed in considerable detail, the length gauge is simply the Eulerian description and, therefore, all we have to do is shift the velocity gauge velocity operator to the Eulerian coordinate $\textbf{p} = \boldsymbol{\Pi}(\textbf{p}_0,t)$ : $V_i^{\text{vel}}(\boldsymbol{\Pi}(\textbf{p},t))\to V_i(\textbf{p})=V_i^{\text{length}}(\textbf{p})$.\\
We point out that neither of the velocity operators are defined as $[\mc{H},r_i]$ where $\mc{H}$ describes the full time evolution with it being \eqref{eq:velHamiltonian} for the velocity gauge and \eqref{eq:lengthHamiltonian} for the length gauge. This is because $\mc{H}$, through the connection $\mc{A}_i$, also describes evolution in the momentum direction, whereas velocity is defined as the change of position in the time direction.\\

\section{Perturbative response}
\label{pertresponse}

Henceforth, in an effort to avoid clutter, we shall stop explicitly distinguishing between abstract operators and their components in a local frame and simply refer to all such objects as `operators' with the understanding that they are component `matrices'. By this point it should not lead to any confusion.\\

Before beginning with the explicit calculations, we briefly summarize some necessary aspects of standard perturbative response theory upto second order in the Kubo formalism and refer to \cite{Bonbien2021a} for details and derivations.\\

Let the total many-body Hamiltonian be $\mathcal{H}=\mathcal{H}_0+H'(t)$ where $\mathcal{H}_0$ is the time-independent equilibrium Hamiltonian and $H'(t)$ is a time-dependent perturbation. Suppose $\textbf{F}(t)$ is a spatially uniform \textit{classical} field that we consider as an external perturbation and let $\mathcal{M}_j^{(0)},\mathcal{M}_{jk}^{(1)}$ denote the components in an array of Hermitian operators that the field couples to. Then the interaction Hamiltonian is 

\begin{equation}
\label{eq:HextOp}
\begin{split}
H'(t) =\sum_{j_1}\mathcal{M}^{(0)}_{j_1}F^{j_1}(t)+\sum_{j_1,j_2}\mathcal{M}^{(1)}_{j_1j_2}F^{j_1}(t)F^{j_2}(t)+O(F^3)
\end{split}
\end{equation}

and the array $\mathcal{M}^{(1)}_{j_1j_2}$ is defined to be completely symmetric in $j_1,j_2$. There is no loss of information, since we are summing over these indices and the array is multiplied by $F^{j_1}(t)F^{j_2}(t)$, a symmetric expression.\\

Similarly, let $\mathcal{O}_i$ be the $i$-th component of a vector of observable operators whose expectation values we are looking for. Keeping in mind that these might depend on the applied field, we have

\begin{equation}
\label{eq:ObsOp}
\begin{split}
\mathcal{O}_{i_0}(\textbf{F}(t)) =&\mathcal{O}^{(0)}_{i_0}+
\sum_{i_1}\mathcal{O}^{(1)}_{i_0i_1}F^{i_1}(t)
\\
&+\sum_{i_1}\sum_{i_2}\mathcal{O}^{(2)}_{i_0i_1i_2}F^{i_1}(t)F^{i_2}(t),
\end{split}
\end{equation}

and take $\mathcal{O}^{(2)}_{i_0i_1i_2}$ to be symmetric in $i_1,i_2$. The expectation value upto second order in the perturbation can be expressed as a sum

\begin{equation}
\label{eq:ObsTotal}
\langle\mathcal{O}_{i_0}(t)\rangle = \sum_{n=0}^{2}\langle\mathcal{O}_{i_0}(t)\rangle_n,
\end{equation}

where

\begin{widetext}
\begin{equation}
\label{eq:Obsn}
\langle\mathcal{O}_{i_0}(t_0)\rangle_n=\sum_{i_1}\dots\sum_{i_n}\int dt_1\dots\int dt_n P_{i_0i_1\dots i_n}(t_0,t_1,\dots,t_n)F^{i_1}(t_1)\dots F^{i_n}(t_n),
\end{equation}

with the first and second order response functions being

\begin{align}
\label{eq:resp1}
&P_{i_0 i_1}(t_0,t_1)=C^r_{\mathcal{O}^{(1)}_{i_0 i_1}}\delta_{t_0t_1}+C^r_{\mathcal{O}^{(0)}_{i_0}\mathcal{M}^{(0)}_{i_1}}(t_0,t_1),
\\\nonumber
\\\nonumber
\label{eq:resp2}
&P_{i_0i_1i_2}(t_0,t_1,t_2)=C^r_{\mathcal{O}^{(2)}_{i_0i_1i_2}}\delta_{t_0t_1}\delta_{t_0t_2}
+C^r_{\mathcal{O}^{(0)}_{i_0}\mathcal{M}^{(1)}_{i_1 i_2}}(t_0,t_2)\delta_{t_1t_2}
+\frac{1}{2}(C^r_{\mathcal{O}^{(1)}_{i_0i_{1}}\mathcal{M}^{(0)}_{i_{2}}}(t_{1},t_{2})\delta_{t_0t_{1}}+C^r_{\mathcal{O}^{(1)}_{i_0i_{2}}\mathcal{M}^{(0)}_{i_{1}}}(t_{2},t_{1})\delta_{t_0t_{2}})
\\
&\qquad\qquad\qquad\qquad+C^r_{\mathcal{O}^{(0)}_{i_0}\mathcal{M}^{(0)}_{i_1}\mathcal{M}^{(0)}_{i_2}}(t_0,t_1,t_2).
\end{align}

Here $\delta_{t_it_j}\equiv \delta(t_i-t_j)$ is Dirac's delta function and we have the retarded 2 and 3-point correlators

\begin{align}
&C^r_{A^0A^1}(t_0,t_1)=-\frac{i}{\hbar}\theta(t_0-t_1)\text{tr}\left(\rho_0\left[A^0_{\mathcal{H}_0}(t_0),A^1_{\mathcal{H}_0}(t_1)\right]\right),
\\
\nonumber
&C^r_{A^0A^1A^2}(t_0,t_1,t_2)=-\frac{1}{2\hbar^2}\bigg(\theta(t_0-t_1)\theta(t_1-t_2)\text{tr}\left(\rho_0\left[\left[A^0_{\mathcal{H}_0}(t_0),A^1_{\mathcal{H}_0}(t_1)\right],A^2_{\mathcal{H}_0}(t_2)\right]\right)
\\
&\qquad\qquad\qquad\qquad\qquad\qquad+\theta(t_0-t_2)\theta(t_2-t_1)\text{tr}\left(\rho_0\left[\left[A^0_{\mathcal{H}_0}(t_0),A^2_{\mathcal{H}_0}(t_2)\right],A^1_{\mathcal{H}_0}(t_1)\right]\right)\bigg),
\end{align}

where $\theta(t_i-t_j)$ is Heaviside's step function and the observable operators $A^i$ are in the interaction picture $A^i_{\mathcal{H}_0}(t)=e^{\frac{i}{\hbar}\mathcal{H}_0t}A^ie^{-\frac{i}{\hbar}\mathcal{H}_0t}$, furthermore $\rho_0$ is the equilibrium density matrix. We also defined the '1-point correlator' $C^r_{A}=C^r_{A}(t)=C_{[A]}(t)=\text{tr}\left(\rho_0 A_{\mathcal{H}_0}(t)\right)=\text{tr}(\rho_0 A)$ to maintain consistency in the notation. Next, we make use of time-translation invariance, move to the frequency domain and express the retarded 2 and 3-point correlators in the spectral representation (see our first paper in this series \cite{Bonbien2021a} for a detailed and rather simple derivation \textit{without} the use of the standard Keldysh or Matsubara formalisms and related discussion)

\begin{align}
\label{eq:Ret2SpecF}
&C^r_{A^0A^1}(\omega_1)=\frac{1}{\pi}\hat{\mathcal{P}}^{(+)}_{\mathcal{K}^*_{\omega}}\,i\int d\varepsilon\,\text{tr}\left(\rho_0A^0G^r_{\omega_1}A^1G^{r-a}\right),
\\
\label{eq:Ret3SpecF}
&C^r_{A^0A^1A^2}(\omega_1,\omega_2)=\frac{1}{\pi}\hat{\mathcal{P}}^{(+)}_{\mathcal{K}^*_{\omega}}\hat{\mathcal{P}}^{(\Gamma_1^+)}_{A^{1}_{\omega_{1}}A^{2}_{\omega_{2}}}i\int d\varepsilon\,\text{tr}\left(\rho_0\left(A^0G^r_{\omega_1+\omega_2}A^1G^r_{\omega_2}A^2+\frac12A^1G^a_{-\omega_1}A^0G^r_{\omega_2}A^2\right)G^{r-a}\right),
\end{align}

where we defined the operations
\begin{equation}
\hat{\mathcal{P}}^{(+)}_{\mathcal{K}^*_{\omega}}f(\omega)=\frac{f(\omega)+f^*(-\omega)}{2},\quad \hat{\mathcal{P}}^{(\Gamma_1^+)}_{A^{1}_{\omega_{1}}A^{2}_{\omega_{2}}}f_{A^1A^2}(\omega_1,\omega_2)=\frac{f_{A^1A^2}(\omega_1,\omega_2)+f_{A^2A^1}(\omega_2,\omega_1)}{2},
\end{equation}
\end{widetext}
for some complex valued functions ($f^*$ denotes complex conjugation). The notation is related to projections to the irreducible representations of certain groups (see \cite{Bonbien2021a} for details). We also have Green's retarded and advanced operators $G^{r(a)}_{\pm\omega}\equiv G^{r(a)}(\varepsilon\pm\hbar\omega)$ as

\begin{equation}
\label{eq:GrGaEps}
G^r(\varepsilon) = \lim_{\eta\to 0}\frac{1}{\varepsilon-\mathcal{H}_0+i\eta},\, G^a(\varepsilon) = \lim_{\eta\to 0}\frac{1}{\varepsilon-\mathcal{H}_0-i\eta},
\end{equation}

and used the short-hand notation $G^{r-a}\equiv G^{r}(\varepsilon)-G^{a}(\varepsilon)$. We shall also make ample use of a number of identities involving Green's operators that can found in Appendix \ref{GreenIdentities}.\\

We remark that we are working in a general, many-body \textit{framework} in which all operators can be considered as being many-body, including the density matrix $\rho_0(\varepsilon)$. However, we can move to the single-particle approximation using the standard recipe of replacing all operators with their one-body counterparts, the density matrix $\rho_0(\varepsilon)$ with either the Fermi-Dirac $f(\varepsilon) = 1/(e^{\beta(\varepsilon-\mu)} + 1)$ or Bose-Einstein distribution $b(\varepsilon) = 1/(e^{\beta(\varepsilon-\mu)}-1)$ (in our case the former, since we are interested in electron transport) and performing the trace over single particle states \cite{Lax1958}.

\subsection{Non-equilibrium couplings}
In order to proceed with the calculation we first have to identify the non-equilibrium part of our evolution and read-off the couplings. The full time evolution is given by velocity gauge Hamiltonian \eqref{eq:velHamiltonian}

\
\begin{equation}
\label{eq:EHamiltonian}
\begin{split}
\mathcal{H}^{\text{vel}}&=\mathcal{H}_0(\boldsymbol{\Pi}(\textbf{p}_0,t))+e\mathcal{A}_j(\boldsymbol{\Pi}(\textbf{p}_0,t))E^j(t)
\\
&=\mathcal{H}_0(\textbf{p}_0-e\textbf{A}(t))+e\mathcal{A}_j(\textbf{p}_0-e\textbf{A}(t))E^j(t).
\end{split}
\end{equation}

We thus have for our driving field $\textbf{F}(t)=(\textbf{A}(t),\textbf{E}(t))$ and need to read off the perturbative couplings $\mathcal{M}_{i_1}^{(0)},\,\mathcal{M}_{i_1i_2}^{(1)}$ defined in \eqref{eq:HextOp}. Since we decomposed our driving field into $\textbf{A}(t)$ and $\textbf{E}(t)$, we can write for \eqref{eq:HextOp}

\begin{equation}
\begin{split}
H'(t)&=\mathcal{M}^{(0)}_{i_1}F^{i_1}(t)+\mathcal{M}^{(1)}_{i_1i_2}F^{i_1}(t)F^{i_2}(t)+O(F^3)
\\
&=\mathcal{M}^{(0)A}_{i_1}A^{i_1}(t)+\mathcal{M}^{(0)E}_{i_1}E^{i_1}(t)
\\
&\quad+\mathcal{M}^{(1)AA}_{i_1i_2}A^{i_1}(t)A^{i_2}(t)+\mathcal{M}^{(1)AE}_{i_1i_2}A^{i_1}(t)E^{i_2}(t)
\\
&\quad+\mathcal{M}^{(1)EA}_{i_1i_2}E^{i_1}(t)A^{i_2}(t)+O(A^3,A^2E),
\end{split}
\end{equation}

where $H'(t)$ corresponds to the non-equilibrium part $\mathcal{H}=\mathcal{H}_0+H'(t)$. Performing the \textit{standard} Taylor expansion (see the discussion in section \ref{Taylor}) of \eqref{eq:EHamiltonian} around $\textbf{p}_0$ to second order we have

\begin{equation}
\begin{split}
H'(t)=&-e(\partial_{p^i}\mathcal{H}_0)A^i(t)+\frac{1}{2}e^2(\partial_{p^i}\partial_{p^j}\mathcal{H}_0)A^i(t)A^j(t)
\\
&+e\mathcal{A}_iE^i(t)-e^2(\partial_{p^i}\mathcal{A}_j)A^i(t)E^j(t)
\\
&+O(A^3,A^2E),
\end{split}
\end{equation}

where we dropped the $\textbf{p}_0$ arguments. Next, we use the flat position operator $x_i=i\hbar\partial_{p^i}$ to write the operator derivatives as commutators $\partial_{p^i}\mathcal{H}_0=-\frac{i}{\hbar}[x_i,\mathcal{H}_0]=\frac{i}{\hbar}[x_i,G^{-1}]$, where $G^{-1}=\varepsilon-\mc{H}_0$ is the inverse of Green's operator, and read off the couplings

\begin{equation}
\label{eq:couplings}
\begin{split}
&\mathcal{M}^{(0)A}_{i_1}=-\frac{ie}{\hbar}[x_{i_1},G^{-1}]\equiv -j_{i_1},\,\, \mathcal{M}^{(0)E}_{i_1}=e\mathcal{A}_{i_1},
\\
&\mathcal{M}^{(1)AA}_{i_1i_2}=-\frac{1}{4}\frac{ie}{\hbar}([x_{i_1},j_{i_2}]+[x_{i_2},j_{i_1}])\equiv -j_{i_1i_2},
\\
&\mathcal{M}^{(1)AE}_{i_1i_2}=e\frac{1}{2}\frac{ie}{\hbar}[x_{i_1},\mathcal{A}_{i_2}],\,\,\mathcal{M}^{(1)EA}_{i_1i_2}=e\frac{1}{2}\frac{ie}{\hbar}[x_{i_2},\mathcal{A}_{i_1}],
\end{split}
\end{equation}

where we defined the flat current operator $j_{i}$ and its second order version $j_{ik}$. Note also that the relevant permutation symmetries were taken into account.\\

The only input left is an observable $\mathcal{O}_{i}$ whose expectation value we want to calculate as a response to the applied electric field. In order to keep with the generality, we suppose that the observable is momentum-dependent and so it obtains an implicit time-dependence as $\mathcal{O}_{i}(\boldsymbol{\Pi}(\textbf{p}_0,t))$. An example of such an observable would be the velocity \eqref{eq:velocityFrame} defined earlier. Performing the standard Taylor expansion around $\textbf{p}_0$ we can express the terms as defined in \eqref{eq:ObsOp}

\begin{equation}
\label{eq:ObsExpansion}
\begin{split}
&\mathcal{O}^{(0)}_{i_0}\equiv \mathcal{O}_{i_0},\,\,\mathcal{O}^{(1)A}_{i_0i_1}=\frac{ie}{\hbar}[x_{i_1},\mathcal{O}_{i_0}],
\\
&\mathcal{O}^{(2)AA}_{i_0i_1i_2}=\frac{1}{4}\left(\frac{ie}{\hbar}\right)^2([x_{i_1},[x_{i_2},\mathcal{O}_{i_0}]]+[x_{i_2},[x_{i_1},\mathcal{O}_{i_0}]]).
\end{split}
\end{equation}

\subsection{The Coulomb gauge and the vector potential}

We now have to discuss one final subtlety related to the vector potential $\textbf{A}(t)$. We have represented the spatially uniform, time-dependent classical electric field in the electric dipole approximation of the Coulomb gauge as $\textbf{E}=-\partial\textbf{A}(t)/\partial t$. However, even though the gauge has been 'fixed', crucially, we still have some remaining freedom. Consider the Coulomb gauge condition $\text{div}\textbf{A}(\textbf{x},t)=0$ in the general, non-uniform, time-dependent case. Now suppose we perform a time-\textit{independent} gauge transformation $\Lambda(\textbf{x})$ under which only the vector potential changes as $\textbf{A}'(\textbf{x},t)=\textbf{A}(\textbf{x},t)+\nabla\Lambda(\textbf{x})$. We would like the transformed $\textbf{A}'(\textbf{x},t)$ to continue satisfying the Coulomb gauge condition, which requires $\text{div}\textbf{A}'(\textbf{x},t)=\text{div}\textbf{A}(\textbf{x},t)+\nabla^2\Lambda(\textbf{x})=\nabla^2\Lambda(\textbf{x})=0$, where we used the fact that $\textbf{A}(\textbf{x},t)$ is already in the Coulomb gauge. We thus see that choosing as gauge function any smooth function that satisfies the condition $\nabla^2\Lambda(\textbf{x})=0$ will not force us to leave the gauge \cite{Jackson}. One particular function that satisfies this condition is a linear function $\Lambda(\textbf{x})=\textbf{c}\cdot \textbf{x}$ where $\textbf{c}$ is some constant vector. In general, such a gauge function is discarded and not considered since it is not compactly supported---a common, mathematically convenient requirement. However, the gauge function is a redundancy and we have to be careful in how the boundary conditions imposed on the physical fields reach the level of the gauge function. In particular, the electromagnetic field $(\textbf{E},\textbf{B})$ is given by a combination of the first derivatives of the scalar and vector potentials $(\phi,\textbf{A})$. Suppose the electromagnetic field satisfies certain asymptotic boundary conditions. Then, since the electromagnetic field is the derivative of the potentials, the corresponding asymptotic boundary conditions that are satisfied by the potentials can only be specified upto a constant. Going one step further to the level of the gauge functions, the asymptotic boundary conditions for these can then only be specified upto a linear function. We can thus always perform a gauge transformation with a linear gauge function and remain compatible with \textit{any} asymptotic boundary conditions imposed on the \textit{physical} electromagnetic field. With the linear gauge function, the vector potential changes to $\textbf{A}'(\textbf{x},t)=\textbf{A}(\textbf{x},t)+\textbf{c}$ and we still remain in the Coulomb gauge. In the electric dipole approximation, the vector potential becomes spatially uniform, but this freedom, that cannot be fixed, still remains since it is just a constant $\textbf{A}'(t)=\textbf{A}(t)+\textbf{c}$. Note how this is connected to $\textbf{p}_0$. Momentum changes in time as $\textbf{p}_0-e\textbf{A}(t)$. Adding a constant to $\textbf{A}(t)$ amounts to a shift of the origin in the coordinate charts used to  cover momentum space.\\

The physical responses cannot be dependent on such an arbitrary constant, consequently we have to make sure that the response functions satisfy appropriate gauge conditions. We will show explicitly that in our framework, these are satisfied for both first and second order responses and are the constraints responsible for the cancellations of unphysical divergences in the static limit.\\

The fact that the cancellation of such divergences, for charge current responses in particular, is closely related to gauge invariance is well-known \cite{Rammer1986,RammerQT} however, the general perspective shown here and the specific technique to be discussed below is, to the best of our knowledge, not used.\\

\subsection{First order response}

We denote the total first order response function to the electric field as $P^{E}_{i_0i_1}(t_0,t_1)$, and the response to the $\textbf{A}(t)$ and $\textbf{E}(t)$ parts as $K^{A}_{i_0i_1}(t_0,t_1)$ and $K^{E}_{i_0i_1}(t_0,t_1)$ respectively. We take the $n=1$ term from \eqref{eq:Obsn} and obtain

\begin{equation}
\label{eq:firstOrderTtotal}
\begin{split}
&\langle\mathcal{O}_{i_0}(t_0)\rangle_1=\int dt_1 P^{E}_{i_0i_1}(t_0,t_1)E^{i_1}(t_1)
\\
&\,=\int dt_1 K^{A}_{i_0i_1}(t_0,t_1)A^{i_1}(t_1)+\int dt_1 K^{E}_{i_0i_1}(t_0,t_1)E^{i_1}(t_1).
\end{split}
\end{equation}

Furthermore, \eqref{eq:resp1} provides the expression for the first order response functions $K^{A}_{i_0i_1}(t_0,t_1)$ and $K^{E}_{i_0i_1}(t_0,t_1)$ with retarded correlators

\begin{equation}
\label{eq:firstOrderT}
\begin{split}
&K^{A}_{i_0i_1}(t_0,t_1)=C^r_{\mathcal{O}^{(1)A}_{i_0 i_1}}\delta_{t_0t_1}-C^r_{\mathcal{O}_{i_0}j_{i_1}}(t_0,t_1),
\\
&K^{E}_{i_0i_1}(t_0,t_1)=C^r_{\mathcal{O}_{i_0}\mathcal{M}^{(0)E}_{i_1}}(t_0,t_1),
\end{split}
\end{equation}

where the minus sign before the second term in $K^{A}_{i_0i_1}(t_0,t_1)$ arose due to the coupling being $\mathcal{M}^{(0)A}_{i_1}=-j_{i_1}$ (cf. Eq. \eqref{eq:couplings}).\\

Recalling our discussion on the remaining freedom in the Coulomb gauge, in order to find the condition that $K^{A}_{i_0i_1}(t_0,t_1)$ has to satisfy, we consider the response to an applied vector potential given by $\textbf{A}(t)+\textbf{c}$. This yields

\begin{equation}
\begin{split}
&\int dt_1 K^{A}_{i_0i_1}(t_0,t_1)(A^{i_1}(t_1)+c^{i_1})
\\
&=\int dt_1 K^{A}_{i_0i_1}(t_0,t_1)A^{i_1}(t_1)+c^{i_1}\int dt_1 K^{A}_{i_0i_1}(t_0,t_1).
\end{split}
\end{equation}

Independence from the constant requires

\begin{equation}
\int dt_1 K^{A}_{i_0i_1}(t_0,t_1)=0.
\end{equation}

To make contact with experimental results we have to express our response functions in the frequency domain. Using the time-translation invariance of the retarded correlator, we take the Fourier transform of \eqref{eq:firstOrderT} and get

\begin{equation}
\label{eq:firstOrderOmega}
\begin{split}
&K^{A}_{i_0i_1}(\omega_1)=C^r_{\mathcal{O}^{(1)A}_{i_0 i_1}}-C^r_{\mathcal{O}_{i_0}j_{i_1}}(\omega_1),
\\
&K^{E}_{i_0i_1}(\omega_1)=C^r_{\mathcal{O}_{i_0}\mathcal{M}^{(0)E}_{i_1}}(\omega_1),
\end{split}
\end{equation}

with the gauge condition becoming

\begin{equation}
\label{eq:firstOrderGauge}
K^{A}_{i_0i_1}(\omega_1=0)=0.
\end{equation}

We now make use of the spectral representation \eqref{eq:Ret2SpecF} of the retarded 2-point correlator

\begin{equation}
\label{eq:firstOrder2Corr}
\begin{split}
C^r_{\mathcal{O}_{i_0}j_{i_1}}(\omega_1)=\frac{i}{2\pi}\int& d\varepsilon\,\rho_0(\varepsilon)\text{tr}((\mathcal{O}_{i_0}G^r_{\omega_1}j_{i_1}
\\
&+j_{i_1}G^a_{-\omega_1}\mathcal{O}_{i_0})G^{r-a}).
\end{split}
\end{equation}

Similarly, we write for the '1-point' correlator:

\begin{equation}
\label{eq:firstOrder1Corr}
\begin{split}
C^r_{\mathcal{O}^{(1)A}_{i_0 i_1}}&=\text{tr}(\rho_0\mathcal{O}^{(1)A}_{i_0 i_1})
\\
&=\frac{i}{2\pi}\frac{ie}{\hbar}\int d\varepsilon\,\rho_0(\varepsilon)\text{tr}([x_{i_1},\mathcal{O}_{i_0}]G^{r-a}),
\end{split}
\end{equation}

where we used identity $G^{r-a}=-i2\pi\delta(\varepsilon-\mc{H}_0)$ for Green's operators and the observable expansion \eqref{eq:ObsExpansion}.\\

We now prove the gauge condition $K^{A}_{i_0i_1}(\omega_1=0)=0$ via the standard method from \cite{Bastin1971}. We take $\omega_1=0$ in \eqref{eq:firstOrder2Corr}, use the cyclicity of the trace and the definition $j_{i_1}=\frac{ie}{\hbar}[x_{i_1},G^{-1}]$ of the flat current. Then we have

\begin{equation}
\begin{split}
&C^r_{\mathcal{O}_{i_0}j_{i_1}}(\omega_1=0)=\frac{i}{2\pi}\int d\varepsilon\,\rho_0(\varepsilon)\text{tr}((\mathcal{O}_{i_0}G^rj_{i_1}
\\
&\qquad\qquad+j_{i_1}G^a\mathcal{O}_{i_0})G^{r-a})
\\
&=\frac{i}{2\pi}\frac{ie}{\hbar}\int d\varepsilon\,\rho_0(\varepsilon)\text{tr}(\mathcal{O}_{i_0}G^r[x_{i_1},G^{-1}]G^r
\\
&\qquad\qquad\qquad-G^{a}[x_{i_1},G^{-1}]G^a\mathcal{O}_{i_0})
\\
&=-\frac{i}{2\pi}\frac{ie}{\hbar}\int d\varepsilon\,\rho_0(\varepsilon)\text{tr}(\mathcal{O}_{i_0}[x_{i_1},G^{r-a}])
\\
&=\frac{i}{2\pi}\frac{ie}{\hbar}\int d\varepsilon\,\rho_0(\varepsilon)\text{tr}([x_{i_1},\mathcal{O}_{i_0}]G^{r-a}),
\end{split}
\end{equation}

which is the same as \eqref{eq:firstOrder1Corr}. Thus $C^r_{\mathcal{O}_{i_0}j_{i_1}}(\omega_1=0)=C^r_{\mathcal{O}^{(1)A}_{i_0 i_1}}$, and looking at \eqref{eq:firstOrderOmega} we see that

\begin{equation}
K^{A}_{i_0i_1}(\omega_1=0)=C^r_{\mathcal{O}^{(1)A}_{i_0 i_1}}-C^r_{\mathcal{O}_{i_0}j_{i_1}}(\omega_1=0)=0,
\end{equation} 

thereby satisfying the condition. Note that $C^r_{\mathcal{O}_{i_0}j_{i_1}}(\omega_1=0)$ is symmetric under the exchange of operators $\mathcal{O}_{i_0},\,j_{i_1}$. In the special case of charge current response, i.e., $\mathcal{O}_{i_0}=j_{i_0}$, this means that the gauge condition only concerns the longitudinal part of the conductivity with the transverse part, responsible for effects such as the anomalous Hall effect \cite{NagaosaAHErev}, remaining unaffected. Crucially, the gauge condition is necessary for a proper discussion of first order longitudinal transport effects such as those occuring in the Boltzmann limit \cite{RammerQT}. The fact that the gauge condition carries such a symmetry is a peculiarity of first order response and does not spill over to higher orders. This results in the unfortunate circumstance that no effects of the latter sort can be analyzed without a thorough overview of the relevant gauge conditions.\\

We are led to the standard result \cite{RammerQT}

\begin{equation}
K^{A}_{i_0i_1}(\omega_1)=-(C^r_{\mathcal{O}_{i_0}j_{i_1}}(\omega_1)-C^r_{\mathcal{O}_{i_0}j_{i_1}}(0)).
\end{equation}

We can now find the frequency domain expression of the total response function $P^{E}_{i_0i_1}(t_0,t_1)$ defined in \eqref{eq:firstOrderTtotal}. Since we have taken care of the gauge ambiguity we can now write $E^i(\omega)=i\omega A^i(\omega)$ and consequently combine the frequency domain version of the two responses $K^{A}_{i_0i_1}(t_0,t_1)$ and $K^{E}_{i_0i_1}(t_0,t_1)$. Indeed, taking the Fourier transform of \eqref{eq:firstOrderTtotal} and making use of $E^i(\omega)=i\omega A^i(\omega)$ we have for the total response $\langle\mathcal{O}_{i_0}(\omega_1)\rangle_1= P^{E}_{i_0i_1}(\omega_1)E^{i_1}(\omega_1)$, where

\begin{equation}
\begin{split}
P^{E}_{i_0i_1}(\omega_1)&=\frac{K^{A}_{i_0i_1}(\omega_1)}{i\omega_1}+K^{E}_{i_0i_1}(\omega_1)
\\
&=i\frac{C^r_{\mathcal{O}_{i_0}j_{i_1}}(\omega_1)-C^r_{\mathcal{O}_{i_0}j_{i_1}}(0)}{\omega_1}+C^r_{\mathcal{O}_{i_0}\mathcal{M}^{(0)E}_{i_1}}(\omega_1).
\end{split}
\end{equation}

Note that this expression is manifestly free of apparent divergences and the underlying mechanism is the gauge condition. This will generalize to second order as we will show in the next section.\\
In order to arrive at our final result we have to apply some identities between Green's operators that can be shown directly from the definitions \eqref{eq:GrGaEps}

\begin{equation}
\label{eq:GrGaOmega}
\begin{split}
&G^r_{\omega}-G^r=-\hbar\omega G^r_{\omega}G^r,
\\
&G^a_{-\omega}-G^a=\hbar\omega G^a_{-\omega}G^a.
\end{split}
\end{equation}

Using these identities with the spectral representation \eqref{eq:firstOrder2Corr}, we obtain
\begin{equation}
\label{eq:CrOmegaCr0}
\begin{split}
&i\frac{C^r_{\mathcal{O}_{i_0}j_{i_1}}(\omega_1)-C^r_{\mathcal{O}_{i_0}j_{i_1}}(0)}{\omega_1}
\\
&=\frac{i}{\omega_1}\frac{i}{2\pi}\int d\varepsilon\,\rho_0(\varepsilon)\text{tr}((\mathcal{O}_{i_0}(G^r_{\omega_1}-G^r)j_{i_1}
\\
&\qquad\quad+j_{i_1}(G^a_{-\omega_1}-G^a)\mathcal{O}_{i_0})G^{r-a})
\\
&=-i\hbar\frac{i}{2\pi}\int d\varepsilon\,\rho_0(\varepsilon)\text{tr}((\mathcal{O}_{i_0}G^r_{\omega_1}G^rj_{i_1}
\\
&\qquad\quad-j_{i_1}G^a_{-\omega_1}G^a\mathcal{O}_{i_0})G^{r-a}).
\end{split}
\end{equation}

Next, we express $K^{E}_{i_0i_1}(\omega_1)$ in the spectral representation and use the relevant coupling from \eqref{eq:couplings} to get

\begin{equation}
\begin{split}
&K^{E}_{i_0i_1}(\omega_1)=e\frac{i}{2\pi}\int d\varepsilon\,\rho_0(\varepsilon)\text{tr}((\mathcal{O}_{i_0}G^r_{\omega_1}\mathcal{A}_{i_1}
\\
&\qquad+\mathcal{A}_{i_1}G^a_{-\omega_1}\mathcal{O}_{i_0})G^{r-a})
\\
&=-e\frac{i}{2\pi}\int d\varepsilon\,\rho_0(\varepsilon)\text{tr}((\mathcal{O}_{i_0}G^r_{\omega_1}G^r[\mathcal{A}_{i_1},G^{-1}]
\\
&\qquad-[\mathcal{A}_{i_1},G^{-1}]G^a_{-\omega_1}G^a\mathcal{O}_{i_0})G^{r-a}),
\end{split}
\end{equation}

where the second line is the result of the insertions $G^rG^{-1}\to 1$ and $G^aG^{-1}\to 1$ between $G^r_{\omega_1}$, $\mathcal{A}_{i_1}$ and $A_{i_1}$, $G^a_{-\omega_1}$ respectively, together with $G^{-1}(G^r-G^a)\to 0$. Combining this with the result of \eqref{eq:CrOmegaCr0} we have

\begin{equation}
\begin{split}
&\frac{K^{A}_{i_0i_1}(\omega_1)}{i\omega_1}+K^{E}_{i_0i_1}(\omega_1)
\\
&=-i\hbar\frac{i}{2\pi}\int d\varepsilon\,\rho_0(\varepsilon)\text{tr}((\mathcal{O}_{i_0}G^r_{\omega_1}G^r\left(j_{i_1}-\frac{ie}{\hbar}[\mathcal{A}_{i_1},G^{-1}]\right)
\\
&\qquad\quad-\left(j_{i_1}-\frac{ie}{\hbar}[\mathcal{A}_{i_1},G^{-1}]\right)G^a_{-\omega_1}G^a\mathcal{O}_{i_0})G^{r-a})
\\
&=\frac{\hbar}{2\pi}\int d\varepsilon\,\rho_0(\varepsilon)\text{tr}((\mathcal{O}_{i_0}G^r_{\omega_1}G^r J_{i_1}
\\
&\qquad\quad-J_{i_1}G^a_{-\omega_1}G^a\mathcal{O}_{i_0})G^{r-a}),
\end{split}
\end{equation}

where we defined the curved current operator $J_{i}=j_{i}-\frac{ie}{\hbar}[\mathcal{A}_{i},G^{-1}]=\frac{ie}{\hbar}[x_i-\mathcal{A}_{i},G^{-1}]=\frac{ie}{\hbar}[r_i,G^{-1}]$ with $r_i=x_i-\mathcal{A}_{i}=i\hbar(\partial_{i}+i/\hbar\mathcal{A}_i)\to i\hbar\mathcal{D}_{i}$ being the curved position operator given by the covariant derivative.\\

Our final result for the first order response is

\begin{equation}
\label{eq:firstOrderFull}
\begin{split}
P^{E}_{i_0i_1}(\omega_1)=\frac{\hbar}{2\pi}&\int d\varepsilon\,\rho_0(\varepsilon)\text{tr}((\mathcal{O}_{i_0}G^r_{\omega_1}G^rJ_{i_1}
\\
&-J_{i_1}G^a_{-\omega_1}G^a\mathcal{O}_{i_0})G^{r-a}).
\end{split}
\end{equation}

This is a fully gauge invariant result, since all the components of all the operators transform covariantly under under a change of frame and the curved current operator $J_i$ , which is the covariant derivative of $\mathcal{H}_0$, emerged naturally in our curved space formalism. This result is a generalization of the widely-used Kubo-Bastin formula \cite{Bastin1971} to frequency dependent responses. Indeed, taking $\omega_1=0$ and $G^2=-dG/d\varepsilon$ we get the standard Kubo-Bastin formula describing linear response in the static case \cite{Bastin1971}.\\
Taking a look at the result of \eqref{eq:CrOmegaCr0}, we see that the only difference from \eqref{eq:firstOrderFull} is that the flat current $j_i$ is replaced by the curved current $J_i$ meaning that we can also write the final result as 

\begin{equation}
\label{eq:firstOrderKubo}
P^{E}_{i_0i_1}(\omega_1)=i\frac{C^r_{\mathcal{O}_{i_0}J_{i_1}}(\omega_1)-C^r_{\mathcal{O}_{i_0}J_{i_1}}(0)}{\omega_1},
\end{equation}

where $J_i$ is now the curved current operator. This has the form of the standard Kubo formula \cite{RammerQT}. Thus, formally, we could have forgone the entire discussion on curved connections, simply exchanged the flat current $j_i$ with the curved current $J_i$ in the standard Kubo formula, and found that the truncation of the Hilbert space does not affect the qualitative aspects of the formula. However, this trickery would only work for first order response and completely break down for higher orders due to subtleties involving the path-dependence of the parallel transporter beyond first order.\\

\subsection{Second order response}
\label{secondorderresponse}

Having warmed-up with the first order response we now apply our formalism to second order response. We denote the total second order response function to the electric field as $P^{E}_{i_0i_1i_2}(t_0,t_1,t_2)$, and the response to the $A^{i}A^j$, $A^iE^j$, $E^iA^j$ and $E^iE^j$ parts as $K^{AA}_{i_0i_1i_2}(t_0,t_1,t_2)$, $K^{AE}_{i_0i_1i_2}(t_0,t_1,t_2)$, $K^{EA}_{i_0i_1i_2}(t_0,t_1,t_2)$ and $K^{EE}_{i_0i_1i_2}(t_0,t_1,t_2)$ respectively. We take the $n=2$ term from \eqref{eq:Obsn} and obtain

\begin{widetext}
\begin{equation}
\label{eq:secondOrderTtotal}
\begin{split}
&\langle\mathcal{O}_{i_0}(t_0)\rangle_2=\int dt_1\int dt_2 P^{E}_{i_0i_1i_2}(t_0,t_1,t_2)E^{i_1}(t_1)E^{i_2}(t_2)
\\
&=\int dt_1\int dt_2 K^{AA}_{i_0i_1i_2}(t_0,t_1,t_2)A^{i_1}(t_1)A^{i_2}(t_2)
+\int dt_1\int dt_2 K^{AE}_{i_0i_1i_2}(t_0,t_1,t_2)A^{i_1}(t_1)E^{i_2}(t_2)
\\
&\quad+\int dt_1\int dt_2 K^{EA}_{i_0i_1i_2}(t_0,t_1,t_2)E^{i_1}(t_1)A^{i_2}(t_2)
+\int dt_1\int dt_2 K^{EE}_{i_0i_1i_2}(t_0,t_1,t_2)E^{i_1}(t_1)E^{i_2}(t_2).
\end{split}
\end{equation}

Even though $K^{AE}_{i_0i_1i_2}(t_0,t_1,t_2)= K^{EA}_{i_0i_2i_1}(t_0,t_2,t_1)$, meaning that the two terms are thus closely related and could be combined by relabeling indices, we keep them separate in order to avoid confusion later on.\\
The retarded correlator expressions \eqref{eq:resp2} for the second order response functions are

\begin{align}
\label{eq:KAATT}
&K^{AA}_{i_0i_1i_2}(t_0,t_1,t_2)=C^r_{\mathcal{O}^{(2)AA}_{i_0i_1i_2}}\delta_{t_0t_1}\delta_{t_0t_2}
-C^r_{\mathcal{O}_{i_0}j_{i_1 i_2}}(t_0,t_2)\delta_{t_1t_2}
-\frac{1}{2}(C^r_{\mathcal{O}^{(1)A}_{i_0i_1}j_{i_2}}(t_{1},t_{2})\delta_{t_0t_{1}}+
C^r_{\mathcal{O}^{(1)A}_{i_0i_2}j_{i_1}}(t_{2},t_{1})\delta_{t_0t_{2}})
\\\nonumber
&\qquad\qquad\qquad\qquad+C^r_{\mathcal{O}_{i_0}j_{i_1}j_{i_2}}(t_0,t_1,t_2),
\\
\label{eq:KAETT}
&K^{AE}_{i_0i_1i_2}(t_0,t_1,t_2)=
C^r_{\mathcal{O}_{i_0}\mathcal{M}^{(1)AE}_{i_1 i_2}}(t_0,t_2)\delta_{t_1t_2}
+\frac12 C^r_{\mathcal{O}^{(1)A}_{i_0i_1}\mathcal{M}^{(0)E}_{i_2}}(t_{1},t_{2})\delta_{t_0t_{1}}
-C^r_{\mathcal{O}_{i_0}j_{i_1}\mathcal{M}^{(0)E}_{i_2}}(t_0,t_1,t_2),
\\
\label{eq:KEATT}
&K^{EA}_{i_0i_1i_2}(t_0,t_1,t_2)=
C^r_{\mathcal{O}_{i_0}\mathcal{M}^{(1)EA}_{i_1 i_2}}(t_0,t_2)\delta_{t_1t_2}
+\frac12 C^r_{\mathcal{O}^{(1)A}_{i_0i_2}\mathcal{M}^{(0)E}_{i_1}}(t_{2},t_{1})\delta_{t_0t_{2}}
-C^r_{\mathcal{O}_{i_0}\mathcal{M}^{(0)E}_{i_1}j_{i_2}}(t_0,t_1,t_2),
\\
\label{eq:KEETT}
&K^{EE}_{i_0i_1i_2}(t_0,t_1,t_2)=C^r_{\mathcal{O}_{i_0}\mathcal{M}^{(0)E}_{i_1}\mathcal{M}^{(0)E}_{i_2}}(t_0,t_1,t_2),
\end{align}

where the minus signs appear due to the coupling definitions $\mathcal{M}^{(0)A}_{i_1}=-j_{i_1}$ and $\mathcal{M}^{(1)AA}_{i_1i_2}=-j_{i_1i_2}$ in \eqref{eq:couplings}.\\

We now find the gauge conditions that have to be satisfied. Suppose the applied vector potential is $\textbf{A}(t)+\textbf{c}$, then the response functions involved are

\begin{equation}
\begin{split}
&\int dt_1\int dt_2 K^{AA}_{i_0i_1i_2}(t_0,t_1,t_2)(A^{i_1}(t_1)+c^{i_1})(A^{i_2}(t_2)+c^{i_2})=\int dt_1\int dt_2 K^{AA}_{i_0i_1i_2}(t_0,t_1,t_2)A^{i_1}(t_1)A^{i_2}(t_2)
\\
&+c^{i_1}\int dt_1\int dt_2 K^{AA}_{i_0i_1i_2}(t_0,t_1,t_2)A^{i_2}(t_2)
+c^{i_2}\int dt_1\int dt_2 K^{AA}_{i_0i_1i_2}(t_0,t_1,t_2)A^{i_1}(t_1)
\\
&+c^{i_1}c^{i_2}\int dt_1\int dt_2 K^{AA}_{i_0i_1i_2}(t_0,t_1,t_2).
\end{split}
\end{equation}
We can read-off the requirements for independence from the arbitrary constant
\begin{equation}
\label{eq:secondOrderGaugeAA}
\int dt_1 K^{AA}_{i_0i_1i_2}(t_0,t_1,t_2)=0,\,\int dt_2 K^{AA}_{i_0i_1i_2}(t_0,t_1,t_2)=0,\,\int dt_1\int dt_2 K^{AA}_{i_0i_1i_2}(t_0,t_1,t_2)=0.
\end{equation}

These conditions are in fact not independent. The first two are related by symmetry since $K^{AA}_{i_0i_1i_2}(t_0,t_1,t_2)=K^{AA}_{i_0i_2i_1}(t_0,t_2,t_1)$ and the third is satisfied if the first two are.\\
Similarly we have for the other two response functions

\begin{equation}
\label{eq:secondOrderGaugeAE}
\begin{split}
&\int dt_1K^{AE}_{i_0i_1i_2}(t_0,t_1,t_2)=0,
\\
&\int dt_2K^{EA}_{i_0i_1i_2}(t_0,t_1,t_2)=0,
\end{split}
\end{equation}
which are likewise related by symmetry.\\

We now move on to the frequency domain. Making use of time-translation invariance and applying the Fourier transform to response functions \eqref{eq:KAATT}-\eqref{eq:KEETT} we find

\begin{align}
\label{eq:KAAomega}
&K^{AA}_{i_0i_1i_2}(\omega_1,\omega_2)=C^r_{\mathcal{O}^{(2)AA}_{i_0i_1i_2}}
-C^r_{\mathcal{O}_{i_0}j_{i_1 i_2}}(\omega_1+\omega_2)
-\frac{1}{2}(C^r_{\mathcal{O}^{(1)A}_{i_0i_1}j_{i_2}}(\omega_2)+
C^r_{\mathcal{O}^{(1)A}_{i_0i_2}j_{i_1}}(\omega_1))
\\\nonumber
&\qquad\qquad\qquad\qquad+C^r_{\mathcal{O}_{i_0}j_{i_1}j_{i_2}}(\omega_1,\omega_2),
\\
\label{eq:KAEomega}
&K^{AE}_{i_0i_1i_2}(\omega_1,\omega_2)=
C^r_{\mathcal{O}_{i_0}\mathcal{M}^{(1)AE}_{i_1 i_2}}(\omega_1+\omega_2)
+\frac12 C^r_{\mathcal{O}^{(1)A}_{i_0i_1}\mathcal{M}^{(0)E}_{i_2}}(\omega_2)
-C^r_{\mathcal{O}_{i_0}j_{i_1}\mathcal{M}^{(0)E}_{i_2}}(\omega_1,\omega_2),
\\
\label{eq:KEAomega}
&K^{EA}_{i_0i_1i_2}(\omega_1,\omega_2)=
C^r_{\mathcal{O}_{i_0}\mathcal{M}^{(1)EA}_{i_1 i_2}}(\omega_1+\omega_2)
+\frac12 C^r_{\mathcal{O}^{(1)A}_{i_0i_2}\mathcal{M}^{(0)E}_{i_1}}(\omega_1)
-C^r_{\mathcal{O}_{i_0}\mathcal{M}^{(0)E}_{i_1}j_{i_2}}(\omega_1,\omega_2),
\\
\label{eq:KEEomega}
&K^{EE}_{i_0i_1i_2}(\omega_1,\omega_2)=C^r_{\mathcal{O}_{i_0}\mathcal{M}^{(0)E}_{i_1}\mathcal{M}^{(0)E}_{i_2}}(\omega_1,\omega_2).
\end{align}
\end{widetext}

The gauge conditions \eqref{eq:secondOrderGaugeAA} and \eqref{eq:secondOrderGaugeAE} become
\begin{equation}
\label{eq:secondOrderGauge}
\begin{split}
&K^{AA}_{i_0i_1i_2}(\omega_1,0)=K^{AA}_{i_0i_1i_2}(0,\omega_2)=K^{AA}_{i_0i_1i_2}(0,0)=0,
\\
&K^{AE}_{i_0i_1i_2}(0,\omega_2)=K^{EA}_{i_0i_1i_2}(\omega_1,0)=0.
\end{split}
\end{equation}

Note that $\omega_1,\,\omega_2$ are \textit{arbitrary}. We use the spectral representation to prove that these conditions, however unlikely, do in fact hold but, due to the length of the calculation, we relegate the proof to Appendix \ref{secondOrderGaugeProof}.\\

The existence of these conditions accounts for an intimate relation between the retarded correlators making up the response functions. Indeed, consider the following. We have

\begin{equation}
\label{eq:secondOrderCondK}
K^{AA}_{i_0i_1i_2}(\omega_1,0)+K^{AA}_{i_0i_1i_2}(0,\omega_2)-K^{AA}_{i_0i_1i_2}(0,0)=0,
\end{equation}
since each term in the sum is itself zero. Plugging in the relevant retarded correlator expressions from \eqref{eq:KAAomega} into this sum and rearranging we obtain

\begin{equation}
\label{eq:secondOrderCondAA}
\begin{split}
&C^r_{\mathcal{O}^{(2)AA}_{i_0i_1i_2}}
-\frac{1}{2}(C^r_{\mathcal{O}^{(1)A}_{i_0i_1}j_{i_2}}(\omega_2)+
C^r_{\mathcal{O}^{(1)A}_{i_0i_2}j_{i_1}}(\omega_1))
\\
&=C^r_{\mathcal{O}_{i_0}j_{i_1 i_2}}(\omega_1)+C^r_{\mathcal{O}_{i_0}j_{i_1 i_2}}(\omega_2)-C^r_{\mathcal{O}_{i_0}j_{i_1 i_2}}(0)
\\
&\quad -C^r_{\mathcal{O}_{i_0}j_{i_1}j_{i_2}}(\omega_1,0)-C^r_{\mathcal{O}_{i_0}j_{i_1}j_{i_2}}(0,\omega_2)+C^r_{\mathcal{O}_{i_0}j_{i_1}j_{i_2}}(0,0).
\end{split}
\end{equation}

Similarly the conditions $K^{AE}_{i_0i_1i_2}(0,\omega_2)=0$ and $K^{EA}_{i_0i_1i_2}(\omega_1,0)=0$ yield the respective relations

\begin{equation}
\label{eq:secondOrderCondAE}
\begin{split}
&\frac12 C^r_{\mathcal{O}^{(1)A}_{i_0i_1}\mathcal{M}^{(0)E}_{i_2}}(\omega_2)=-C^r_{\mathcal{O}_{i_0}\mathcal{M}^{(1)AE}_{i_1 i_2}}(\omega_2)
+C^r_{\mathcal{O}_{i_0}j_{i_1}\mathcal{M}^{(0)E}_{i_2}}(0,\omega_2),
\\
&\frac12 C^r_{\mathcal{O}^{(1)A}_{i_0i_2}\mathcal{M}^{(0)E}_{i_1}}(\omega_1)=-C^r_{\mathcal{O}_{i_0}\mathcal{M}^{(1)EA}_{i_1 i_2}}(\omega_1)
+C^r_{\mathcal{O}_{i_0}\mathcal{M}^{(0)E}_{i_1}j_{i_2}}(\omega_1,0).
\end{split}
\end{equation}

We can use these relations to rewrite the responses in \eqref{eq:KAAomega}-\eqref{eq:KEAomega} as

\begin{widetext}
\begin{align}
\label{eq:KAAomega1}
&K^{AA}_{i_0i_1i_2}(\omega_1,\omega_2)=
-(C^r_{\mathcal{O}_{i_0}j_{i_1 i_2}}(\omega_1+\omega_2)
-C^r_{\mathcal{O}_{i_0}j_{i_1 i_2}}(\omega_1)-C^r_{\mathcal{O}_{i_0}j_{i_1 i_2}}(\omega_2)+C^r_{\mathcal{O}_{i_0}j_{i_1 i_2}}(0))
\\\nonumber
&\qquad\qquad\qquad\qquad+C^r_{\mathcal{O}_{i_0}j_{i_1}j_{i_2}}(\omega_1,\omega_2)-C^r_{\mathcal{O}_{i_0}j_{i_1}j_{i_2}}(\omega_1,0)-C^r_{\mathcal{O}_{i_0}j_{i_1}j_{i_2}}(0,\omega_2)+C^r_{\mathcal{O}_{i_0}j_{i_1}j_{i_2}}(0,0),
\\
\label{eq:KAEomega1}
&K^{AE}_{i_0i_1i_2}(\omega_1,\omega_2)=
C^r_{\mathcal{O}_{i_0}\mathcal{M}^{(1)AE}_{i_1 i_2}}(\omega_1+\omega_2)
-C^r_{\mathcal{O}_{i_0}\mathcal{M}^{(1)AE}_{i_1 i_2}}(\omega_2)
-(C^r_{\mathcal{O}_{i_0}j_{i_1}\mathcal{M}^{(0)E}_{i_2}}(\omega_1,\omega_2)-C^r_{\mathcal{O}_{i_0}j_{i_1}\mathcal{M}^{(0)E}_{i_2}}(0,\omega_2)),
\\
\label{eq:KEAomega1}
&K^{EA}_{i_0i_1i_2}(\omega_1,\omega_2)=
C^r_{\mathcal{O}_{i_0}\mathcal{M}^{(1)EA}_{i_1 i_2}}(\omega_1+\omega_2)
-C^r_{\mathcal{O}_{i_0}\mathcal{M}^{(1)EA}_{i_1 i_2}}(\omega_1)
-(C^r_{\mathcal{O}_{i_0}\mathcal{M}^{(0)E}_{i_1}j_{i_2}}(\omega_1,\omega_2)-C^r_{\mathcal{O}_{i_0}\mathcal{M}^{(0)E}_{i_1}j_{i_2}}(\omega_1,0)).
\end{align}

Having applied the gauge conditions, we take the Fourier transform of the total response \eqref{eq:secondOrderTtotal}

\begin{equation}
\label{eq:secondOrderTotalOmega1}
\langle\mathcal{O}_{i_0}(\omega)\rangle_2=\int \frac{d\omega_1}{2\pi}\int \frac{d\omega_2}{2\pi}2\pi\delta(\omega-\omega_1-\omega_2) P^{E}_{i_0i_1i_2}(\omega_1,\omega_2)E^{i_1}(\omega_1)E^{i_2}(\omega_2),
\end{equation}

and use $E^i(\omega)=i\omega A^{i}(\omega)$ to find

\begin{equation}
\label{eq:secondOrderStart}
\begin{split}
P^{E}_{i_0i_1i_2}(\omega_1,\omega_2)=&-\frac{K^{AA}_{i_0i_1i_2}(\omega_1,\omega_2)}{\omega_1\omega_2}+\frac{K^{AE}_{i_0i_1i_2}(\omega_1,\omega_2)}{i\omega_1}+\frac{K^{EA}_{i_0i_1i_2}(\omega_1,\omega_2)}{i\omega_2}+K^{EE}_{i_0i_1i_2}(\omega_1,\omega_2)
\\
=&\frac{C^r_{\mathcal{O}_{i_0}j_{i_1 i_2}}(\omega_1+\omega_2)
-C^r_{\mathcal{O}_{i_0}j_{i_1 i_2}}(\omega_1)-C^r_{\mathcal{O}_{i_0}j_{i_1 i_2}}(\omega_2)+C^r_{\mathcal{O}_{i_0}j_{i_1 i_2}}(0)}{\omega_1\omega_2}
\\
&-\frac{C^r_{\mathcal{O}_{i_0}j_{i_1}j_{i_2}}(\omega_1,\omega_2)-C^r_{\mathcal{O}_{i_0}j_{i_1}j_{i_2}}(\omega_1,0)-C^r_{\mathcal{O}_{i_0}j_{i_1}j_{i_2}}(0,\omega_2)+C^r_{\mathcal{O}_{i_0}j_{i_1}j_{i_2}}(0,0)}{\omega_1\omega_2}
\\
&+\frac{C^r_{\mathcal{O}_{i_0}\mathcal{M}^{(1)AE}_{i_1 i_2}}(\omega_1+\omega_2)
-C^r_{\mathcal{O}_{i_0}\mathcal{M}^{(1)AE}_{i_1 i_2}}(\omega_2)}{i\omega_1}
+\frac{C^r_{\mathcal{O}_{i_0}\mathcal{M}^{(1)EA}_{i_1 i_2}}(\omega_1+\omega_2)
-C^r_{\mathcal{O}_{i_0}\mathcal{M}^{(1)EA}_{i_1 i_2}}(\omega_1)}{i\omega_2}
\\
&-\frac{C^r_{\mathcal{O}_{i_0}j_{i_1}\mathcal{M}^{(0)E}_{i_2}}(\omega_1,\omega_2)-C^r_{\mathcal{O}_{i_0}j_{i_1}\mathcal{M}^{(0)E}_{i_2}}(0,\omega_2)}{i\omega_1}
-\frac{C^r_{\mathcal{O}_{i_0}\mathcal{M}^{(0)E}_{i_1}j_{i_2}}(\omega_1,\omega_2)-C^r_{\mathcal{O}_{i_0}\mathcal{M}^{(0)E}_{i_1}j_{i_2}}(\omega_1,0)}{i\omega_2}
\\
&+C^r_{\mathcal{O}_{i_0}\mathcal{M}^{(0)E}_{i_1}\mathcal{M}^{(0)E}_{i_2}}(\omega_1,\omega_2).
\end{split}
\end{equation}

It is clear that this expression is manifestly free of apparent divergences in the zero frequency limit. Finally, through a lengthy calculation described in Appendix \ref{secondOrderKuboProof}, we combine the terms and arrive at

\begin{equation}
\label{eq:secondOrderFinal}
\begin{split}
P^{E}_{i_0i_1i_2}(\omega_1,\omega_2)
=&\frac{C^r_{\mathcal{O}_{i_0}J_{i_1 i_2}}(\omega_1+\omega_2)
-C^r_{\mathcal{O}_{i_0}J_{i_1 i_2}}(\omega_1)-C^r_{\mathcal{O}_{i_0}J_{i_1 i_2}}(\omega_2)+C^r_{\mathcal{O}_{i_0}J_{i_1 i_2}}(0)}{\omega_1\omega_2}
\\
&-\frac{C^r_{\mathcal{O}_{i_0}J_{i_1}J_{i_2}}(\omega_1,\omega_2)-C^r_{\mathcal{O}_{i_0}J_{i_1}J_{i_2}}(\omega_1,0)-C^r_{\mathcal{O}_{i_0}J_{i_1}J_{i_2}}(0,\omega_2)+C^r_{\mathcal{O}_{i_0}J_{i_1}J_{i_2}}(0,0)}{\omega_1\omega_2}
\\
&-\frac{e^2}{4\pi}\mathcal{P}^{(+)}_{\mathcal{K}^{*}_{\omega}}i\int d\varepsilon \rho_0(\varepsilon)\text{tr}(\mathcal{O}_{i_0}G^r_{\varepsilon+\hbar(\omega_1+\omega_2)}(G^r_{\varepsilon+\hbar\omega_1}-G^r_{\varepsilon+\hbar\omega_2})[r_{i_1},r_{i_2}]G^{r-a}_{\varepsilon}),
\end{split}
\end{equation} 

where $J_i=\frac{ie}{\hbar}[r_i,G^{-1}]$ is the curved current, and $J_{ik}=\frac{1}{4}\frac{ie}{\hbar}([r_{i},J_{k}]+[r_{k},J_{i}])$ is the second order symmetrized version of the former. Furthermore, $[r_i,r_k]=-\hbar^2[\mathcal{D}_{i},\mathcal{D}_{k}]=-i\hbar\mathcal{F}_{ik}$, where $\mathcal{F}_{ik}$, expressed in \eqref{eq:curvature}, are the curvature components of the connection. \\

Formula \eqref{eq:secondOrderFinal} is one of the main results of this paper and a fully consistent generalization of the first order Kubo formula \eqref{eq:firstOrderKubo} that is manifestly free of apparent divergences, and accurately describes second order effects in a Hilbert bundle with a curved connection, such as one arrived at through truncation or projection to an arbitrary eigenspace of $\mc{H}_0$. The formula is invariant under a unitary change of frame since only covariantly transforming operators appear within the trace. Crucially, its structure is fundamentally different from the first order version. For the latter, we could simply perform the calculation with the flat currents $j_i$ and describe evolution by the standard Schrödinger equation before going on to exchange $j_i$ with the curved current $J_i$ in the final result. However, this does not work for second order and an extra term  containing the curvature of the connection appears. This is a direct reflection of the parallel transporter’s path-dependence---a consequence of the connection’s curvature---that becomes important beyond first order. This might prove to be useful for numerical procedures aimed at calculating second order responses, since should a Hilbert space truncation be performed in a controlled manner and the truncation's effects summarized in a curved connection (see Eq. \eqref{eq:AiProj}), our formula would allow a consistent way of arriving at the response coefficient while working in the truncated subspace. We show in Appendix \ref{secondOrderGreenProof}, that the curvature term can be combined with the term on the first line of\eqref{eq:secondOrderFinal} and we finally descend to the realm of practical calculations by rewriting our result in the spectral representation \eqref{eq:Ret2SpecF} and \eqref{eq:Ret3SpecF}

\begin{equation}
\label{eq:secondOrderGreenFinal}
\begin{split}
P^{E}_{i_0i_1i_2}(\omega_1,\omega_2)=&-\frac{e\hbar^2}{2\pi}\hat{\mathcal{P}}^{(+)}_{\mathcal{K}^*_{\omega}}\,i\int d\varepsilon\,\rho_0(\varepsilon)\text{tr}\left(\mathcal{O}_{i_0}G^r_{\omega_1+\omega_2}G^r(G^r_{\omega_1}[\mathcal{D}_{i_2},J_{i_1}]+G^r_{\omega_2}[\mathcal{D}_{i_1},J_{i_2}])G^{r-a}\right)
\\
&-\frac{\hbar^2}{\pi}\hat{\mathcal{P}}^{(+)}_{\mathcal{K}^*_{\omega}}\hat{\mathcal{P}}^{(\Gamma_1^+)}_{(i_1,\omega_1)(i_2,\omega_{2})}i\int d\varepsilon\,\rho_0(\varepsilon)\text{tr}\bigg(\bigg(
\mathcal{O}_{i_0}G^r_{\omega_1+\omega_2}(G^r_{\omega_1}+G^r_{\omega_2})G^rJ_{i_1}G^rJ_{i_2}
\\
&\qquad\qquad+\mathcal{O}_{i_0}G^r_{\omega_1+\omega_2}G^r_{\omega_2}J_{i_1}G^r_{\omega_2}G^rJ_{i_2}
-\frac12 J_{i_1}G^a_{-\omega_1}G^a\mathcal{O}_{i_0}G^r_{\omega_2}G^rJ_{i_2}\bigg)G^{r-a}\bigg).
\end{split}
\end{equation}
\end{widetext}

where $[\mathcal{D}_i,J_k]=\partial_i J_k+i/\hbar[\mathcal{A}_i,J_k]$ is the covariant derivative of the curved current operator and recall that  $\hat{\mathcal{P}}^{(\pm)}_{\mathcal{K}^*_{\omega}}$ acts as $\hat{\mathcal{P}}^{(\pm)}_{\mathcal{K}^*_{\omega}}f(\omega)=(f(\omega)\pm f^*(-\omega))/2$, whereas $\hat{\mathcal{P}}^{(\Gamma_1^+)}_{(i_1,\omega_1)(i_2,\omega_{2})}f_{i_1,i_2}(\omega_1,\omega_2)=(f_{i_1,i_2}(\omega_1,\omega_2)+f_{i_2,i_1}(\omega_2,\omega_1))/2$. The term on the first line is the result of combining the first and third lines of \eqref{eq:secondOrderFinal}. The procedure of this section has also been generalized to third order responses and a general velocity gauge formula free of apparent divergences has also been derived \cite{Bonbien3rdOrder}.\\

Formula \eqref{eq:secondOrderGreenFinal} expresses the second order response to a spatially uniform time-varying electric field in terms of the curved current operator $J_i$ and its covariant derivative $[\mathcal{D}_i,J_k]$, thereby providing a firm and robust platform for computing such responses even in the presence of controlled Hilbert space truncations. The static limit of \eqref{eq:secondOrderGreenFinal} can be arrived at in a straightforward manner by taking $\omega_1 = \omega_2 = 0$ and there is no need for resorting to expansions in terms of the frequency.\\

 During the final stages of this work, it came to our attention that very recently the static limit of the flat case was derived \cite{Michishita2021} and also used for a diagrammatic analysis of disorder contributions to the nonlinear Hall effect \cite{Du2021}. However, we highlight that the subtleties involving the gauge conditions \eqref{eq:secondOrderCondK} and the resulting intimate relation between retarded correlators \eqref{eq:secondOrderCondAA} leading to the cancellations of apparent divergences was not realized, furthermore, the truncation effects and deep connection to curved connections on Hilbert bundles were not discussed. Finally, the fact that the finite-frequency correlators can be `collapsed', as in \eqref{eq:secondOrderGreenFinal}, and the static limit can thereby be arrived at without the performance of an expansion in terms of the frequency, failed to be mentioned.\\

We can arrive at the simpler, flat version of \eqref{eq:secondOrderGreenFinal} as follows. In this case, the curvature $[\mc{D}_i , \mc{D}_k] = 0$ of $\mc{A}_i$ vanishes meaning that we can gauge the latter away and find an expression in terms of the flat currents $j_i \propto \partial_i \mc{H}_0$. Decomposing the first line of \eqref{eq:secondOrderGreenFinal} into parts symmetric and anti-symmetric in $i_1,i_2$ and using the Jacobi identity allows us to recognize that the latter part contains the curvature, and, since this vanishes, only the symmetric part will prove to be sustained. Equivalently, we could look directly at \eqref{eq:secondOrderFinal}, discard the curvature term, restrict to the flat current operators, and rewrite the remaining terms in the spectral representation. Overall, we are left with

\begin{widetext}
\begin{equation}
\label{eq:secondOrderGreenFlat}
\begin{split}
P^{E,\text{flat}}_{i_0i_1i_2}(\omega_1,\omega_2)=
&\frac{C^r_{\mathcal{O}_{i_0}j_{i_1 i_2}}(\omega_1+\omega_2)
-C^r_{\mathcal{O}_{i_0}j_{i_1 i_2}}(\omega_1)-C^r_{\mathcal{O}_{i_0}j_{i_1 i_2}}(\omega_2)+C^r_{\mathcal{O}_{i_0}j_{i_1 i_2}}(0)}{\omega_1\omega_2}
\\
&-\frac{C^r_{\mathcal{O}_{i_0}j_{i_1}j_{i_2}}(\omega_1,\omega_2)-C^r_{\mathcal{O}_{i_0}j_{i_1}j_{i_2}}(\omega_1,0)-C^r_{\mathcal{O}_{i_0}j_{i_1}j_{i_2}}(0,\omega_2)+C^r_{\mathcal{O}_{i_0}j_{i_1}j_{i_2}}(0,0)}{\omega_1\omega_2}
\\
=&-\frac{\hbar^2}{4\pi}\hat{\mathcal{P}}^{(+)}_{\mathcal{K}^*_{\omega}}\,i\int d\varepsilon\,\rho_0(\varepsilon)\text{tr}\left(\mathcal{O}_{i_0}G^r_{\omega_1+\omega_2}(G^r_{\omega_1}+G^r_{\omega_2})G^r(e(\partial_{i_1}j_{i_2}+\partial_{i_2}j_{i_1})+2(j_{i_1}G^rj_{i_2}+j_{i_2}G^rj_{i_1}))G^{r-a}\right)
\\
&-\frac{\hbar^2}{\pi}\hat{\mathcal{P}}^{(+)}_{\mathcal{K}^*_{\omega}}\hat{\mathcal{P}}^{(\Gamma_1^+)}_{(i_1,\omega_1)(i_2,\omega_{2})}i\int d\varepsilon\,\rho_0(\varepsilon)\text{tr}\bigg(\bigg(\mathcal{O}_{i_0}G^r_{\omega_1+\omega_2}G^r_{\omega_2}j_{i_1}G^r_{\omega_2}G^rj_{i_2}-\frac12 j_{i_1}G^a_{-\omega_1}G^a\mathcal{O}_{i_0}G^r_{\omega_2}G^rj_{i_2}\bigg)G^{r-a}\bigg).
\end{split}
\end{equation}
\end{widetext}

Putting $\mc{O}_i = j_i$ provides the charge current responses, or conductivities, which are the subjects of paper III \cite{Bonbien2021c} in our series. In the latter paper, we apply the group theory based decompositions introduced in paper I \cite{Bonbien2021a} and find basis-independent formulae for a large number of transport effects and light-induced current responses.\\

\section{Discussion}
\label{discussion}
\subsection{Geometric framework}
Our general, integrated geometric framework developed throughout Section \ref{curvedSpace} has highlighted the necessity of distinguishing between `abstract' states/operators $|\psi(P)\rangle/\mc{O}(P)$ $(P \equiv (t, \textbf{p}))$ and their components $\ul{\psi}(P)/\dul{\mc{O}}(P)$ in a frame. The former can be thought of as sections of a Hilbert bundle over $P$-space, i.e., maps from $P$-space to the Hilbert spaces at each $P$, whereas the latter as collections of scalar-valued functions on $P$-space. Since the abstract states (operators) live in (act on) \textit{different} copies of the same Hilbert space at different points $P$ and $P'$, the only way to compare them is by transporting them to the same copy of the Hilbert space. This is performed by means of a connection on the Hilbert bundle, which provides a natural way for transporting objects between Hilbert spaces at different points without `change', i.e., it arms us with a parallel transporter \eqref{eq:TransporterExp}. For points infinitesimally close, the comparison following the parallel transport yields an intuitive notion of the corresponding covariant derivative on the abstract states/operators (Eq. \eqref{eq:DTransportNoFrame}/\eqref{eq:DopTransporterNoFrame}). On the other hand, the components of the abstract states/operators in a (local) frame are simply collections of scalar-valued functions, hence they can always be compared at different $P$ and expanded around a particular $P$. This subtle distinction has not been appreciated in the quantum transport literature, since, in general, the connection (on the total Hilbert bundle) in the $\textbf{p}$-direction has mostly been considered as flat. Note that we identify the Hamiltonian as being the components of a connection in the time-direction (see Eq. \eqref{eq:DtDef}). The flatness allows us to choose a local frame along $\textbf{p}$---that we refer to as the `F-frame' throughout the paper---in which the components of the connection vanish $\dul{\mc{A}^{\text{F}}_i} = 0$. In this particular case we can then
identify the abstract state/operator with its components in the F-frame and freely perform the usual comparisons and Taylor expansions at different points---we are trivially identifying the Hilbert spaces at each \textbf{p}. However, as soon as we change frames via a $\textbf{p}$-dependent unitary transformation, for example by moving to the frame in which the Hamiltonian is diagonal, this identification between abstract states/operators and their components no longer holds since the bases or frames become $\textbf{p}$-dependent and we have to utilize our general framework. Should the connection become curved, a scenario achievable through a truncation of the ‘total’ Hilbert spaces at each \textbf{p} (see \eqref{eq:AiProj} and the surrounding discussion), we would go on to lose our ability to choose the F-frame and, consequently, under no circumstance can we identify the abstract quantities with their components. This case demands taking complete advantage of our framework. The neglect of this distinction in the flat case has led to inconsistent Taylor expansions \cite{Ventura2017,Passos2018,Parker2019} (which we discuss in section II.2), confusion between parallel transport of abstract quantities and translation of their components in a frame \cite{Wilhelm2021} (we discuss this in section III) and overall underappreciation of the geometric subtleties involved in the cancellation of apparent divergences and the transition between the velocity and length gauges. We note that the issues in the Taylor expansions do not affect the results of the response calculation in the flat case due to the appearance of a compensating error, however, this self-correction process breaks down in the curved case. We believe that these inconsistencies are rooted in an abuse of notation concerned with different notions of differentiation widespread within the condensed matter literature and we devote Appendix \ref{diffgeo}.\ref{covDerAbuse} to a thorough examination of this issue.\\

Our integrated geometric framework handles all of the discussed subtleties in a natural manner and provides a consistent response formalism that is valid even when working within a Hilbert bundle with a non-flat connection. This is particularly important if we would like to perform our calculations only within a truncated Hilbert space, obtained, for example via a projection onto an isolated set of bands, because this case can be described via a curved-connection. The derived response formula \eqref{eq:secondOrderFinal} is naturally capable of handling this case.\\

\subsection{The sum rules of Aversa and Sipe}
\label{sumRules}
Currently, the prevalent procedure used to eliminate the static limit spurious or apparent divergences appearing in the response functions to an electric field in the velocity gauge is to utilize so-called sum rules resulting from the expression of the response function in the energy eigenbasis, i.e., a frame in which $\mc{H}_0(\textbf{p}) = U(\textbf{p})\mc{E}(\textbf{p})U^{\dagger}(\textbf{p})$, where $\mc{E}$ is diagonal at each \textbf{p} \cite{Ventura2017,Passos2018,Parker2019}. An example of a sum rule responsible for the removal of a spurious divergence at second order is $[x_i,x_j] = 0$ (here $x_i = i\hbar\partial_i$) when written in the energy eigenbasis $[U^{\dagger}x_i U, U^{\dagger}x_j U] = 0$ \cite{AversaSipeSumRules}. After utilizing such sum rules, the final result is the response function that we would obtain should we have done our calculations in the length gauge. This approach via sum rules went through an arduous process \cite{Aspnes1972,Ghahramani1991} before crystallizing in the general form just presented through the work of Aversa and Sipe \cite{AversaSipeSumRules}. More recently \cite{Ventura2017,Passos2018,Parker2019}, interest in these sum rules has experienced a renewal and they were adapted to a modern treatment with an understanding that they are the vehicles for moving between the velocity and length gauges.\\

This point of view can be contrasted with our approach, during the course of which we first show that the fundamental mechanism behind the cancellation of apparent divergences in the velocity gauge is, in fact, the necessity of the physical result being independent from the shift of the vector potential by a constant. We formulate this in terms of gauge conditions---Eq. \eqref{eq:secondOrderGauge} for second order---and, for the second order case, show that the trivial expression $[x_i , x_j ] = 0$ is precisely what lets the gauge condition hold. We then prove that the gauge conditions result in intimate relations, such as \eqref{eq:secondOrderCondAA}, between the retarded correlators giving the second order response, and that these lead to the \textit{manifest} cancellation of apparent divergences in our general response formula \eqref{eq:secondOrderFinal}, still in the velocity gauge expressed in terms of the derivatives of the equilibrium Hamiltonian $\mc{H}_0$. Crucially, we do not actually move to the length gauge, moreover, we do not even need to work in the energy eigenbasis and talk about bands! Our gauge conditions can thus be thought of as being the `soul' of the sum rules in the sense that the purpose of the latter is to cancel the spurious divergences, which they do accomplish, but at the cost of desertion to the length gauge, whereas our gauge conditions also bring with them the removal of spurious divergences, once and for all, but without incurring the cost of leaving the velocity gauge.\\

Due to our formalism being completely intertwined with geometry at a fundamental level, we can also provide a geometric interpretation of why terms such as
$[x_i,x_j]$ start to appear at second order. The curvature effects of the connection on the Hilbert bundle over momentum space only appear through the connection component operator $\mc{A}_i$ which is treated separately in the total velocity gauge `Hamiltonian' \eqref{eq:velHamiltonian}. Thus, the response $K^{AA}_{i_0i_1i_2}$ arising purely from the equilibrium  Hamiltonian $\mc{H}_0$, can be considered as the response in the case of a flat connection, with the responses due to the connection's curvature summarized
in $K^{AE}_{i_0i_1i_2},\,K^{EA}_{i_0i_1i_2}$, and $K^{EE}_{i_0i_1i_2}$  (see the beginning of section \ref{secondorderresponse}). The commutator $[x_i, x_j]$ appears in the gauge condition for $K^{AA}_{i_0i_1i_2}$ (see Eq. \eqref{eq:gaugeConditionFinal} in Appendix \ref{secondOrderGaugeProof}) and is, in fact, the curvature of a flat connection written in the F-frame. As discussed in section \ref{densmat}, the density matrix is evolved by a parallel transporter and from \eqref{eq:TransporterExp} it is clear that, unless the connection is flat, the parallel transporter becomes path-dependent beyond first order. The curvature of the connection can thus be thought of as a measure of the path-dependence
of the parallel transporter and given that $K^{AA}_{i_0i_1i_2}$ is responsible for the `flat' response, the appearance of $[x_i , x_j ]$---the curvature of a flat connection---is precisely the reflection of path-independence or `flatness'. On the other hand, bringing $\mc{A}_i$ into the mix, through Hilbert space truncation for example, gives rise to the additional `$K^{EA}, K^{AE}, K^{EE}$' responses, makes the total parallel transporter path-dependent, and this latter fact is reflected by the appearance of the curvature term in the total response \eqref{eq:secondOrderFinal}.\\

\subsection{Limitations and perspectives}

In the end, we believe that our approach is rather robust, since sum rules are completely bypassed and our final formula for the second order response is rather general. Moreover, the extension to responses of arbitrary order is straightforward and has been done for third order \cite{Bonbien3rdOrder}. Naturally, we have to keep in mind the limitations of this formalism. We have assumed that the external forcing can be described by a classical field and that the resulting non-equilibrium aspects of the system under study can be handled in a perturbative manner via an expansion around the equilibrium configuration. This is the fundamental tenet of Kubo’s formalism \cite{Kubo1957} and, over the decades, has proven to be rather useful for the understanding of quantum transport phenomena. We have also completely forgone discussions of disorder related effects and effects resulting from the spatial non-uniformity of the driving electric field. Our formalism can accommodate the former in a straightforward manner via impurity averaging leading to a diagrammatic expansion akin to recent work for the flat, static case \cite{Du2021}, whereas the latter can be handled along lines similar to the first order flat, static case that has been receiving attention recently \cite{Kozii2021}. Finally, we remark that we have not included the effects of a static magnetic field since a discussion of this aspect within the integrated geometric viewpoint considered in this paper would require the introduction of further mathematical refinements in the form of non-commutative geometry \cite{Grensing,NonCommRev}. The situation can then be handled by considering (crystal) momentum space as non-commutative \cite{Grensing} and looking at fibre bundles over a non-commutative base space. Non-commutative geometry has been successful in providing a general, Kubo formalism-based framework for the first order quantized Hall effect \cite{NonCommQHall,Yu2012} and we believe that this framework can be generalized to higher orders in the spirit of our paper.\\

\section{Conclusion}
\label{conclusion}

Throughout this paper, we have developed a fully geometric framework to describe linear and nonlinear perturbative responses to a spatially uniform, time-varying electric field in the velocity gauge. We highlighted numerous subtleties and several inconsistencies contained within the literature. We discussed the transition between the velocity and length gauges in great detail and provided a geometric perspective. We showed that the static limit spurious divergences that have plagued velocity gauge calculations can be completely eliminated without the use of sum rules while still remaining within the velocity gauge and this allowed us to arrive at a finite frequency response formula that is \textit{manifestly} free of such divergences. We further highlighted that controlled band truncations can be handled by working within a curved space, a case that is accurately described by our response formula. We believe that the results of this paper will open the path towards a comprehensive analysis of nonlinear transport effects and provide a firm, robust platform for both numerical and analytical investigations. In the next paper of our series \cite{Bonbien2021c} we apply the decompositions discussed in paper I \cite{Bonbien2021a} to the flat response formula just derived and obtain relatively simple, exact velocity gauge expressions for a large number of electric field and light-induced transport effects.\\

\begin{acknowledgments}
This research was supported by the King Abdullah University of Science and Technology (KAUST). A. M. acknowledges support from the Excellence Initiative of Aix-Marseille Université—A*Midex, a French ‘Investissements d’Avenir’ program.
\end{acknowledgments}

\appendix

\section{Basic differential geometric tools}
\label{diffgeo}
\subsection{Hilbert bundles, local frames, and connections}
\label{diffgeo1}
In this appendix, we briefly summarize basic aspects of Hilbert (vector) bundles, frames and connections using Dirac’s notation and highlight concepts applied throughout the main text. All of this material can be found with varying notation in standard texts \cite{Nakahara,FrankelGOPh,TuDiffGeo}.\\

We consider a smooth Hilbert manifold $\mathsf{H}^{\text{tot}}$, i.e., a smooth manifold whose locally defined charts map to Hilbert spaces, a Hilbert space $\mathsf{H}$ and a smooth manifold $\mc{B}$. We define the smooth projection $\pi : \mathsf{H}^{\text{tot}}\to\mc{B}$ such that for an open covering $\{V_i\}$ of $\mc{B}$, i.e., the union of open sets $V_i$ covers $\mc{B}$, we have the pre-image $\pi^{-1}(V_i) = V_i\times\mathsf{H}$. This yields a local trivialization of the fibred space $\pi$ and imbues it with the structure of a vector bundle with typical fibre being the Hilbert space $\mathsf{H}$. If $\mathsf{H}$ is infinite dimensional, then this local trivialization can be extended to the entire bundle and the Hilbert bundle is, in fact, trivial $\pi^{\infty} : \mc{B}\times\mathsf{H} \to \mc{B}$; a consequence of Kuiper’s theorem from functional analysis, roughly stating that the infinite-dimensional unitary group is contractible \cite{BoosTopAnal}. On the other hand, if $\mathsf{H}$ is finite-dimensional, then the Hilbert bundle is not necessary trivial if $\mc{B}$ is not contractible. This latter case is of interest in applications of topology to condensed matter since, when $\mc{B}$ is the Brillouin torus of a periodic crystal, it is non-contractible and projecting from the corresponding Hilbert space spanned by an infinite number of bands to a finite-dimensional subspace $\mathsf{H}$ spanned by only a finite number of bands results in a non-trivial Hilbert bundle.\\
Let us now treat both cases in unison and look at a local trivialization (that can always be extended to a global one in the infinite-dimensional case but not always in the finite-dimensional case). Fix local coordinates on $\mc{B}$ and consider local sections $|\psi^V(\textbf{p})\rangle\in\pi^{-1}(V) \subset \mathsf{H}^{\text{tot}}$. Quantum states are interpreted as these local sections. A local frame over $V$ consists of a set $\{|e^V_a (\textbf{p})\rangle\}$ of local sections that form a local basis for section in $\pi^{-1}(V)$, i.e., any local section can be written as

\begin{equation}
|\psi^V(\textbf{p})\rangle = \sum_a\psi^V_a(\textbf{p})|e^V_a(\textbf{p})\rangle.
\end{equation}

\begin{figure}
\begin{center}
\includegraphics[width=11cm]{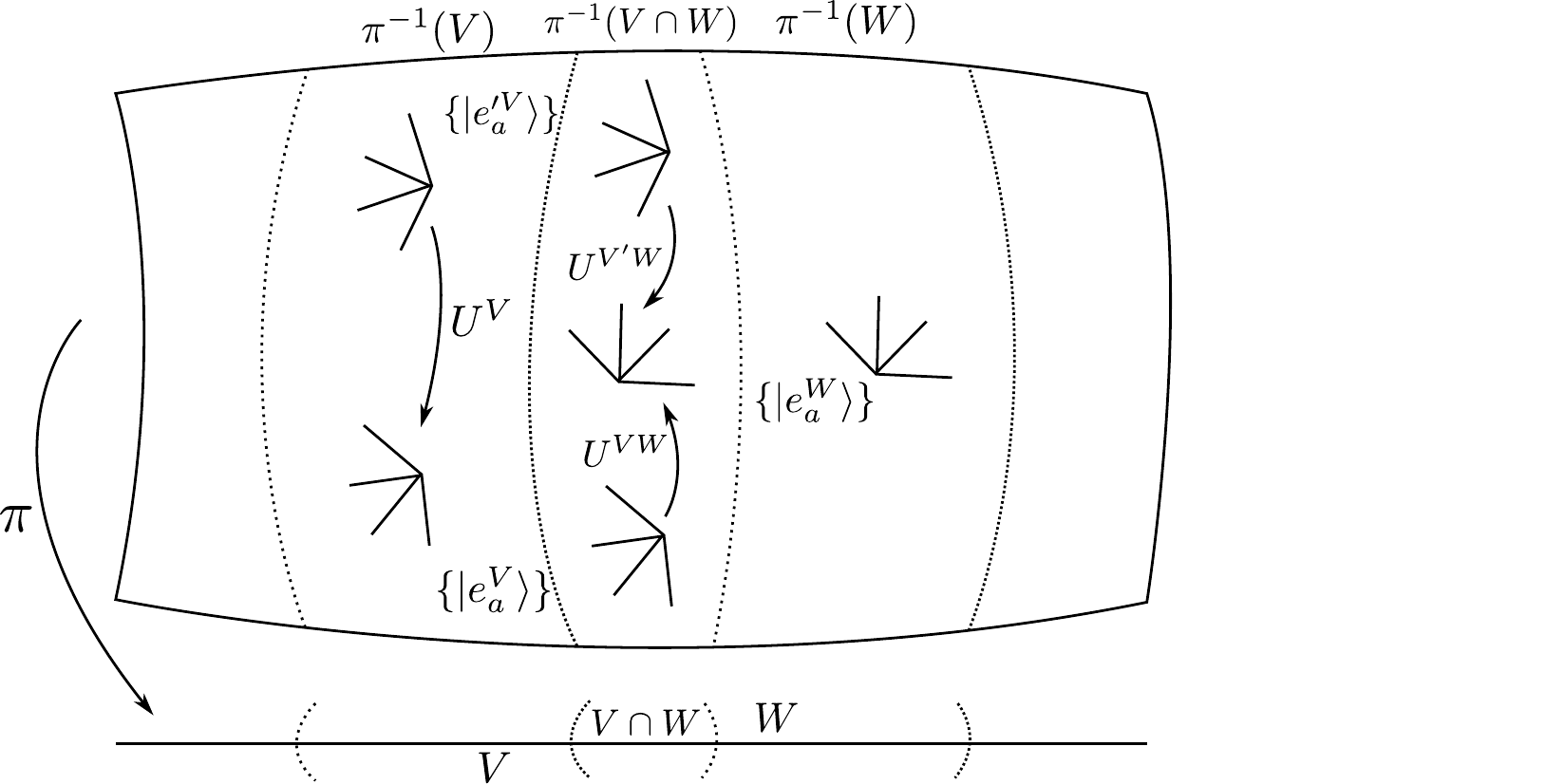}
\caption{Local frames on a Hilbert bundle. $V$ and $W$ are two
sets in an open cover of the base with intersection $V \cap W$ . The
sets $\pi^{-1}(V),\,\pi^{-1}(W)$, and $\pi^{-1}(V\cap W)$ are their pre-images
with respect to the bundle projection $\pi$. $\{|e^V_a\rangle\}$ and $\{|e'^V_a \rangle\}$ 
are two frames over $V$ connected by unitary $U^V$ , whereas $\{|e^W_a \rangle\}$ is a frame over $W$. Within the intersection $V\cap W$, the frames over $V$ and $W$ are connected by the unitaries $U^{VW}$ and $U^{V'W}$. \label{fig:frameChange}}
\end{center}
\end{figure}

We take the local frames to be orthonormal and label the collection of components as $\ul{\psi}^V (\textbf{p})$. Now let us look at the intersection $V\cap W$ between two sets $V$ and $W$ in the open covering with corresponding local frames $\{|e^V_a (\textbf{p})\rangle\}$ and $\{|e^W_a (\textbf{p})\rangle\}$ (see FIG. \ref{fig:frameChange}). Both of the latter are frames over the intersection, hence they must related by a unitary transformation

\begin{equation}
\label{eq:VWFrame}
|e^V_a(\textbf{p})\rangle = \sum_b U^{WV}_{ba}(\textbf{p})|e^W_b(\textbf{p})\rangle.
\end{equation}

This means that the components of local sections over $V$ and $W$ satisfy

\begin{equation}
\ul{\psi}^V(\textbf{p})=\dul{U}^{WV}(\textbf{p})\ul{\psi}^V(\textbf{p}),
\end{equation}

where $\dul{U}^{WV}(\textbf{p})$ is the unitary matrix formed from the components $U^{WV}_{ba} (\textbf{p})$. Note that the change from $W$ to $V$ is given by $\dul{U}^{VW}(\textbf{p})=(\dul{U}^{WV})^{\dagger}(\textbf{p})$ which is the inverse of $\dul{U}^{WV}(\textbf{p})$.

We now introduce a connection on the Hilbert bundle and express it locally via a covariant derivative $\mc{D}_i^V$ of its local sections $|\psi^V(\textbf{p})\rangle$ with respect to local tangent vector fields over $V$ , i.e., $i$ refers to a direction on the tangent spaces $T_{\textbf{p}}(\mc{B})$, where $\textbf{p}\in V$, of $\mc{B}$. Let $\mathsf{H}_{\textbf{p}}$ be a fibre (Hilbert space) at $\textbf{p}\in V \subset \mc{B}$. Then, we have for the covariant derivative of a local section at $\textbf{p}$

\begin{equation}
\mc{D}^{\textbf{p}}:\mathsf{H}_{\textbf{p}}\to\mathsf{H}_{\textbf{p}},\,|\psi^V(\textbf{p})\rangle \mapsto \mc{D}^{\textbf{p}}_i|\psi^V(\textbf{p})\rangle,
\end{equation}

that acts on scalar-valued functions $f:V\to\mathbb{C}$ on the base space as a partial derivative $\mc{D}^V_i f(\textbf{p})=\partial_i f(\textbf{p})$ and satisfies the properties of a derivative, such as the Leibniz rule

\begin{equation}
\mc{D}^V_i(f(\textbf{p})|\psi^V(\textbf{p})\rangle)=\partial_i f(\textbf{p})|\psi^V(\textbf{p})\rangle+f(\textbf{p})\mc{D}^V_i|\psi^V(\textbf{p})\rangle.
\end{equation}

We omitted some properties, such as linearity in the tangent vector fields with respect to which it differentiates the local sections, (we refer to \cite{TuDiffGeo} for a more thorough and precise presentation of connections on vector
bundles) because we will not be needing them explicitly. Since $\mc{D}_i^{\textbf{p}}$ maps a local section at $\textbf{p}\in V$ to another local section at $\textbf{p}$, its action on an element of a local frame at $\textbf{p}$ should be expressible as a linear combination of frame elements at $\textbf{p}$. Extending over $V$ we then have

\begin{equation}
\label{eq:covDerV}
\mc{D}^V_i|e^V_a(\textbf{p})\rangle=\frac{i}{\hbar}\sum_b \mc{A}^V_{iba}(\textbf{p})|e^V_b(\textbf{p})\rangle,
\end{equation}

where $\mc{A}^V_{iba}(\textbf{p})$ are the components over $V$ of the local form of the connection. They are analogues of the Christoffel symbols of the local form of a metric compatible connection on the tangent bundle of a Riemannian manifold. Similarly, we can move over to $W$ and also obtain components $\mc{A}^W_{iba}(\textbf{p})$ in a local frame $\{|e^W_a (\textbf{p})\rangle\}$ over $W$. To find how the connection components are related in an overlap $V\cap W$, we simply apply the relation \eqref{eq:VWFrame} between the local frames and use the Leibniz rule for the covariant derivative as follows. We have the two expressions

\begin{equation}
\begin{split}
\mc{D}^V_i|e^V_a(\textbf{p})\rangle&=\frac{i}{\hbar}\sum_b \mc{A}^V_{iba}(\textbf{p})|e^V_b(\textbf{p})\rangle
\\
&=\frac{i}{\hbar}\sum_c\sum_b U^{WV}_{cb}(\textbf{p}) \mc{A}^V_{iba}(\textbf{p})|e^W_c(\textbf{p})\rangle,
\end{split}
\end{equation}

and

\begin{equation}
\begin{split}
\mc{D}^V_i&|e^V_a(\textbf{p})\rangle=\mc{D}^V_i\sum_c U^{WV}_{ca}(\textbf{p})|e^W_c(\textbf{p})\rangle
\\
&=\frac{i}{\hbar}\sum_c\left(\mc{A}^W_{icb}(\textbf{p})U^{WV}_{ba}(\textbf{p})-i\hbar\partial_iU^{WV}_{ba}(\textbf{p})\right)|e^W_c(\textbf{p})\rangle.
\end{split}
\end{equation}

We can read-off the compatibility relation between the components of the local form of the connection over $V$ and $W$

\begin{equation}
\label{eq:AiTransformVW}
\dul{\mc{A}^W_i}=\dul{U}^{WV}\dul{\mc{A}^V_i}(\dul{U}^{WV})^{\dagger}+i\hbar(\partial_i\dul{U}^{WV})(\dul{U}^{WV})^{\dagger},
\end{equation}

which is the standard rule for the transformation of a gauge potential under a gauge transformation. In this case we described a `passive' gauge transformation via a change of frame.\\

Till now, we have looked at the relation between local sections, frames and connections over different subsets $V$ and $W$ of the open cover. What if we only looked at $V$, say, and wanted to change the frame over $V$ via a unitary transformation (see FIG. \ref{fig:frameChange}). How would the components over $V$ transform? The two frames on $V$ are related by

\begin{equation}
\label{eq:VVpFrame}
|e^V_a(\textbf{p})\rangle = \sum_b U^{V}_{ba}(\textbf{p})|e^{'V}_b(\textbf{p})\rangle,
\end{equation}

and the same analysis applies as for the intersection, i.e., the components of a local section over $V$ transform as $\ul{\psi}(\textbf{p})=\dul{U}^V(\textbf{p})\psi(\textbf{p})$ and the components of the local form of the connection over $V$ transform as

\begin{equation}
\label{eq:AiTransformVVp}
\dul{\mc{A}^{'V}_i}=\dul{U}^{V}\dul{\mc{A}^V_i}(\dul{U}^{V})^{\dagger}+i\hbar(\partial_i\dul{U}^{V})(\dul{U}^{V})^{\dagger}.
\end{equation}

Note that the difference from \eqref{eq:AiTransformVW} is that in the present case $\mc{A}^{'V}_i$ and $\mc{A}^V_i$ are defined over the \textit{same} set $V$, whereas in the former case, $\mc{A}^W_i$ and $\mc{A}^V_i$ are defined over \textit{different} sets $V$ and $W$, with \eqref{eq:AiTransformVW} being interpreted as a compatibility relation on the overlap $V\cap W$. In both cases the transformation rules relate components in different frames but the interpretation is different: in the former case, it quantifies a mismatch between frames over $V$ and $W$, whereas in the present case it simply measures the change in components when moving to another frame over $V$.\\
To illustrate this, suppose that the components over $V$ of the local form of the connection are given by $\dul{\mc{A}^V_i} = i\hbar(\partial_i(\dul{U^V_1})^{\dagger})\dul{U^V_1}$, where $\dul{U^V_1}$ is a unitary matrix over $V$. By the compatibility relations \eqref{eq:AiTransformVW}, the corresponding components over $W$ are

\begin{equation}
\begin{split}
\dul{\mc{A}^W_i}&=i\hbar\dul{U}^{WV}(\partial_i(\dul{U^V_1})^{\dagger})\dul{U^V_1}(\dul{U}^{WV})^{\dagger}+i\hbar(\partial_i\dul{U}^{WV})(\dul{U}^{WV})^{\dagger}
\\
&=i\hbar\partial_i(\dul{U}^{WV}(\dul{U^V_1})^{\dagger})\dul{U^V_1}(\dul{U}^{WV})^{\dagger}.
\end{split}
\end{equation}

Now let us change the frame over $V$ with $\dul{U^V_1}$ according to \eqref{eq:VVpFrame}. Then, by \eqref{eq:AiTransformVVp}, we have $\dul{\mc{A}^{'V}_i} = 0$, meaning that the components over $W$ become

\begin{equation}
\dul{\mc{A}^W_i}=i\hbar\partial_i(\dul{U}^{WV'})(\dul{U}^{WV'})^{\dagger},
\end{equation}

where $\dul{U}^{WV'}=\dul{U}^{WV}(\dul{U^V_1})^{\dagger}$. The point is, that even
though, for this specific case, we could choose a frame over $V$ in which the connection components vanish, we might fail with this over $W$. Such a connection is known as a flat connection, since its curvature \eqref{eq:curvature} over any set $V$ within the open cover vanishes. In the main text, we do not label the sets $V$ explicitly to avoid clutter but we repeatedly refer to `local' frames. Among the latter, the local frame in which the components of a flat connection vanish is dubbed the `F-frame' and another frame in which they do not, such as the one in which the Hamiltonian is diagonal, is the `$U$-frame'.\\

We wrap up this subsection by summarizing the different notions of triviality with respect to Hilbert (vector) bundles and connections on them. In the bundle context, (topological) triviality always refers to whether the bundle is a product of the base space $\mc{B}$ and a typical fibre $\mathsf{H}$, i.e., whether it be written as $\mc{B}\times\mathsf{H}$. Every bundle is \textit{locally} trivial (trivializable), meaning that we can always write locally $V \times \mathsf{H}$ with $V\subset\mc{B}$ being part of an open covering \cite{TuDiffGeo}. If the base space $\mc{B}$ is contractible, e.g., a plane but not a torus or sphere, then the bundle is trivial \cite{Nakahara}. If the base space is arbitrary and the typical fibre is an \textit{infinite}-dimensional Hilbert space, then the bundle is trivial \cite{BoosTopAnal}. If the base space is non-contractible and the typical fibre is a \textit{finite}-dimensional Hilbert space, then the bundle \textit{can be} non-trivial. Moving on to connections, they carry a different meaning of triviality. A connection on a vector bundle is trivial if we can choose a global frame in which all of its components vanish. This means that only trivial bundles admit trivial connections. A connection is flat if its curvature vanishes and consequently a trivial connection is always flat, however, a flat connection is not always trivial. If the base space is non-contractible, then a trivial bundle can admit flat connections that are non-trivial, but a non-trival bundle can only admit curved connections, which, by definition, cannot be trivial. On the other hand, if the base space is contractible and its dimension is greater than 1, then the trivial bundle cannot admit non-trivial flat connections, hence all non-trivial connections on a trivial bundle over a contractible base (with dim$>$1) are necessarily curved.\\

\subsection{Covariant derivatives and partial derivatives}
\label{covDerAbuse}

In this appendix we discuss an abuse of notation, ubiquitous within the condensed matter literature, that resulted in the appearance of the inconsistencies related to Taylor expansions discussed in section \ref{Taylor} of the main text.\\
Henceforth, we dispose of the label keeping track of which set of the open cover we are working over and should just keep in mind that, in general, all of our objects are defined locally. As usual, states are interpreted as local sections of the Hilbert bundle.\\

From the local form of the covariant derivative \eqref{eq:covDerV} in the previous subsection, we can express the components of the local form of the connection as

\begin{equation}
\label{eq:AiCompD}
\mc{A}_{iab}(\textbf{p})=-i\hbar\langle e_a(\textbf{p})|\mc{D}_i|e_b(\textbf{p})\rangle.
\end{equation}

Note how it is the \textit{covariant} derivative that is acting on the \textit{states}, since the partial derivative \textit{only} acts on the \textit{components} of the state with respect to a local frame. Indeed, in general, it is only the covariant derivative that makes sense without reference to a basis or a frame (we show the special case of the trivial connection, i.e., exterior derivative, below). We can thus find the action of $\mc{D}_i$ on a state to be

\begin{equation}
\begin{split}
\langle e_a(\textbf{p})|\mc{D}_i|\psi(\textbf{p})\rangle&=\langle e_a(\textbf{p})|\sum_b\mc{D}_i(\psi_b(\textbf{p})|e_b(\textbf{p})\rangle)
\\
&=\partial_i\psi_a(\textbf{p})+\frac{i}{\hbar}\sum_b \mc{A}_{iab}(\textbf{p})\psi_b(\textbf{p}).
\end{split}
\end{equation}

Now suppose that the connection is flat. This means that we can choose a local frame $\{|e^{\text{F}}_a(\textbf{p})\rangle\}$, the `F-frame' in which all of the components of its local form vanish: $\mc{D}_i|e^{\text{F}}_a (\textbf{p})\rangle = 0 \leftrightarrow \mc{A}^{\text{F}}_{iab}(\textbf{p}) = 0$. This follows from the fact that the curvature of the connection \eqref{eq:curvature} transforms covariantly under a change of frame, meaning that if it vanishes in one frame it has to vanish in all frames. Thus, we can simply move to a frame (locally) in which the $\mc{A}^{\text{F}}_{iab}(\textbf{p})$ are zero. The state $|\psi(\textbf{p})\rangle$ can also be expanded in this frame with the covariant derivative becoming

\begin{equation}
\mc{D}_i|\psi(\textbf{p})\rangle=\sum_a \partial_i\psi^{\text{F}}_a|e^{\text{F}}_a(\textbf{p})\rangle,
\end{equation}

and we can see that it acts as the partial derivative of the components in the F-frame. In this particular case we can then abuse notation and write `$\partial_i|\psi(\textbf{p})\rangle$’ since we can consider the basis as `constant' in the locality we are looking at and, consequently, we are allowed to identify the abstract state $|\psi(\textbf{p})\rangle$ with its components. However, this cannot be done in any `non-constant' frame. Indeed, let us move to another frame $\{|u_a(\textbf{p})\rangle\}$ via

\begin{equation}
\label{eq:Uframe}
\begin{split}
&|u_a(\textbf{p})\rangle = \sum_b U_{ba}(\textbf{p})|e^{\text{F}}_a(\textbf{p})\rangle 
\\
&\leftrightarrow \psi^{\text{F},U^{\dagger}}_a(\textbf{p})=\sum_b U^*_{ab}(\textbf{p})\psi^{\text{F}}_b(\textbf{p}),
\end{split}
\end{equation}

where $U_{ba}(\textbf{p})$ are the elements of a unitary matrix. By \eqref{eq:AiCompD}, the components of the covariant derivative in this frame are

\begin{equation}
\label{eq:AiCompDFU}
\begin{split}
\mc{A}^{\text{F},U^{\dagger}}_{iab}(\textbf{p})&=-i\hbar\langle u_a(\textbf{p})|\mc{D}_i|u_b(\textbf{p})\rangle
\\
&=-i\hbar\sum_c U^*_{ca}(\textbf{p})\partial_iU_{cb}(\textbf{p}).
\end{split}
\end{equation}

The partial derivative only acts on the components $U_{ab}(\textbf{p})$ of the $U$-frame with respect to the F-frame, whereas it is the covariant derivative that acts on the states themselves. In the case of a periodic crystal, $|u_a(\textbf{p})\rangle$ are taken to be the periodic Bloch states and the elements $U_{ab}(\textbf{p})$ of the unitary matrix are the components of $|u_a(\textbf{p})\rangle$ in a countable basis $|e^{\text{F}}_a \rangle$ of the total Hilbert space of states. The components $\mc{A}^{\text{F},U^{\dagger}}_{iab}(\textbf{p})$ are considered as those of the flat Berry connection and written as $\langle u_a(\textbf{p})|\partial_i u_b(\textbf{p})\rangle$---a blatant abuse of notation permeating the literature.\\

Using \eqref{eq:AiCompDFU}, the components of the flat covariant derivative of a state in this $U$-frame become

\begin{equation}
\label{eq:DFUPsi}
\begin{split}
\langle u_a(\textbf{p})|\mc{D}_i|\psi(\textbf{p})\rangle
=\partial_i\psi^{\text{F},U^{\dagger}}_a(\textbf{p})+\frac{i}{\hbar}\sum_b\mc{A}^{\text{F},U^{\dagger}}_{iab}(\textbf{p})\psi^{\text{F},U^{\dagger}}_a(\textbf{p}).
\end{split}
\end{equation}

We can now present the standard notational abuse leading to inconsistent Taylor expansions (for example, \citet{Ventura2017} do this for operators (see below) instead of states, but the same applies). Consider

\begin{equation}
\label{eq:psiDabuse}
\begin{split}
` \langle u_a(\textbf{p})|\partial_i|\psi(\textbf{p})\rangle&=\langle u_a(\textbf{p})|\sum_b\partial_i(\psi_b(\textbf{p})|u_b(\textbf{p})\rangle
\\
&=\partial_i\psi_a(\textbf{p})+\sum_b \mc{A}^{\text{F},U^{\dagger}}_{iab}(\textbf{p})\psi_b(\textbf{p})
\\
&=\langle u_a(\textbf{p})|\mc{D}_i|\psi(\textbf{p})\rangle\text{'},
\end{split}
\end{equation}

which means

\begin{equation}
` \partial_i|\psi(\textbf{p})\rangle = \mc{D}_i|\psi(\textbf{p})\rangle\text{'}.
\end{equation}

We perform a Taylor expansion of the state $|\psi(\textbf{p}+d\textbf{p})\rangle$ around $\textbf{p}$ and apply this to get

\begin{equation}
\begin{split}
` |\psi(\textbf{p}+d\textbf{p})\rangle&=|\psi(\textbf{p})\rangle+dp^i\partial_i|\psi(\textbf{p})\rangle+O(dp^2)
\\
&=|\psi(\textbf{p})\rangle+dp^i\mc{D}_i|\psi(\textbf{p})\rangle+O(dp^2)\text{'}.
\end{split}
\end{equation}

By now, it should be clear why this procedure is not appropriate. First of all, on the left hand side of \eqref{eq:psiDabuse} we assumed that the partial derivative $\partial_i$ acts on the state. This can just be considered as an innocent abuse of notation and would still be fine if we kept in mind its meaning: it represents the covariant derivative. However, after applying it on a state expanded in the $U$-frame and noting that the components of the flat connection $\mc{A}^{\text{F},U^{\dagger}}_{iab}$ in the $U$-frame appear, thereby manifesting the structure of a covariant derivative, we cannot simply change the meaning of the $\partial_i$ to that of $\mc{D}_i$, since it meant $\mc{D}_i$ all along! The covariant derivative is a frame-independent object. Moving on to the Taylor expansion, note that since the states at different $\textbf{p}$ live in different spaces they cannot just be compared: we have to use a connection to transport them to the same space before we can. Their components in a frame, on the other hand, are just scalar-valued functions and can be compared; consequently, they afford a standard Taylor expansion. However, in the special case of a flat connection, we can always choose (locally) a frame in which the components of the connection vanish (the F-frame above) and can consider the abstract state as equivalent to its components in this frame. In this particular case, with heavy abuse of notation, the expansion of the state with the partial derivatives makes sense, as long as we keep in mind that they are being identified with their components in the F-frame. As soon as we change to the $U$-frame, this does not make sense anymore (we show what goes wrong in section \ref{Taylor} of the main text).\\

The exact same arguments can be repeated for `abstract' operators $\mc{O}(\textbf{p})$ acting on the abstract states $|\psi(\textbf{p})\rangle$ at each $\textbf{p}$. However, there is an important subtlety that we would like to emphasize. We can have such abstract operators that can be defined in a frame-independent manner, an example being the velocity operator \eqref{eq:velocityNoFrame}, and, upon introducing a frame, we can go on to define the `matrix' of their components in the particular frame. On the other hand, we also have a fundamental `operator' that cannot be defined in a frame-independent manner: the Hamiltonian. The Hamiltonian represents the time-direction components of a connection in a frame (see Eq. \eqref{eq:DtDef}); it is linked to a frame by definition! It can be looked at as an `operator' (a finite or infinite dimensional `countable matrix' of components in the energy eigenbasis, or an `uncountable matrix', i.e., integral kernel, in the position basis) but not as an abstract operator. Of course, this means that the Hamiltonian always affords a standard Taylor expansion around a particular point in parameter space: it is just a collection of components in a frame.\\

Going forward, we have the covariant derivative $\mc{D}_i\mc{O}(\textbf{p}) \equiv [\mc{D}_i,\mc{O}(\textbf{p})]$ (see Eq. \eqref{eq:DopDef}) of an abstract operator defined via the Leibniz rule and can look at its components in a frame

\begin{equation}
\begin{split}
&\langle e_a(\textbf{p})|[\mc{D}_i,\mc{O}(\textbf{p})]|e_b(\textbf{p})\rangle
\\
&=\partial_i\mc{O}_{ab}(\textbf{p})+\frac{i}{\hbar}\sum_c(\mc{A}_{iac}(\textbf{p})\mc{O}_{cb}(\textbf{p})-\mc{O}_{ac}(\textbf{p})\mc{A}_{icb}(\textbf{p})).
\end{split}
\end{equation}

Suppose the connection is flat and let us choose the F-frame in which $\mc{A}^{\text{F}}_{iab}(\textbf{p})=0$. Then,

\begin{equation}
\langle e^{\text{F}}_a(\textbf{p})|[\mc{D}_i,\mc{O}(\textbf{p})]|e^{\text{F}}_b(\textbf{p})\rangle=\partial_i\mc{O}^{\text{F}}_{ab}(\textbf{p}).
\end{equation}

Moving to the $U$-frame via \eqref{eq:Uframe}, we find

\begin{equation}
\begin{split}
&\langle u_a(\textbf{p})|[\mc{D}_i,\mc{O}(\textbf{p})]|u_b(\textbf{p})\rangle
=\partial_i\mc{O}^{\text{F},U^{\dagger}}_{ab}(\textbf{p})
\\
&+\frac{i}{\hbar}\sum_c(\mc{A}^{\text{F},U^{\dagger}}_{iac}(\textbf{p})\mc{O}^{\text{F},U^{\dagger}}_{cb}(\textbf{p})-\mc{O}^{\text{F},U^{\dagger}}_{ac}(\textbf{p})\mc{A}^{\text{F},U^{\dagger}}_{icb}(\textbf{p})),
\end{split}
\end{equation}

where we used \eqref{eq:AiCompDFU}.\\
We can also abuse notation as follows

\begin{equation}
\begin{split}
` &\langle u_a(\textbf{p})|\partial_i\mc{O}(\textbf{p})|u_b(\textbf{p})\rangle
=\partial_i\mc{O}_{ab}(\textbf{p})
\\
&\qquad-\langle \partial_i u_a(\textbf{p})|\mc{O}(\textbf{p})|u_b(\textbf{p})\rangle
-\langle u_a(\textbf{p})|\mc{O}(\textbf{p})|\partial_i u_b(\textbf{p})\rangle
\\
&=\partial_i\mc{O}_{ab}(\textbf{p})
\\
&\qquad+\frac{i}{\hbar}\sum_c(\mc{A}^{\text{F},U^{\dagger}}_{iac}(\textbf{p})\mc{O}^{\text{F},U^{\dagger}}_{cb}(\textbf{p})-\mc{O}^{\text{F},U^{\dagger}}_{ac}(\textbf{p})\mc{A}^{\text{F},U^{\dagger}}_{icb}(\textbf{p}))
\\
&=\langle u_a(\textbf{p})|[\mc{D}_i,\mc{O}(\textbf{p})]|u_b(\textbf{p})\rangle\text{'},
\end{split}
\end{equation}

which means

\begin{equation}
` \partial_i\mc{O}_{ab}(\textbf{p}) = [\mc{D}_i,\mc{O}(\textbf{p})]\text{'},
\end{equation}

and then go on to perform a Taylor expansion using the covariant derivative. This is precisely the procedure adopted by \cite{Ventura2017} and the inconsistencies should be apparent by now. While the abuse of notation can be handled, should we keep in mind that when acting with the partial derivative on the operator, we are, in fact, applying a covariant derivative, it might lead to confusion regarding the fact that the covariant derivative acting on abstract operators and states is defined in a frame-independent manner. However, as mentioned above, the Hamiltonian is, by definition, attached to a frame and we \textit{cannot} define its `abstract' covariant derivative. What we \textit{can} do, is define the frame-independent velocity operator as $[\mc{D}_i, \mc{D}_0]$ (see Eq. \eqref{eq:velocityNoFrame} and the surrounding discussion), which is a part of the abstract curvature of the connection, and then neglect explicit time-dependence which leads to the `frame-dependent' covariant derivative of the Hamiltonian \eqref{eq:velocityFrame}.\\

\section{Dynamics in a periodic crystal}
\label{crystal}
In this appendix, we recount how a crystal under the influence of a spatially uniform, time-varying electric field provides a specific case in which the general, parameter-dependent standard Schrödinger equation \eqref{eq:SchParFrameF} is realized and briefly overview the definition of Blount’s position operator \cite{Blount}.

\subsection{Time-dependent Schrödinger equation in a periodic crystal}
\label{dynamics}
The single particle crystal Hamiltonian; which we write in the position basis as $\mc{H}^{(\textbf{r})}(\nabla;\textbf{r})$, where the superscript $^{(\textbf{r})}$ labels the fact that we are in the position basis and the real-space gradient $\nabla$ labels the `true' momentum operator in the position basis; is periodic in real space. By Bloch’s theorem, the eigenfunctions $\phi_a(\textbf{k};\textbf{r})$ of $\mc{H}^{(\textbf{r})}(\nabla;\textbf{r})$ can be labelled by two parameters, the crystal momentum $\textbf{k}$ and band index $a$, and are of the form

\begin{equation}
\label{eq:BlochPhiu}
\phi_a(\textbf{k};\textbf{r})=e^{i\textbf{k}\cdot \textbf{r}} u_a(\textbf{k};\textbf{r}),
\end{equation}

where $u_a(\textbf{k};\textbf{r})=u_a(\textbf{k};\textbf{r}+\textbf{R})$ are periodic with lattice vector $\textbf{R}$. These periodic functions satisfy the eigenvalue equation

\begin{equation}
\label{eq:Hkeigenvalue}
\mc{H}^{(\textbf{r})}(\nabla+i\textbf{k};\textbf{r})u_a(\textbf{k};\textbf{r})=\varepsilon_a(\textbf{k})u_a(\textbf{k};\textbf{r}),
\end{equation}

where $\mc{H}^{(\textbf{r})}(\nabla+i\textbf{k};\textbf{r})=e^{-i\textbf{k}\cdot \textbf{r}}\mc{H}^{(\textbf{r})}(\nabla;\textbf{r})e^{i\textbf{k}\cdot \textbf{r}}$ is the Bloch Hamiltonian and $\varepsilon_a(\textbf{k})$ are the bands.\\

Now let $|\psi(t)\rangle$ be an `abstract' state satisfying the standard Schrödinger equation $\mc{D}^{\text{tot}}_t|\psi(t)\rangle = 0$, where $\mc{D}^{\text{tot}}_t$ is the covariant derivative in the time direction whose action on the position basis `elements' provides the components $\mc{H}^{(\textbf{r})}(\nabla;\textbf{r})$ of the crystal Hamiltonian in the position basis, and consider expanding $|\psi(t)\rangle$ in two bases: the position basis and the basis provided by the eigenstates of the Hamiltonian

\begin{equation}
\begin{split}
&|\psi(t)\rangle = \int d\textbf{r}\psi^{(\textbf{r})}(\textbf{r};t)|\textbf{r}\rangle,
\\
&|\psi(t)\rangle = \int d\textbf{k}\sum_a\psi^{(\mc{E})}_a(\textbf{k};t)|\phi_a(\textbf{k})\rangle,
\end{split}
\end{equation}

where the superscripts on the components, e.g. $\psi^{(\textbf{r})}$, are merely labels to indicate the basis. We can thus express the components in the position basis with the com- ponents in the energy eigenbasis as

\begin{equation}
\label{eq:rToEnergy}
\psi^{(\textbf{r})}(\textbf{r};t) = \int d\textbf{k}\sum_a\psi^{(\mc{E})}_a(\textbf{k};t)\langle\textbf{r}|\phi_a(\textbf{k})\rangle,
\end{equation}

and can interpret the Bloch states $\langle\textbf{r}|\phi_a(\textbf{k})\rangle = \phi_a(\textbf{k};\textbf{r})$ as the components of a basis transformation used to move between the two bases. Using \eqref{eq:BlochPhiu} in \eqref{eq:rToEnergy} to rewrite $\langle\textbf{r}|\phi_a(\textbf{k})\rangle$ with the periodic function $\langle\textbf{r}|u_a(\textbf{k})\rangle$, we have 

\begin{equation}
\label{eq:PsirE}
\begin{split}
\psi^{(\textbf{r})}(\textbf{r};t) &= \int d\textbf{k}\,e^{i\textbf{k}\cdot\textbf{r}}\langle\textbf{r}|\sum_a\psi^{(\mc{E})}_a(\textbf{k};t)|u_a(\textbf{k})\rangle
\\
&=\int d\textbf{k}\,e^{i\textbf{k}\cdot\textbf{r}}\langle\textbf{r}|\psi(\textbf{k},t)\rangle,
\end{split}
\end{equation}

where we defined

\begin{equation}
\label{eq:Psik}
|\psi(\textbf{k},t)\rangle=\sum_a\psi^{(\mc{E})}_a(\textbf{k};t)|u_a(\textbf{k})\rangle,
\end{equation}

and can interpret it as the expansion of a $\textbf{k}$-dependent abstract state $|\psi(\textbf{k},t)\rangle$ in a $\textbf{k}$-dependent frame $\{|u_a(\textbf{k})\rangle\}$ at each $\textbf{k}$. We label the Hilbert spaces spanned by this frame at each $\textbf{k}$ as $\mathsf{H}_{\textbf{k}}$ and consider them as the fibres of a Hilbert bundle over $\textbf{k}$-space. The standard Schrödinger equation $\mc{D}^{\text{tot}}_t|\psi(t)\rangle = 0$ satisfied by the `total' state $|\psi(t)\rangle$ then reduces to 

\begin{equation}
\label{eq:SchParKNoFrame}
\mc{D}^{\text{tot},\textbf{k}}_t|\psi(\textbf{k},t)\rangle = 0,
\end{equation}

for the abstract states $|\psi(\textbf{k},t)\rangle$ at each $\textbf{k}$. Here, the superscript $\textbf{k}$ on $\mc{D}^{\text{tot},\textbf{k}}_t$ refers to the fact that the action of the latter covariant derivative on elements of a frame in each $\mathsf{H}_{\textbf{k}}$ provides the $\textbf{k}$-dependent Bloch Hamiltonian in the frame.\\

Now suppose that we apply a spatially uniform, time-varying electric field $\textbf{E}(t) = -\partial\textbf{A}(t)/\partial t$ and implement it on the Hamiltonian via the minimal coupling prescription $\nabla\to\nabla-\frac{ie}{\hbar}\textbf{A}(t)$ applied to the `true' momentum operator $\propto\nabla$. Then, the Hamiltonian begets time-dependence and becomes $\mc{H}^{(\textbf{r})}(\nabla;\textbf{r})\to \mc{H}^{(\textbf{r})}(\nabla-\frac{ie}{\hbar}\textbf{A}(t);\textbf{r})$, which is still periodic in real-space meaning that Bloch's theorem continues to apply. By noting that the Bloch Hamiltonian gets modified as $\mc{H}^{(\textbf{r})}(\nabla+i\textbf{k};\textbf{r})\to \mc{H}^{(\textbf{r})}(\nabla-\frac{ie}{\hbar}\textbf{A}(t)+i\textbf{k};\textbf{r})=\mc{H}^{(\textbf{r})}(\nabla+i\textbf{k}(t);\textbf{r})$, where we defined $\textbf{k}(t)=\textbf{k}-\frac{e}{\hbar}\textbf{A}(t)$, we can rewrite the eigenvalue equation \eqref{eq:Hkeigenvalue} as

\begin{equation}
\label{eq:HkeigenvalueT}
\mc{H}^{(\textbf{r})}(\nabla+i\textbf{k}(t);\textbf{r})u_a(\textbf{k};\textbf{r})=\varepsilon_a(\textbf{k}(t))u_a(\textbf{k}(t);\textbf{r}),
\end{equation}

with the periodic functions becoming instantaneous eigenfunctions of the Bloch Hamiltonian. The $\textbf{k}$-dependent abstract states \eqref{eq:Psik} are modified by instantaneously adapting to the frame $|u_a(\textbf{k}(t))\rangle$ at each $\textbf{k}$ and become

\begin{equation}
\label{eq:PsikT}
|\psi(\textbf{k}(t),t)\rangle=\sum_a\psi^{(\mc{E})}_a(\textbf{k}(t);t)|u_a(\textbf{k}(t))\rangle.
\end{equation}

The applied vector potential then modifies the components of the total state \eqref{eq:PsirE}  to

\begin{equation}
\label{eq:PsirEA}
\begin{split}
\psi^{(\textbf{r}),\textbf{A}}(\textbf{r};t)
=\int d\textbf{k}\,e^{i\textbf{k}\cdot\textbf{r}}\langle\textbf{r}|\psi(\textbf{k}(t),t)\rangle,
\end{split}
\end{equation}

where the $\textbf{A}$ superscript indicates that we have applied the vector potential. The position basis components $\psi^{(\textbf{r}),\textbf{A}}(\textbf{r};t)$  of the total state satisfy the standard Schrödinger equation with minimal-coupling, hence the `reduced' evolution equation \eqref{eq:SchParKNoFrame} gets modified to

\begin{equation}
\label{eq:SchParKTNoFrame}
\begin{split}
&\mc{D}^{\text{tot},\textbf{k}(t)}_t|\psi(\textbf{k}(t),t)\rangle
\\
&=\mc{D}^{\textbf{k}(t)}_t|\psi(\textbf{k}(t),t)\rangle+\frac{e}{\hbar}E^i(t)\mc{D}_i|\psi(\textbf{k}(t),t)\rangle=0,
\end{split}
\end{equation}

where the second term arises from the time-dependence of $\textbf{k}(t)$. Choosing a frame $\{|e^{\text{F}}_a(\textbf{k})\rangle\}$ of the Hilbert spaces $\mathsf{H}_{\textbf{k}}$ spanned by the periodic Bloch states that is `constant' with respect to $\mc{D}_i$, i.e., a frame in which the components of the connection vanish in the $\textbf{k}$ direction, we have

\begin{equation}
\begin{split}
&\mc{D}^{\textbf{k}(t)}_t|e^{\text{F}}_a(\textbf{k}(t))\rangle=\sum_b\mc{H}^{\text{F}}_{ba}(\textbf{k}(t))|e^{\text{F}}_b(\textbf{k}(t))\rangle,
\\
&\mc{D}_i|\psi(\textbf{k}(t),t)\rangle=\sum_a\partial_i\psi^{\text{F}}_{a}(\textbf{k}(t),t)|e^{\text{F}}_a(\textbf{k}(t))\rangle,
\end{split}
\end{equation}

where $\mc{H}^{\text{F}}_{ba}(\textbf{k})$ are the components of the Bloch Hamiltonian in the standard F-frame. In this frame we then get the `Bloch' form of the standard Schrödinger equation

\begin{equation}
i\hbar\frac{d}{dt}\ul{\psi}^{\text{F}}(\textbf{k}(t),t) = \dul{\mc{H}}^{\text{F}}(\textbf{k}(t))\ul{\psi}^{\text{F}}(\textbf{k}(t),t).
\end{equation}

Note that the Bloch Hamiltonian will only be diagonal if we choose the frame to consist of the periodic Bloch states themselves, in which case the action of $\mc{D}_i$ on the frame elements will no longer vanish and will provide the components of the flat Berry connection (see below). We have thus arrived at a special case of equation \eqref{eq:SchParFrameF}, wherein the parameter $\textbf{p}$ corresponds to $\hbar\textbf{k}$, the fibres $\mathsf{H}_{(t,\textbf{p})}$ of the Hilbert bundle over time-parameter space correspond to the spaces $\mathsf{H}_{\textbf{k}}$ spanned by the periodic Bloch states with copies at each $t$, and the Hamiltonian is not dependent on time explicitly, meaning that we do not need to consider local frames with explicit time-dependence.\\

We can transfer the Bloch form of the standard Schrödinger equation to the Bloch form of the von Neumann equation describing the evolution of the reduced density matrix as

\begin{equation}
i\hbar\frac{d}{dt}\dul{\rho}^{\text{F}}(\textbf{k}(t),t) = \left[\dul{\mc{H}}^{\text{F}}(\textbf{k}(t)),\dul{\rho}^{\text{F}}(\textbf{k}(t),t)\right],
\end{equation}

which are known as the semiconductor Bloch equations. We can move to a frame $\{|u_a(\textbf{k})\rangle\}$ given by the periodic Bloch states in which the Bloch Hamiltonian is diagonal at each $\textbf{k}$, i.e., $\dul{\mc{H}}^{\text{F}}(\textbf{k})=\dul{U}(\textbf{k})\dul{\mc{E}}(\textbf{k})\dul{U}^{\dagger}(\textbf{k})$ where $\dul{\mc{E}}(\textbf{k})$ is diagonal, via \eqref{eq:Uframe} and the equation becomes

\begin{equation}
\begin{split}
i\hbar\frac{d}{dt}\dul{\rho}^{\text{F},U^{\dagger}}(\textbf{k}(t),t) =& \left[\dul{\mc{E}}(\textbf{k}(t)),\dul{\rho}^{\text{F},U^{\dagger}}(\textbf{k}(t),t)\right]
\\
&+eE^i\left[\dul{\mc{A}^{B}_i}(\textbf{k}(t)),\dul{\rho}^{\text{F},U^{\dagger}}(\textbf{k}(t),t)\right],
\end{split}
\end{equation}

where $\dul{\rho}^{\text{F},U^{\dagger}}(\textbf{k}(t),t) = \dul{U}^{\dagger}(\textbf{k}(t))\dul{\rho}^{\text{F}}(\textbf{k}(t),t)\dul{U}(\textbf{k}(t))$ and $\dul{\mc{A}^{B}_i}(\textbf{k}(t))=-i\dul{U}^{\dagger}(\textbf{k})\partial_i\dul{U}(\textbf{k})$ are the components of a flat Berry connection (see also the next subsection for their relation to the position operator). This is the more familiar form of the semiconductor Bloch equations \cite{Wilhelm2021,HaugSemiCond} and note how, as required by the covariance of the von Neumann equation under a change of frame, the components of Berry's flat connection arose naturally (see also ref. \cite{Li2019} for a description of the importance of Berry’s connection terms in this form of the semiconductor Bloch equations).\\

\subsection{Blount's position operator}
\label{Blount}

We briefly discuss the form of the position operator in different bases of the Hilbert space.\\
Just as in \eqref{eq:PsirE}, we expand an abstract state $|\psi\rangle \in \mathsf{H}$ in two different bases of the total Hilbert space $\mathsf{H}$; the position basis and the energy eigenbasis. The components $\dul{\textbf{r}}$ of the position operator $\hat{\textbf{r}}$ in the position basis are simply the position `vector' $\textbf{r}$

\begin{equation}
\langle\textbf{r}|\hat{\textbf{r}}|\psi\rangle = \textbf{r}\psi^{(\textbf{r})}(\textbf{r}).
\end{equation}

How would the components of the position operator look like in the energy eigenbasis? To find out, we use \eqref{eq:rToEnergy} to write $\psi^{(\textbf{r})}(\textbf{r})$ in the latter basis

\begin{equation}
\langle\textbf{r}|\hat{\textbf{r}}|\psi\rangle = \int d\textbf{k}\sum_a\textbf{r}\psi_a^{(\mc{E})}(\textbf{k})\langle\textbf{r}|\phi_a(\textbf{k})\rangle.
\end{equation}

Next, we use $\langle\textbf{r}|\phi_a(\textbf{k})\rangle = e^{i\textbf{k}\cdot\textbf{r}}\langle \textbf{r}|u_a(\textbf{k})\rangle$ and, following in the footsteps of Blount \cite{Blount}, obtain

\begin{equation}
\label{eq:RopE}
\begin{split}
\langle\textbf{r}|\hat{\textbf{r}}|\psi\rangle &= \int d\textbf{k}\,\textbf{r}e^{i\textbf{k}\cdot\textbf{r}}\langle\textbf{r}|\sum_a\psi_a^{(\mc{E})}(\textbf{k})|u_a(\textbf{k})\rangle
\\
&=\int d\textbf{k}\,(-i\partial_{\textbf{k}}e^{i\textbf{k}\cdot\textbf{r}})\langle\textbf{r}|\sum_a\psi_a^{(\mc{E})}(\textbf{k})|u_a(\textbf{k})\rangle
\\
&=\int d\textbf{k}\,e^{i\textbf{k}\cdot\textbf{r}}i\partial_{\textbf{k}}\left(\langle\textbf{r}|\sum_a\psi_a^{(\mc{E})}(\textbf{k})|u_a(\textbf{k})\rangle\right).
\end{split}
\end{equation}

We can read-off the abstract states $|\psi(\textbf{k})\rangle \in \mathsf{H}_{\textbf{k}}$ in the spaces spanned by $\{|u_a(k)\rangle\}$ at each $\textbf{k}$ as

\begin{equation}
|\psi(\textbf{k})\rangle = \sum_a \psi^{(\mc{E})}_a(\textbf{k})|u_a(\textbf{k})\rangle.
\end{equation}

With this, the result of \eqref{eq:RopE} can be written as

\begin{equation}
\label{eq:posop}
\begin{split}
\langle\textbf{r}|\hat{\textbf{r}}|\psi\rangle &= \int d\textbf{k}\,e^{i\textbf{k}\cdot\textbf{r}}i\partial_{\textbf{k}}\left(\langle\textbf{r}|\psi(\textbf{k})\rangle\right)
\\
&=\int d\textbf{k}\,e^{i\textbf{k}\cdot\textbf{r}}\langle\textbf{r}|i\mc{D}_{\textbf{k}}|\psi(\textbf{k})\rangle,
\end{split}
\end{equation}

where we noted that $\langle\textbf{r}|\psi(\textbf{k})\rangle$ are the components of $|\psi(\textbf{k})\rangle$ in a basis---the position basis---in which the components of the connection inducing the covariant derivative $\mc{D}_{\textbf{k}}$ over $\textbf{k}$ vanish, i.e., we have the trivial connection on the total space of the Hilbert bundle---which is just the total Hilbert space fibred with respect to $\textbf{k}$---with infinite-dimensional fibres $\mathsf{H}_{\textbf{k}}$ spanned by the periodic Bloch states. Thus, we have the result that the position operator corresponds to a covariant derivative induced by a flat (trivial) connection acting on states in the spaces spanned by the periodic Bloch states. The covariant derivative can be defined in a frame-independent manner, and this is precisely what allows us to also consider the position operator for the finite-dimensional case---we can simply choose a countable basis for each fibre. By natural restriction, we can then \textit{define} the position operator for a \textit{finite}-dimensional \textit{total} Hilbert space, i.e., a finite-band model, as the covariant derivative corresponding to a flat connection. Note that a Hilbert bundle with \textit{infinite}-dimensional fibres is always trivial regardless of the base-space being contractible or not \cite{BoosTopAnal}, hence the flat connection in the general case discussed above is just the trivial one, however, in the finite-dimensional case, a flat connection can be non-trivial for a non-contractible base, hence we can only choose frames in which the connection components vanish, locally. Indeed, since the connection is flat, let us choose a local frame $\{|e^{\text{F}}_a\rangle\}$ that is countable and is `constant', i.e, $\mc{D}_{\textbf{k}}|e^{\text{F}}_a\rangle = 0$. In this frame, the position operator simply acts as the partial derivative on the components of states.\\

Now we look at a distinguished frame. Suppose we choose the frame of the spaces $\mathsf{H}_{\textbf{k}}$ spanned by the periodic Bloch states to be these Bloch states themselves, i.e., $\{|u_a(\textbf{k})\rangle\}$. This frame is no longer `constant' and the components of the flat connection then provide the components of the flat Berry connection

\begin{equation}
\mc{D}_i|u_a(\textbf{k})\rangle = i\sum_b \mc{A}^{B}_{iba}|u_b(\textbf{k})\rangle.
\end{equation}

We can find the expression for these components by changing the frame from $|e^{\text{F}}_a\rangle$ as

\begin{equation}
|u_a(\textbf{k})\rangle = \sum_b U_{ba}(\textbf{k})|e^{\text{F}}_b\rangle,
\end{equation}

and can interpret $U_{ba}(\textbf{k})$, apart from being components of a unitary matrix, as the components of the periodic Bloch states in the `constant' countable basis. For example, in the case of a 2-band model we have

\begin{equation}
|u_1(\textbf{k})\rangle\to\begin{pmatrix}U_{11}(\textbf{k})\\U_{21}(\textbf{k})\end{pmatrix}
,\quad 
|u_2(\textbf{k})\rangle\to\begin{pmatrix}U_{12}(\textbf{k})\\U_{22}(\textbf{k})\end{pmatrix}.
\end{equation}

We then obtain for the covariant derivative

\begin{equation}
\begin{split}
\mc{D}_i|u_a(\textbf{k})\rangle &=\sum_b\partial_i U_{ba}(\textbf{k})|e^{\text{F}}_b\rangle
\\
&=i\sum_{b,c}\mc{A}^{B}_{ica}(\textbf{k})U_{bc}(\textbf{k})|e^{\text{F}}_b\rangle,
\end{split}
\end{equation}

from which we can read-off

\begin{equation}
\mc{A}^{B}_{iab}(\textbf{k})=-i\sum_c U^*_{ca}(\textbf{k})\partial_i U_{cb}(\textbf{k}).
\end{equation}

This is the standard expression for the flat Berry connection. The covariant derivative acting on a state written in the $\{|u_a(\textbf{k})\rangle\}$ frame is then

\begin{equation}
\label{eq:blountpos}
\mc{D}_i|\psi(\textbf{k})\rangle=\sum_a\left(\partial_i\psi^{(\mc{E})}_a(\textbf{k})+i\sum_b\mc{A}^{B}_{iab}(\textbf{k})\psi^{(\mc{E})}_b(\textbf{k})\right)|u_a(\textbf{k})\rangle,
\end{equation}

and we have arrived at Blount’s position operator since recall that the position operator corresponds to the covariant derivative. Note how we did not abuse notation: partial derivatives only acted on components in a frame!\\

While our focus was on the position operator and hence the periodic Bloch states $|u_a(\textbf{k}\rangle$, the spaces of Bloch states $|\phi_a(\textbf{k}\rangle$ at each $\textbf{k}$ provide another fibration of the total Hilbert space. However, a discussion of this aspect is beyond our goals and we refer to \cite{Carpentier2014} for a detailed overview.\\

The key point is that Blount’s expression for the position operator is closely linked to the local frame given by the periodic Bloch states, however, it is best to think of the position operator as a covariant derivative acting on abstract states which can be written in any local frame (ipso facto \eqref{eq:posop}). Then, since this covariant derivative is induced by a flat connection, we can always choose a local frame in which its components vanish and proceed to perform our calculations in this frame. We can, of course, move back to the Bloch frame at any time, provided we do it consistently. On the other hand, should we project to a subspace of the Hilbert space and work within this subspace, the connection will cease to be flat (see section \ref{curvedSpace} in the main text) and we can no longer choose a frame in which the connection components vanish. There are two general situations in which this usually occurs. In the first one, we take our model to have an infinite number of bands and project to a subspace spanned by a finite number of bands. The position operator in the total space of the model corresponds to a flat connection, whereas in the projected subspace corresponds to a curved connection. In the second case, we take our model to have only a finite number of bands and project to a subspace, such as the space spanned by the valence bands, that contains a lesser number of bands. The position operator in the total space, yet again, corresponds to a flat connection, whereas in the projected subspace, corresponds to a curved connection.\\

\section{Identities involving Green's operators}
\label{GreenIdentities}
We present several identities that are used throughout the paper.\\

Recall the definitions from \eqref{eq:GrGaEps}

\begin{equation}
\label{eq:GrGaEpsApp}
G^r(\varepsilon) = \lim_{\eta\to 0^+}\frac{1}{\varepsilon-\mathcal{H}_0+i\eta},\, G^a(\varepsilon) = \lim_{\eta\to 0^+}\frac{1}{\varepsilon-\mathcal{H}_0-i\eta}.
\end{equation}

Henceforth we shall not write out the limit explicitly and use the short-hand notation $G^{r(a)}_{\pm\omega}\equiv G^{r(a)}(\varepsilon+\pm\hbar\omega)$, as introduced in the main text.\\

Following straight from the definitions, we have

\begin{equation}
\label{eq:greenId2Point}
\begin{split}
&G^r_{\omega}-G^r_{\omega'}
\\
&=\frac{1}{\varepsilon+\hbar\omega-\mathcal{H}_0+i\eta}-\frac{1}{\varepsilon+\hbar\omega'-\mathcal{H}_0+i\eta}
\\
&=\frac{-\hbar(\omega-\omega')}{(\varepsilon+\hbar\omega-\mathcal{H}_0+i\eta)(\varepsilon+\hbar\omega'-\mathcal{H}_0+i\eta)}
\\
&=-\hbar(\omega-\omega') G^r_{\omega}G^r_{\omega'},
\end{split}
\end{equation}

and similarly for the advanced correlator.\\

Next, we consider the following combination of retarded Green's operators

\begin{equation}
\label{eq:greenId3Point1}
\begin{split}
&G^r_{\omega_1+\omega_2}-G^r_{\omega_1}-G^r_{\omega_2}+G^r
\\
&=-\hbar\omega_2(G^r_{\omega_1+\omega_2}G^r_{\omega_1}-G^r_{\omega_2}G^r)
\\
&=-\hbar\omega_1(G^r_{\omega_1+\omega_2}G^r_{\omega_2}-G^r_{\omega_1}G^r),
\end{split}
\end{equation}

where we used the identity in \eqref{eq:greenId2Point} with two different groupings. For the first equality we grouped the (first, second) and (third, fourth) terms of the first line, whereas for the second equality we grouped the (first, third) and (second, fourth) terms. We can manipulate further by taking common denominators and arrive at

\begin{equation}
\label{eq:greenId3Point1_5}
\begin{split}
&G^r_{\omega_1+\omega_2}G^r_{\omega_2}-G^r_{\omega_1}G^r=
\\
&=-\hbar\omega_2G^r_{\omega_1+\omega_2}(G^r_{\omega_1}+G^r_{\omega_2})G^r.
\end{split}
\end{equation}

Combining with \eqref{eq:greenId3Point1} we find

\begin{equation}
\label{eq:greenId3Point2}
\begin{split}
&G^r_{\omega_1+\omega_2}-G^r_{\omega_1}-G^r_{\omega_2}+G^r
\\
&=\hbar^2\omega_1\omega_2G^r_{\omega_1+\omega_2}(G^r_{\omega_1}+G^r_{\omega_2})G^r.
\end{split}
\end{equation}

Naturally, the exact same relation can be found for the advanced correlator.\\

In a very similar manner, by straightforward application of the definitions, we can go further and show the following identity for the case of three frequency arguments

\begin{equation}
\begin{split}
&G^r_{\omega_1+\omega_2+\omega_3}-G^r_{\omega_1+\omega_2}-G^r_{\omega_2+\omega_3}
-G^r_{\omega_3+\omega_1}
\\
&\qquad +G^r_{\omega_1}+G^r_{\omega_2}+G^r_{\omega_3}-G^r
\\
&=-\hbar^3\omega_1\omega_2\omega_3G^r_{\omega_1+\omega_2+\omega_3}G^r(G^r_{\omega_1+\omega_2}(G^r_{\omega_1}+G^r_{\omega_2})
\\
&\qquad\quad+G^r_{\omega_2+\omega_3}(G^r_{\omega_2}+G^r_{\omega_3})+G^r_{\omega_3+\omega_1}(G^r_{\omega_3}+G^r_{\omega_1})),
\end{split}
\end{equation}

that is useful for third order responses.\\

\section{Proof of the second order gauge conditions}
\label{secondOrderGaugeProof}
The second order gauge conditions \eqref{eq:secondOrderGauge} are
\begin{equation}
\begin{split}
&K^{AA}_{i_0i_1i_2}(\omega_1,0)=K^{AA}_{i_0i_1i_2}(0,\omega_2)=K^{AA}_{i_0i_1i_2}(0,0)=0,
\\
&K^{AE}_{i_0i_1i_2}(0,\omega_2)=K^{EA}_{i_0i_1i_2}(\omega_1,0)=0.
\end{split}
\end{equation}

\subsection{The case of $K^{AA}_{i_0i_1i_2}(0,\omega)$}

We start with the conditions for $K^{AA}$. Due to the intrinsic permutation symmetry, $K^{AA}_{i_0i_1i_2}(\omega_1,0)=K^{AA}_{i_0i_2i_1}(0,\omega_1)$, if $K^{AA}_{i_0i_1i_2}(0,\omega)=0$, then $K^{AA}_{i_0i_1i_2}(\omega,0)=0$ for arbitrary $\omega$ this naturally imples $K^{AA}_{i_0i_1i_2}(0,0)=0$. We thus only need to prove $K^{AA}_{i_0i_1i_2}(0,\omega)=0$.\\

$K^{AA}_{i_0i_1i_2}(0,\omega)$ is given in terms of retarded correlators by \eqref{eq:KAAomega}

\begin{widetext}
\begin{equation}
\label{eq:respKAA}
K^{AA}_{i_0i_1i_2}(0,\omega)=C^r_{\mathcal{O}^{(2)AA}_{i_0i_1i_2}}
-C^r_{\mathcal{O}_{i_0}j_{i_1 i_2}}(\omega)
-\frac{1}{2}(C^r_{\mathcal{O}^{(1)A}_{i_0i_1}j_{i_2}}(\omega)+
C^r_{\mathcal{O}^{(1)A}_{i_0i_2}j_{i_1}}(0))+C^r_{\mathcal{O}_{i_0}j_{i_1}j_{i_2}}(0,\omega).
\end{equation}

Next, we make use of the couplings \eqref{eq:couplings}, observable expansion \eqref{eq:ObsExpansion} and spectral representations \eqref{eq:Ret2SpecF} and \eqref{eq:Ret3SpecF}. Furthermore, we have the flat current definition $j_i=\frac{ie}{\hbar}[x_i,G^{-1}]$ and shall repeatedly apply the cyclic property of the trace. The retarded correlators are

\begin{equation}
\label{eq:fullCorr1}
C^r_{\mathcal{O}^{(2)AA}_{i_0 i_1i_2}}=\text{tr}(\rho_0\mathcal{O}^{(2)AA}_{i_0 i_1i_2})
=\left(\frac{ie}{\hbar}\right)^2\frac{i}{8\pi}\int d\varepsilon\,\rho_0(\varepsilon)\text{tr}(([x_{i_1},[x_{i_2},\mathcal{O}_{i_0}]]+[x_{i_2},[x_{i_1},\mathcal{O}_{i_0}]])G^{r-a}),
\end{equation}
\begin{equation}
\label{eq:fullCorr2}
C^r_{\mathcal{O}_{i_0}j_{i_1i_2}}(\omega)=\frac{ie}{\hbar}\frac{i}{8\pi}\int d\varepsilon\,\rho_0(\varepsilon)\text{tr}((\mathcal{O}_{i_0}G^r_{\omega}([x_{i_1},j_{i_2}]+[x_{i_2},j_{i_1}])+([x_{i_1},j_{i_2}]+[x_{i_2},j_{i_1}])G^a_{-\omega}\mathcal{O}_{i_0})G^{r-a}),
\end{equation}

\begin{equation}
\label{eq:fullCorr31}
\frac12 C^r_{\mathcal{O}^{(1)A}_{i_0i_1}j_{i_2}}(\omega)=\frac{ie}{\hbar}\frac{i}{4\pi}\int d\varepsilon\,\rho_0(\varepsilon)\text{tr}(([x_{i_1},\mathcal{O}_{i_0}]G^r_{\omega}j_{i_2}+j_{i_2}G^a_{-\omega}[x_{i_1},\mathcal{O}_{i_0}])G^{r-a}),
\end{equation}

\begin{equation}
\label{eq:fullCorr32}
\begin{split}
\frac12 C^r_{\mathcal{O}^{(1)A}_{i_0i_2}j_{i_1}}(0)&=\frac{ie}{\hbar}\frac{i}{4\pi}\int d\varepsilon\,\rho_0(\varepsilon)\text{tr}(([x_{i_2},\mathcal{O}_{i_0}]G^rj_{i_1}+j_{i_1}G^a [x_{i_2},\mathcal{O}_{i_0}])G^{r-a})
\\
&=\left(\frac{ie}{\hbar}\right)^2\frac{i}{4\pi}\int d\varepsilon\,\rho_0(\varepsilon)\text{tr}([x_{i_2},\mathcal{O}_{i_0}]G^r[x_{i_1},G^{-1}]G^r-G^a[x_{i_1},G^{-1}]G^a [x_{i_2},\mathcal{O}_{i_0}])
\\
&=-\left(\frac{ie}{\hbar}\right)^2\frac{i}{4\pi}\int d\varepsilon\,\rho_0(\varepsilon)\text{tr}([x_{i_2},\mathcal{O}_{i_0}][x_{i_1},G^{r-a}])
\\
&=\left(\frac{ie}{\hbar}\right)^2\frac{i}{4\pi}\int d\varepsilon\,\rho_0(\varepsilon)\text{tr}([x_{i_1},[x_{i_2},\mathcal{O}_{i_0}]]G^{r-a}),
\end{split}
\end{equation}

and

\begin{equation}
\label{eq:fullCorr4}
\begin{split}
C^r_{\mathcal{O}_{i_0}j_{i_1}j_{i_2}}(0,\omega)=\frac{i}{4\pi}\int d\varepsilon\,\rho_0(\varepsilon)\text{tr}((
&\mathcal{O}_{i_0}G^r_{\omega}j_{i_1}G^r_{\omega}j_{i_2}
+\mathcal{O}_{i_0}G^r_{\omega}j_{i_2}G^r j_{i_1}
+j_{i_2}G^a_{-\omega}j_{i_1}G^a_{-\omega}\mathcal{O}_{i_0}
+j_{i_1}G^aj_{i_2}G^a_{-\omega}\mathcal{O}_{i_0}
\\
&+j_{i_1}G^a\mathcal{O}_{i_0}G^r_{\omega}j_{i_2}
+j_{i_2}G^a_{-\omega}\mathcal{O}_{i_0}G^r j_{i_1})
G^{r-a}),
\end{split}
\end{equation}
where for \eqref{eq:fullCorr4}, $\omega_1=0$ is taken in \eqref{eq:Ret3SpecF} only \textit{after} the application of the projector $\hat{\mathcal{P}}^{(\Gamma_1^+)}_{(i_1,\omega_1)(i_1,\omega_{2})}$ has been completed.\\

We focus on the trace part of \eqref{eq:fullCorr4}. By the cyclic property of the trace, several terms cancel and we are left with
\begin{equation}
\label{eq:fullCorr4Manip}
\begin{split}
&\text{tr}(\mathcal{O}_{i_0}G^r_{\omega}j_{i_1}G^r_{\omega}j_{i_2}G^{r-a}+G^{r-a}j_{i_2}G^a_{-\omega}j_{i_1}G^a_{-\omega}\mathcal{O}_{i_0}
+\mathcal{O}_{i_0}G^r_{\omega}j_{i_2}G^r j_{i_1}G^r-\mathcal{O}_{i_0}G^r_{\omega}j_{i_2}G^aj_{i_1}G^a
\\
&\quad+G^r j_{i_1}G^rj_{i_2}G^a_{-\omega}\mathcal{O}_{i_0}-G^aj_{i_1}G^aj_{i_2}G^a_{-\omega}\mathcal{O}_{i_0}
)
\\
&=\frac{ie}{\hbar}\text{tr}(\mathcal{O}_{i_0}G^r_{\omega}[x_{i_1},G^{-1}_{+\omega}]G^r_{\omega}j_{i_2}G^{r-a}+G^{r-a}j_{i_2}G^a_{-\omega}[x_{i_1},G^{-1}_{-\omega}]G^a_{-\omega}\mathcal{O}_{i_0}
\\
&\qquad\quad+\mathcal{O}_{i_0}G^r_{\omega}j_{i_2}G^r[x_{i_1},G^{-1}]G^r-\mathcal{O}_{i_0}G^r_{\omega}j_{i_2}G^a[x_{i_1},G^{-1}]G^a
\\
&\qquad\quad+G^r[x_{i_1},G^{-1}]G^rj_{i_2}G^a_{-\omega}\mathcal{O}_{i_0}-G^a[x_{i_1},G^{-1}]G^aj_{i_2}G^a_{-\omega}\mathcal{O}_{i_0}
)
\\
&=\frac{ie}{\hbar}\text{tr}(\mathcal{O}_{i_0}G^r_{\omega}x_{i_1}j_{i_2}G^{r-a}-\mathcal{O}_{i_0}x_{i_1}G^r_{\omega}j_{i_2}G^{r-a}+G^{r-a}j_{i_2}G^a_{-\omega}x_{i_1}\mathcal{O}_{i_0}-G^{r-a}j_{i_2}x_{i_1}G^a_{-\omega}\mathcal{O}_{i_0}
\\
&\qquad\quad+\mathcal{O}_{i_0}G^r_{\omega}j_{i_2}G^{r-a}x_{i_1}-\mathcal{O}_{i_0}G^r_{\omega}j_{i_2}x_{i_1}G^{r-a}
+G^{r-a}x_{i_1}j_{i_2}G^a_{-\omega}\mathcal{O}_{i_0}-x_{i_1}G^{r-a}j_{i_2}G^a_{-\omega}\mathcal{O}_{i_0})
\\
&=\frac{ie}{\hbar}\text{tr}((\mathcal{O}_{i_0}G^r_{\omega}[x_{i_1},j_{i_2}]+[x_{i_1},j_{i_2}]G^a_{\omega}\mathcal{O}_{i_0})G^{r-a})
\\
&\quad-\frac{ie}{\hbar}\text{tr}(\mathcal{O}_{i_0}[x_{i_1},G^r_{\omega}j_{i_2}G^{r-a}]+[x_{i_1},G^{r-a}j_{i_2}G^a_{-\omega}]\mathcal{O}_{i_0})
\\
&=\frac{ie}{\hbar}\text{tr}((\mathcal{O}_{i_0}G^r_{\omega}[x_{i_1},j_{i_2}]+[x_{i_1},j_{i_2}]G^a_{-\omega}\mathcal{O}_{i_0})G^{r-a})
\\
&\quad+\frac{ie}{\hbar}\text{tr}(([x_{i_1},\mathcal{O}_{i_0}]G^r_{\omega}j_{i_2}+j_{i_2}G^a_{-\omega}[x_{i_1},\mathcal{O}_{i_0}])G^{r-a}),
\end{split}
\end{equation}

where, after the first equality, we defined $j_i=\frac{ie}{\hbar}[x_i,G^{-1}]=\frac{ie}{\hbar}[x_i,G^{-1}\pm\hbar\omega]\equiv\frac{ie}{\hbar}[x_i,G^{-1}_{\pm\omega}]$. After attaching the relevant coefficients and reinstating the integration, we combine the result of \eqref{eq:fullCorr4Manip} with the correlators in \eqref{eq:fullCorr1}-\eqref{eq:fullCorr32} according to \eqref{eq:respKAA} and obtain

\begin{equation}
\label{eq:gaugeConditionFinal}
\begin{split}
K^{AA}_{i_0i_1i_2}(0,\omega)=&\left(\frac{ie}{\hbar}\right)^2\frac{i}{8\pi}\int d\varepsilon\,\rho_0(\varepsilon)\text{tr}(([x_{i_2},[x_{i_1},\mathcal{O}_{i_0}]]-[x_{i_1},[x_{i_2},\mathcal{O}_{i_0}]])G^{r-a})
\\
&+\frac{ie}{\hbar}\frac{i}{8\pi}\int d\varepsilon\,\rho_0(\varepsilon)\text{tr}((\mathcal{O}_{i_0}G^r_{\omega}([x_{i_1},j_{i_2}]-[x_{i_2},j_{i_1}])+([x_{i_1},j_{i_2}]-[x_{i_2},j_{i_1}])G^a_{-\omega}\mathcal{O}_{i_0})G^{r-a})
\\
=&\left(\frac{ie}{\hbar}\right)^2\frac{i}{8\pi}\int d\varepsilon\,\rho_0(\varepsilon)\bigg\{\text{tr}([[x_{i_2},x_{i_1}],\mathcal{O}_{i_0}]]G^{r-a})
\\
&\qquad+\text{tr}((\mathcal{O}_{i_0}G^r_{\omega}([x_{i_1},[x_{i_2},G^{-1}]-[x_{i_2},[x_{i_1},G^{-1}])
+([x_{i_1},[x_{i_2},G^{-1}]]-[x_{i_2},[x_{i_1},G^{-1}]])G^a_{-\omega}\mathcal{O}_{i_0})G^{r-a})\bigg\}
\\
=&\left(\frac{ie}{\hbar}\right)^2\frac{i}{8\pi}\int d\varepsilon\,\rho_0(\varepsilon)\bigg\{\text{tr}([[x_{i_2},x_{i_1}],\mathcal{O}_{i_0}]G^{r-a})
\\
&\qquad\qquad\qquad+\text{tr}((\mathcal{O}_{i_0}G^r_{\omega}[[x_{i_1},x_{i_2}],G^{-1}]+[[x_{i_1},x_{i_2}],G^{-1}]G^a_{-\omega}\mathcal{O}_{i_0})G^{r-a})\bigg\}=0,
\end{split}
\end{equation}

where we used the Jacobi identity multiple times and, finally, the fact that $x_{i_1}$ is just the partial derivative, so $[x_{i_1},x_{i_2}]=0$.\\

\subsection{The case of $K^{AE}_{i_0i_1i_2}(0,\omega)$}

We now show the other gauge conditions $K^{AE}_{i_0i_1i_2}(0,\omega_2)=K^{EA}_{i_0i_1i_2}(\omega_1,0)=0$. By the symmetry condition $K^{AE}_{i_0i_1i_2}(\omega_1,\omega_2)=K^{EA}_{i_0i_2i_1}(\omega_2,\omega_1)$, if  $K^{AE}_{i_0i_1i_2}(0,\omega)=0$ holds then so does $K^{EA}_{i_0i_1i_2}(\omega,0)=0$, hence we only need to prove the former.\\

The response function $K^{AE}_{i_0i_1i_2}(0,\omega)$ is given in terms of retarded correlators by \eqref{eq:KAEomega}

\begin{equation}
\label{eq:respKAE}
K^{AE}_{i_0i_1i_2}(0,\omega)=
C^r_{\mathcal{O}_{i_0}\mathcal{M}^{(1)AE}_{i_1 i_2}}(\omega)
+\frac12 C^r_{\mathcal{O}^{(1)A}_{i_0i_1}\mathcal{M}^{(0)E}_{i_2}}(\omega)
-C^r_{\mathcal{O}_{i_0}j_{i_1}\mathcal{M}^{(0)E}_{i_2}}(0,\omega).
\end{equation}

Just as for $K^{AA}$ we make use of the relevant couplings from \eqref{eq:couplings}, the observable expansion \eqref{eq:ObsExpansion} and spectral representations to get

\begin{equation}
\label{eq:fullCorr2AE}
C^r_{\mathcal{O}_{i_0}\mathcal{M}^{(1)AE}_{i_1 i_2}}(\omega)=e\frac{1}{2}\frac{ie}{\hbar}\frac{i}{2\pi}\int d\varepsilon\,\rho_0(\varepsilon)\text{tr}((\mathcal{O}_{i_0}G^r_{\omega}[x_{i_1},\mathcal{A}_{i_2}]+[x_{i_1},\mathcal{A}_{i_2}]G^a_{-\omega}\mathcal{O}_{i_0})G^{r-a}),
\end{equation}
\begin{equation}
\label{eq:fullCorr3AE}
\frac12 C^r_{\mathcal{O}^{(1)A}_{i_0i_1}\mathcal{M}^{(0)E}_{i_2}}(\omega)=e\frac12\frac{ie}{\hbar}\frac{i}{2\pi}\int d\varepsilon\,\rho_0(\varepsilon)\text{tr}(([x_{i_1},\mathcal{O}_{i_0}]G^r_{\omega}\mathcal{A}_{i_2}+\mathcal{A}_{i_2}G^a_{-\omega}[x_{i_1},\mathcal{O}_{i_0}])G^{r-a}),
\end{equation}
and
\begin{equation}
\label{eq:fullCorr4AE}
\begin{split}
C^r_{\mathcal{O}_{i_0}j_{i_1}\mathcal{M}^{(0)E}_{i_2}}(0,\omega)=e\frac12\frac{i}{2\pi}\int d\varepsilon\,\rho_0(\varepsilon)\text{tr}((&\mathcal{O}_{i_0}G^r_{\omega}j_{i_1}G^r_{\omega}\mathcal{A}_{i_2}+\mathcal{O}_{i_0}G^r_{\omega}\mathcal{A}_{i_2}G^r j_{i_1}+\mathcal{A}_{i_2}G^a_{-\omega}j_{i_1}G^a_{-\omega}\mathcal{O}_{i_0}+j_{i_1}G^a\mathcal{A}_{i_2}G^a_{-\omega}\mathcal{O}_{i_0}
\\
&+j_{i_1}G^a\mathcal{O}_{i_0}G^r_{\omega}\mathcal{A}_{i_2}+\mathcal{A}_{i_2}G^a_{-\omega}\mathcal{O}_{i_0}G^r j_{i_1})G^{r-a}),
\end{split}
\end{equation}

for the retarded correlators on the right hand side of \eqref{eq:respKAE}.\\

We only need to manipulate \eqref{eq:fullCorr4AE}. Using the definition of the flat current $j_{i}=\frac{ie}{\hbar}[x_i,G^{-1}]$ and performing the same operations as we did for $K^{AA}_{i_0i_1i_2}(0,\omega)$, we obtain

\begin{equation}
\begin{split}
C^r_{\mathcal{O}_{i_0}j_{i_1}\mathcal{M}^{(0)E}_{i_2}}(0,\omega)=e\frac12\frac{ie}{\hbar}\frac{i}{2\pi}\int d\varepsilon\,\rho_0(\varepsilon)\bigg(&\text{tr}((\mathcal{O}_{i_0}G^r_{\omega}[x_{i_1},\mathcal{A}_{i_2}]+[x_{i_1},\mathcal{A}_{i_2}]G^a_{-\omega}\mathcal{O}_{i_0})G^{r-a})
\\
&+\text{tr}(([x_{i_1},\mathcal{O}_{i_0}]G^r_{\omega}\mathcal{A}_{i_2}+\mathcal{A}_{i_2}G^a_{-\omega}[x_{i_1},\mathcal{O}_{i_0}])G^{r-a})\bigg).
\end{split}
\end{equation}

This is the same as the sum of \eqref{eq:fullCorr2AE} and \eqref{eq:fullCorr3AE}, thus, when combining all three correlators according to \eqref{eq:respKAE}, we get $K^{AE}_{i_0i_1i_2}(0,\omega)=0$, as claimed.

\section{Proof of formula \eqref{eq:secondOrderFinal}}
\label{secondOrderKuboProof}
We show that combining the terms in \eqref{eq:secondOrderStart} leads to the second order response formula \eqref{eq:secondOrderFinal}.\\
Our goal is to pair the flat current operators $j_i=\frac{ie}{\hbar}[x_i,G^{-1}]$ with the connection component operators $\mathcal{A}_i$ to get the curved current operators $J_i=j_i-\frac{ie}{\hbar}[\mathcal{A}_i,G^{-1}]=\frac{ie}{\hbar}[x_i-\mathcal{A}_i,G^{-1}]=\frac{ie}{\hbar}[r_i,G^{-1}]$. The latter is a covariant derivative, meaning that its components transform covariantly under a change of frame which is what we need in order to obtain a frame-independent expectation value.\\
The strategy will be to first use the coupling definitions \eqref{eq:couplings} and write the retarded 2 and 3-point correlators contained in \eqref{eq:secondOrderStart} in their respective spectral representations \eqref{eq:Ret2SpecF} and \eqref{eq:Ret3SpecF}, then apply Green's operator identities discussed in Appendix \ref{GreenIdentities} line-by-line after the second equality of \eqref{eq:secondOrderStart}. Henceforth, when referring to line numbers we mean the lines after the second equality. For brevity, we shall use the short-hand $\hat{\mathcal{P}}^{(\Gamma_1^+)}_{12}\equiv\hat{\mathcal{P}}^{(\Gamma_1^+)}_{(i_1,\omega_1)(i_2,\omega_2)}$. Furthermore, we will be needing the frequency-dependent inverse $G^{-1}_{\pm\omega}=\varepsilon-\mathcal{H}_0+\hbar\omega$.\\

The first line is
\begin{equation}
\label{eq:2ndOrdFirstLine}
\begin{split}
&\frac{C^r_{\mathcal{O}_{i_0}j_{i_1 i_2}}(\omega_1+\omega_2)
-C^r_{\mathcal{O}_{i_0}j_{i_1 i_2}}(\omega_1)-C^r_{\mathcal{O}_{i_0}j_{i_1 i_2}}(\omega_2)+C^r_{\mathcal{O}_{i_0}j_{i_1 i_2}}(0)}{\omega_1\omega_2}
\\
&\quad = \frac{1}{\omega_1\omega_2}\frac{1}{\pi}\hat{\mathcal{P}}^{(+)}_{\mathcal{K}^*_{\omega}}\,i\int d\varepsilon\,\rho_0(\varepsilon)\frac{ie}{4\hbar}\text{tr}\left(\mathcal{O}_{i_0}(G^r_{\omega_1+\omega_2}-G^r_{\omega_1}-G^r_{\omega_2}+G^r)([x_{i_1},j_{i_2}]+[x_{i_2},j_{i_1}])G^{r-a}\right),
\end{split}
\end{equation}

with the second line being

\begin{equation}
\label{eq:2ndOrdSecondLine}
\begin{split}
&-\frac{C^r_{\mathcal{O}_{i_0}j_{i_1}j_{i_2}}(\omega_1,\omega_2)-C^r_{\mathcal{O}_{i_0}j_{i_1}j_{i_2}}(\omega_1,0)-C^r_{\mathcal{O}_{i_0}j_{i_1}j_{i_2}}(0,\omega_2)+C^r_{\mathcal{O}_{i_0}j_{i_1}j_{i_2}}(0,0)}{\omega_1\omega_2}
\\
&=-\frac{1}{\omega_1\omega_2}\frac{1}{\pi}\hat{\mathcal{P}}^{(+)}_{\mathcal{K}^*_{\omega}}\hat{\mathcal{P}}^{(\Gamma_1^+)}_{12}i\int d\varepsilon\,\rho_0(\varepsilon)\text{tr}(\mathcal{O}_{i_0}(G^r_{\omega_1+\omega_2}-G^r_{\omega_2})j_{i_1}G^r_{\omega_2}j_{i_2}-\mathcal{O}_{i_0}(G^r_{\omega_1}-G^r)j_{i_1}G^rj_{i_2}
\\
&\qquad\qquad\qquad\qquad+\frac12 j_{i_1}(G^a_{-\omega_1}-G^a)\mathcal{O}_{i_0}(G^r_{\omega_2}-G^r)j_{i_2})G^{r-a}).
\end{split}
\end{equation}

The third line is

\begin{equation}
\label{eq:2ndOrdThirdLine}
\begin{split}
&\frac{C^r_{\mathcal{O}_{i_0}\mathcal{M}^{(1)AE}_{i_1 i_2}}(\omega_1+\omega_2)
-C^r_{\mathcal{O}_{i_0}\mathcal{M}^{(1)AE}_{i_1 i_2}}(\omega_2)}{i\omega_1}
+\frac{C^r_{\mathcal{O}_{i_0}\mathcal{M}^{(1)EA}_{i_1 i_2}}(\omega_1+\omega_2)
-C^r_{\mathcal{O}_{i_0}\mathcal{M}^{(1)EA}_{i_1 i_2}}(\omega_1)}{i\omega_2}
\\
&= -\frac{e^2}{2\pi}\hat{\mathcal{P}}^{(+)}_{\mathcal{K}^*_{\omega}}i\int d\varepsilon\,\rho_0(\varepsilon)\text{tr}\left(\mathcal{O}_{i_0}G^r_{\omega_1+\omega_2}(G^r_{\omega_2}[x_{i_1},\mathcal{A}_{i_2}]+G^r_{\omega_1}[x_{i_2},\mathcal{A}_{i_1}])G^{r-a}\right)
\\
&=-\frac{e^2}{4\pi}\hat{\mathcal{P}}^{(+)}_{\mathcal{K}^*_{\omega}}i\int d\varepsilon\,\rho_0(\varepsilon)\text{tr}\left(\mathcal{O}_{i_0}G^r_{\omega_1+\omega_2}(G^r_{\omega_1}+G^r_{\omega_2})([x_{i_1},\mathcal{A}_{i_2}]+[x_{i_2},\mathcal{A}_{i_1}])G^{r-a}\right)\to (*)
\\
&\quad+\frac{e^2}{4\pi}\hat{\mathcal{P}}^{(+)}_{\mathcal{K}^*_{\omega}}i\int d\varepsilon\,\rho_0(\varepsilon)\text{tr}\left(\mathcal{O}_{i_0}G^r_{\omega_1+\omega_2}(G^r_{\omega_1}-G^r_{\omega_2})([x_{i_1},\mathcal{A}_{i_2}]-[x_{i_2},\mathcal{A}_{i_1}])G^{r-a}\right),
\end{split}
\end{equation}
where, for the first equality, we used \eqref{eq:greenId2Point} to combine Green's operators and symmetrized to get the second equality. We leave the last integral as it is, and continue to manipulate the second to last one, as indicated by the asterisk

\begin{equation}
\label{eq:2ndOrdThirdLine1}
\begin{split}
&(*)\to \frac{e^2}{4\pi}\hat{\mathcal{P}}^{(+)}_{\mathcal{K}^*_{\omega}}i\int d\varepsilon\,\rho_0(\varepsilon)\text{tr}(\mathcal{O}_{i_0}G^r_{\omega_1+\omega_2}(G^r_{\omega_1}+G^r_{\omega_2})G^r(\underbrace{[[x_{i_1},\mathcal{A}_{i_2}],G^{-1}]+[[x_{i_2},\mathcal{A}_{i_1}],G^{-1}]}_{[[x_{i_1},G^{-1}],\mathcal{A}_{i_2}]+[x_{i_1},[\mathcal{A}_{i_2},G^{-1}]]+(i_1\leftrightarrow i_2)})G^{r-a})
\\
&\quad=-\frac{1}{\omega_1\omega_2}\frac{1}{4\pi}\hat{\mathcal{P}}^{(+)}_{\mathcal{K}^*_{\omega}}i\int d\varepsilon\,\rho_0(\varepsilon)\frac{ie}{\hbar}\text{tr}\left(\mathcal{O}_{i_0}(G^r_{\omega_1+\omega_2}-G^r_{\omega_1}-G^r_{\omega_2}+G^r)\left([j_{i_1},\mathcal{A}_{i_2}]+[j_{i_2},\mathcal{A}_{i_1}]\right)G^{r-a}\right)
\\
&\qquad-\frac{1}{\omega_1\omega_2}\frac{1}{4\pi}\hat{\mathcal{P}}^{(+)}_{\mathcal{K}^*_{\omega}}i\int d\varepsilon\,\rho_0(\varepsilon)\left(\frac{ie}{\hbar}\right)^2\text{tr}\left(\mathcal{O}_{i_0}(G^r_{\omega_1+\omega_2}-G^r_{\omega_1}-G^r_{\omega_2}+G^r)\left([x_{i_1},[\mathcal{A}_{i_2},G^{-1}]]+[x_{i_2},[\mathcal{A}_{i_1},G^{-1}]]\right)G^{r-a}\right),
\end{split}
\end{equation}

where we used \eqref{eq:greenId3Point2}, the Jacobi identity as depicted by the underbraces and the definition $j_i=\frac{ie}{\hbar}[x_i,G^{-1}]$ of the flat current operator.\\

The fourth line becomes
\begin{equation}
\label{eq:2ndOrdFourthLine}
\begin{split}
&-\frac{C^r_{\mathcal{O}_{i_0}j_{i_1}\mathcal{M}^{(0)E}_{i_2}}(\omega_1,\omega_2)-C^r_{\mathcal{O}_{i_0}j_{i_1}\mathcal{M}^{(0)E}_{i_2}}(0,\omega_2)}{i\omega_1}
-\frac{C^r_{\mathcal{O}_{i_0}\mathcal{M}^{(0)E}_{i_1}j_{i_2}}(\omega_1,\omega_2)-C^r_{\mathcal{O}_{i_0}\mathcal{M}^{(0)E}_{i_1}j_{i_2}}(\omega_1,0)}{i\omega_2}
\\
&=\frac{1}{\omega_1}\frac{ie}{2\pi}\hat{\mathcal{P}}^{(+)}_{\mathcal{K}^*_{\omega}}i\int d\varepsilon\,\rho_0(\varepsilon)\text{tr}(\mathcal{O}_{i_0}(G^r_{\omega_1+\omega_2}-G^r_{\omega_2})j_{i_1}G^r_{\omega_2}\mathcal{A}_{i_2}+\mathcal{O}_{i_0}G^r_{\omega_1+\omega_2}\mathcal{A}_{i_2} G^r_{\omega_1}j_{i_1}-\mathcal{O}_{i_0}G^r_{\omega_1}\mathcal{A}_{i_2} G^r j_{i_1})G^{r-a})
\\
&\quad+\frac{1}{\omega_2}\frac{ie}{2\pi}\hat{\mathcal{P}}^{(+)}_{\mathcal{K}^*_{\omega}}i\int d\varepsilon\,\rho_0(\varepsilon)\text{tr}(\mathcal{O}_{i_0}(G^r_{\omega_1+\omega_2}-G^r_{\omega_1})j_{i_2}G^r_{\omega_2}\mathcal{A}_{i_1}+\mathcal{O}_{i_0}G^r_{\omega_1+\omega_2}\mathcal{A}_{i_1} G^r_{\omega_2}j_{i_2}-\mathcal{O}_{i_0}G^r_{\omega_1}\mathcal{A}_{i_1} G^r j_{i_2})G^{r-a})
\\
&\quad+\frac{ie\hbar}{\pi}\hat{\mathcal{P}}^{(-)}_{\mathcal{K}^*_{\omega}}\hat{\mathcal{P}}^{(\Gamma_1^+)}_{12}i\int d\varepsilon\,\rho_0(\varepsilon)\text{tr}\left(\frac12(j_{i_1}G^a_{-\omega_1}G^a\mathcal{O}_{i_0}G^r_{\omega_2}\mathcal{A}_{i_2}-\mathcal{A}_{i_1}G^a_{-\omega_1}\mathcal{O}_{i_0}G^r_{\omega_2}G^rj_{i_2})G^{r-a}\right)
\\
&=\frac{ie}{\pi}\hat{\mathcal{P}}^{(-)}_{\mathcal{K}^*_{\omega}}\hat{\mathcal{P}}^{(\Gamma_1^+)}_{12}i\int d\varepsilon\,\rho_0(\varepsilon)\bigg(\frac{1}{\omega_1}\text{tr}(\mathcal{O}_{i_0}(G^r_{\omega_1+\omega_2}-G^r_{\omega_2})j_{i_1}G^r_{\omega_2}\mathcal{A}_{i_2}G^{r-a})
\\
&\qquad\qquad\qquad\qquad\qquad\qquad+\frac{1}{\omega_2}\text{tr}(\mathcal{O}_{i_0}G^r_{\omega_1+\omega_2}\mathcal{A}_{i_1} G^r_{\omega_2}j_{i_2}-\mathcal{O}_{i_0}G^r_{\omega_1}\mathcal{A}_{i_1} G^r j_{i_2}G^{r-a})
\bigg)
\\
&\quad-\frac{ie\hbar}{\pi}\hat{\mathcal{P}}^{(-)}_{\mathcal{K}^*_{\omega}}\hat{\mathcal{P}}^{(\Gamma_1^+)}_{12}i\int d\varepsilon\,\rho_0(\varepsilon)\text{tr}\left(\frac12(j_{i_1}G^a_{-\omega_1}G^a\mathcal{O}_{i_0}G^r_{\omega_2}G^r[\mathcal{A}_{i_2},G^{-1}]+[\mathcal{A}_{i_1},G^{-1}]G^a_{-\omega_1}G^a\mathcal{O}_{i_0}G^r_{\omega_2}G^rj_{i_2})G^{r-a}\right)
\\
&=\frac{ie}{\pi}\hat{\mathcal{P}}^{(-)}_{\mathcal{K}^*_{\omega}}\hat{\mathcal{P}}^{(\Gamma_1^+)}_{12}i\int d\varepsilon\,\rho_0(\varepsilon)\bigg(\frac{1}{\hbar\omega_1\omega_2}\text{tr}(\mathcal{O}_{i_0}(G^r_{\omega_1+\omega_2}-G^r_{\omega_2})j_{i_1}(1-G^r_{\omega_2}G^{-1})\mathcal{A}_{i_2}G^{r-a})
\\
&\qquad\qquad\qquad\qquad\qquad\qquad+\frac{1}{\omega_2}\text{tr}(\mathcal{O}_{i_0}G^r_{\omega_1+\omega_2}G^r_{\omega_2}G^{-1}_{\omega_2}\mathcal{A}_{i_1} G^r_{\omega_2}j_{i_2}-\mathcal{O}_{i_0}G^r_{\omega_1}G^rG^{-1}\mathcal{A}_{i_1} G^r j_{i_2}G^{r-a})
\bigg) \to (\star)
\\
&\quad+\frac{1}{\omega_1\omega_2}\frac{1}{\pi}\hat{\mathcal{P}}^{(+)}_{\mathcal{K}^*_{\omega}}\hat{\mathcal{P}}^{(\Gamma_1^+)}_{12}i\int d\varepsilon\,\rho_0(\varepsilon)\text{tr}\bigg(\frac12\frac{ie}{\hbar}\bigg(j_{i_1}(G^a_{-\omega_1}-G^a)\mathcal{O}_{i_0}(G^r_{\omega_2}-G^r)[\mathcal{A}_{i_2},G^{-1}]
\\
&\qquad\qquad\qquad\qquad\qquad\qquad\qquad\qquad\qquad+[\mathcal{A}_{i_1},G^{-1}](G^a_{-\omega_1}-G^a)\mathcal{O}_{i_0}(G^r_{\omega_2}-G^r)j_{i_2}\bigg)G^{r-a}\bigg).
\end{split}
\end{equation}

We continue manipulating the second to last integral indicated by the star

\begin{equation}
\label{eq:2ndOrdFourthLine1}
\begin{split}
(\star)\to &\frac{1}{\omega_1\omega_2}\frac{ie}{\hbar}\frac{1}{\pi}\hat{\mathcal{P}}^{(-)}_{\mathcal{K}^*_{\omega}}\hat{\mathcal{P}}^{(\Gamma_1^+)}_{12}i\int d\varepsilon\,\rho_0(\varepsilon)\text{tr}(\mathcal{O}_{i_0}(G^r_{\omega_1+\omega_2}-G^r_{\omega_2})(j_{i_1}\mathcal{A}_{i_2}-\mathcal{A}_{i_1}j_{i_2})
\\
&\qquad\qquad\qquad\qquad\qquad\qquad\qquad\quad+\mathcal{O}_{i_0}(G^r_{\omega_1}-G^r)\mathcal{A}_{i_1}j_{i_2}-\mathcal{O}_{i_0}(G^r_{\omega_1}-G^r)[\mathcal{A}_{i_1},G^{-1}]G^rj_{i_2}
\\
&\qquad\qquad\qquad\qquad\qquad\qquad\qquad\quad+\mathcal{O}_{i_0}(G^r_{\omega_1+\omega_2}-G^r_{\omega_2})(j_{i_1}G^r_{\omega_2}[\mathcal{A}_{i_2},G^{-1}]+[\mathcal{A}_{i_1},G^{-1}]G^r_{\omega_2}j_{i_2})
\\
&\qquad\qquad\qquad\qquad\qquad\qquad\qquad\quad \underbrace{-\mathcal{O}_{i_0}(G^r_{\omega_1}-G^r)j_{i_1}\mathcal{A}_{i_2}+\mathcal{O}_{i_0}(G^r_{\omega_1}-G^r)j_{i_1}\mathcal{A}_{i_2}}_{=0})G^{r-a})
\\
&=\frac{1}{\omega_1\omega_2}\frac{ie}{\hbar}\frac{1}{\pi}\hat{\mathcal{P}}^{(-)}_{\mathcal{K}^*_{\omega}}\hat{\mathcal{P}}^{(\Gamma_1^+)}_{12}i\int d\varepsilon\,\rho_0(\varepsilon)\text{tr}(\mathcal{O}_{i_0}(G^r_{\omega_1+\omega_2}-G^r_{\omega_2}-G^r_{\omega_1}+G^r)(j_{i_1}\mathcal{A}_{i_2}-\mathcal{A}_{i_1}j_{i_2})
\\
&\qquad\qquad\qquad\qquad\qquad\qquad\qquad\qquad+\mathcal{O}_{i_0}(G^r_{\omega_1+\omega_2}-G^r_{\omega_2})(j_{i_1}G^r_{\omega_2}[\mathcal{A}_{i_2},G^{-1}]+[\mathcal{A}_{i_1},G^{-1}]G^r_{\omega_2}j_{i_2})
\\
&\qquad\qquad\qquad\qquad\qquad\qquad\qquad\qquad -\mathcal{O}_{i_0}(G^r_{\omega_1}-G^r)(j_{i_1}G^r[\mathcal{A}_{i_2},G^{-1}]+[\mathcal{A}_{i_1},G^{-1}]G^rj_{i_2}))G^{r-a})
\\
&=\frac{1}{\omega_1\omega_2}\frac{1}{2\pi}\hat{\mathcal{P}}^{(+)}_{\mathcal{K}^*_{\omega}}i\int d\varepsilon\,\rho_0(\varepsilon)\frac{ie}{\hbar}\text{tr}(\mathcal{O}_{i_0}(G^r_{\omega_1+\omega_2}-G^r_{\omega_2}-G^r_{\omega_1}+G^r)([j_{i_1},\mathcal{A}_{i_2}]+[j_{i_2},\mathcal{A}_{i_1}]))G^{r-a})
\\
&\quad+\frac{1}{\omega_1\omega_2}\frac{1}{\pi}\hat{\mathcal{P}}^{(+)}_{\mathcal{K}^*_{\omega}}\hat{\mathcal{P}}^{(\Gamma_1^+)}_{12}i\int d\varepsilon\,\rho_0(\varepsilon)\frac{ie}{\hbar}\text{tr}(\mathcal{O}_{i_0}(G^r_{\omega_1+\omega_2}-G^r_{\omega_2})(j_{i_1}G^r_{\omega_2}[\mathcal{A}_{i_2},G^{-1}]+[\mathcal{A}_{i_1},G^{-1}]G^r_{\omega_2}j_{i_2})
\\
&\qquad\qquad\qquad\qquad\qquad\qquad\qquad\qquad\quad-\mathcal{O}_{i_0}(G^r_{\omega_1}-G^r)(j_{i_1}G^r[\mathcal{A}_{i_2},G^{-1}]+[\mathcal{A}_{i_1},G^{-1}]G^rj_{i_2}))G^{r-a}).
\end{split}
\end{equation}

Finally, the fifth and last line is

\begin{equation}
\label{eq:2ndOrdFifthLine}
C^r_{\mathcal{O}_{i_0}\mathcal{M}^{(0)E}_{i_1}\mathcal{M}^{(0)E}_{i_2}}(\omega_1,\omega_2)=\frac{e^2}{\pi}\hat{\mathcal{P}}^{(+)}_{\mathcal{K}^*_{\omega}}\hat{\mathcal{P}}^{(\Gamma_1^+)}_{12}i\int d\varepsilon\,\rho_0(\varepsilon)\text{tr}((\mathcal{O}_{i_0}G^r_{\omega_1+\omega_2}\mathcal{A}_{i_1}G^r_{\omega_2}\mathcal{A}_{i_2}+\frac12\mathcal{A}_{i_1}G^a_{-\omega_1}\mathcal{O}_{i_0}G^r_{\omega_2}\mathcal{A}_{i_2})G^{r-a}).
\end{equation}

We manipulate the first term in the trace

\begin{equation}
\label{eq:2ndOrdFifthLine1}
\begin{split}
&\text{tr}(\mathcal{O}_{i_0}G^r_{\omega_1+\omega_2}\mathcal{A}_{i_1}G^r_{\omega_2}\mathcal{A}_{i_2}G^{r-a})=\frac{1}{\hbar\omega_2}\text{tr}(\mathcal{O}_{i_0}G^r_{\omega_1+\omega_2}\mathcal{A}_{i_1}(1-G^r_{\omega_2}G^{-1})\mathcal{A}_{i_2}G^{r-a})
\\
&=\frac{1}{\hbar\omega_2}\text{tr}(\mathcal{O}_{i_0}G^r_{\omega_1+\omega_2}\mathcal{A}_{i_1}\mathcal{A}_{i_2}G^{r-a})
+\frac{1}{\hbar\omega_2}\text{tr}(\mathcal{O}_{i_0}G^r_{\omega_1+\omega_2}G^r_{\omega_2}G^{-1}_{\omega_2}\mathcal{A}_{i_1}G^r_{\omega_2}[\mathcal{A}_{i_2},G^{-1}]G^{r-a}).
\end{split}
\end{equation}

Focusing on the second term after the last equality

\begin{equation}
\label{eq:2ndOrdFifthLine2}
\begin{split}
&\frac{1}{\hbar\omega_2}\text{tr}(\mathcal{O}_{i_0}G^r_{\omega_1+\omega_2}G^r_{\omega_2}G^{-1}_{\omega_2}\mathcal{A}_{i_1}G^r_{\omega_2}[\mathcal{A}_{i_2},G^{-1}]G^{r-a})
\\
&=\frac{1}{\hbar^2\omega_1\omega_2}\text{tr}(\mathcal{O}_{i_0}(G^r_{\omega_1+\omega_2}-G^r_{\omega_2})[\mathcal{A}_{i_1},G^{-1}]G^r_{\omega_2}[\mathcal{A}_{i_2},G^{-1}]G^{r-a})
\\
&\quad-\frac{1}{\hbar^2\omega_1\omega_2}\text{tr}(\mathcal{O}_{i_0}(G^r_{\omega_1+\omega_2}-G^r_{\omega_2})\mathcal{A}_{i_1}[\mathcal{A}_{i_2},G^{-1}]G^{r-a})
\\
&\quad+\underbrace{\frac{1}{\hbar^2\omega_1\omega_2}\text{tr}(\mathcal{O}_{i_0}(G^r_{\omega_1}-G^r)\mathcal{A}_{i_1}[\mathcal{A}_{i_2},G^{-1}]G^{r-a})-\frac{1}{\hbar^2\omega_1\omega_2}\text{tr}(\mathcal{O}_{i_0}(G^r_{\omega_1}-G^r)\mathcal{A}_{i_1}[\mathcal{A}_{i_2},G^{-1}]G^{r-a})}_{=0}
\\
&=-\frac{1}{\hbar^2\omega_1\omega_2}\text{tr}(\mathcal{O}_{i_0}(G^r_{\omega_1+\omega_2}-G^r_{\omega_1}-G^r_{\omega_2}+G^r)\mathcal{A}_{i_1}[\mathcal{A}_{i_2},G^{-1}]G^{r-a})
\\
&\quad+\frac{1}{\hbar^2\omega_1\omega_2}\text{tr}(\mathcal{O}_{i_0}(G^r_{\omega_1+\omega_2}-G^r_{\omega_2})[\mathcal{A}_{i_1},G^{-1}]G^r_{\omega_2}[\mathcal{A}_{i_2},G^{-1}]G^{r-a})
\\
&\quad-\frac{1}{\hbar^2\omega_1\omega_2}\text{tr}(\mathcal{O}_{i_0}(G^r_{\omega_1}-G^r)\mathcal{A}_{i_1}[\mathcal{A}_{i_2},G^{-1}]G^{r-a}),
\end{split}
\end{equation}

with the last term being

\begin{equation}
\label{eq:2ndOrdFifthLine3}
\begin{split}
&-\frac{1}{\hbar^2\omega_1\omega_2}\text{tr}(\mathcal{O}_{i_0}(G^r_{\omega_1}-G^r)\mathcal{A}_{i_1}[\mathcal{A}_{i_2},G^{-1}]G^{r-a})
\\
&=\frac{1}{\hbar^2\omega_1\omega_2}\text{tr}(\mathcal{O}_{i_0}(G^r_{\omega_1}-G^r)[\mathcal{A}_{i_1},G^{-1}]\mathcal{A}_{i_2}G^{r-a})+\frac{1}{\hbar^2\omega_1\omega_2}\text{tr}(\mathcal{O}_{i_0}(G^r_{\omega_1}-G^r)G^{-1}\mathcal{A}_{i_1}\mathcal{A}_{i_2}G^{r-a})
\\
&=-\frac{1}{\hbar^2\omega_1\omega_2}\text{tr}(\mathcal{O}_{i_0}(G^r_{\omega_1}-G^r)[\mathcal{A}_{i_1},G^{-1}]G^r[\mathcal{A}_{i_2},G^{-1}]G^{r-a})-\frac{1}{\hbar\omega_2}\text{tr}(\mathcal{O}_{i_0}G^r_{\omega_1}\mathcal{A}_{i_1}\mathcal{A}_{i_2}G^{r-a}).
\end{split}
\end{equation}

The final term of \eqref{eq:2ndOrdFifthLine3} can be combined with the first term after the last equality of \eqref{eq:2ndOrdFifthLine1} yielding

\begin{equation}
\label{eq:2ndOrdFifthLine4}
\begin{split}
&\frac{1}{\hbar\omega_2}\text{tr}(\mathcal{O}_{i_0}(G^r_{\omega_1+\omega_2}-G^r_{\omega_1})\mathcal{A}_{i_1}\mathcal{A}_{i_2}G^{r-a})=-\text{tr}(\mathcal{O}_{i_0}G^r_{\omega_1+\omega_2}G^r_{\omega_1}\mathcal{A}_{i_1}\mathcal{A}_{i_2}G^{r-a})
\\
&=-\frac{1}{2}\text{tr}(\mathcal{O}_{i_0}G^r_{\omega_1+\omega_2}(G^r_{\omega_1}+G^r_{\omega_2})\mathcal{A}_{i_1}\mathcal{A}_{i_2}G^{r-a})-\frac{1}{2}\text{tr}(\mathcal{O}_{i_0}G^r_{\omega_1+\omega_2}(G^r_{\omega_1}-G^r_{\omega_2})\mathcal{A}_{i_1}\mathcal{A}_{i_2}G^{r-a}),
\end{split}
\end{equation}
with the first term after the last equality becoming

\begin{equation}
\label{eq:2ndOrdFifthLine5}
\begin{split}
&-\frac{1}{2}\text{tr}(\mathcal{O}_{i_0}G^r_{\omega_1+\omega_2}(G^r_{\omega_1}+G^r_{\omega_2})G^rG^{-1}\mathcal{A}_{i_1}\mathcal{A}_{i_2}G^{r-a})
\\
&=\frac{1}{2\hbar^2\omega_1\omega_2}\text{tr}(\mathcal{O}_{i_0}(G^r_{\omega_1+\omega_2}-G^r_{\omega_1}-G^r_{\omega_2}+G^r)(\mathcal{A}_{i_1}[\mathcal{A}_{i_2},G^{-1}]+[\mathcal{A}_{i_1},G^{-1}]\mathcal{A}_{i_2})G^{r-a}),
\end{split}
\end{equation}

where we used \eqref{eq:greenId3Point2}. The final tally contains the first two terms after the last equality of \eqref{eq:2ndOrdFifthLine2}, the first term after the last equality of \eqref{eq:2ndOrdFifthLine3}, the second term after the last equality of \eqref{eq:2ndOrdFifthLine4} and the result of \eqref{eq:2ndOrdFifthLine5}. Combining all these terms and attaching the coefficients, the projectors and integral, we get for \eqref{eq:2ndOrdFifthLine}

\begin{equation}
\label{eq:2ndOrdFifthLine6}
\begin{split}
&\frac{1}{\omega_1\omega_2}\frac{1}{4\pi}\hat{\mathcal{P}}^{(+)}_{\mathcal{K}^*_{\omega}}i\int d\varepsilon\,\rho_0(\varepsilon)\left(\frac{ie}{\hbar}\right)^2\text{tr}(\mathcal{O}_{i_0}(G^r_{\omega_1+\omega_2}-G^r_{\omega_1}-G^r_{\omega_2}+G^r)([\mathcal{A}_{i_1},[\mathcal{A}_{i_2},G^{-1}]]+(i_1\leftrightarrow i_2))G^{r-a})
\\
&-\frac{1}{\omega_1\omega_2}\frac{1}{\pi}\hat{\mathcal{P}}^{(+)}_{\mathcal{K}^*_{\omega}}\hat{\mathcal{P}}^{(\Gamma_1^+)}_{12}i\int d\varepsilon\,\rho_0(\varepsilon)\left(\frac{ie}{\hbar}\right)^2\text{tr}((\mathcal{O}_{i_0}(G^r_{\omega_1+\omega_2}-G^r_{\omega_2})[\mathcal{A}_{i_1},G^{-1}]G^r_{\omega_2}[\mathcal{A}_{i_2},G^{-1}]
\\
&\qquad\qquad\qquad\qquad\qquad\qquad\qquad\qquad\qquad-\mathcal{O}_{i_0}(G^r_{\omega_1}-G^r)[\mathcal{A}_{i_1},G^{-1}]G^r[\mathcal{A}_{i_2},G^{-1}]
\\
&\qquad\qquad\qquad\qquad\qquad\qquad\qquad\qquad\qquad+\frac12[\mathcal{A}_{i_1},G^{-1}](G^a_{-\omega_1}-G^a)\mathcal{O}_{i_0}(G^r_{\omega_2}-G^r)[\mathcal{A}_{i_2},G^{-1}])G^{r-a})
\\
&-\frac{e^2}{4\pi}\hat{\mathcal{P}}^{(+)}_{\mathcal{K}^*_{\omega}}i\int d\varepsilon\,\rho_0(\varepsilon)\text{tr}\left(\mathcal{O}_{i_0}G^r_{\omega_1+\omega_2}(G^r_{\omega_1}-G^r_{\omega_2})[\mathcal{A}_{i_1},\mathcal{A}_{i_2}]G^{r-a}\right),
\end{split}
\end{equation}

where the last term within the trace after the second integral is the second term in \eqref{eq:2ndOrdFifthLine} rewritten with the usual identities.\\

Now we combine \eqref{eq:2ndOrdFirstLine}, the result of \eqref{eq:2ndOrdThirdLine1}, the first integral after the last equality of \eqref{eq:2ndOrdFourthLine1} and the first integral of \eqref{eq:2ndOrdFifthLine6}:

\begin{equation}
\label{eq:2ndOrdProofFinal1}
\begin{split}
&\frac{1}{\omega_1\omega_2}\frac{1}{\pi}\hat{\mathcal{P}}^{(+)}_{\mathcal{K}^*_{\omega}}\,i\int d\varepsilon\,\rho_0(\varepsilon)\frac{ie}{4\hbar}\text{tr}(\mathcal{O}_{i_0}(G^r_{\omega_1+\omega_2}-G^r_{\omega_1}-G^r_{\omega_2}+G^r)
\\
&\qquad\times([x_{i_1},j_{i_2}]+[j_{i_1},\mathcal{A}_{i_2}]-\frac{ie}{\hbar} [x_{i_1},[\mathcal{A}_{i_2},G^{-1}]]+\frac{ie}{\hbar} [\mathcal{A}_{i_1},[\mathcal{A}_{i_2},G^{-1}]+ (i_1\leftrightarrow i_2))G^{r-a})
\\
&=\frac{1}{\omega_1\omega_2}\frac{1}{\pi}\hat{\mathcal{P}}^{(+)}_{\mathcal{K}^*_{\omega}}\,i\int d\varepsilon\,\rho_0(\varepsilon)\frac{ie}{4\hbar}\text{tr}(\mathcal{O}_{i_0}(G^r_{\omega_1+\omega_2}-G^r_{\omega_1}-G^r_{\omega_2}+G^r)([r_{i_1},J_{i_2}]+[r_{i_2},J_{i_1}])G^{r-a})
\\
&=\frac{C^r_{\mathcal{O}_{i_0}J_{i_1 i_2}}(\omega_1+\omega_2)
-C^r_{\mathcal{O}_{i_0}J_{i_1 i_2}}(\omega_1)-C^r_{\mathcal{O}_{i_0}J_{i_1 i_2}}(\omega_2)+C^r_{\mathcal{O}_{i_0}J_{i_1 i_2}}(0)}{\omega_1\omega_2}.
\end{split}
\end{equation}

Next we combine \eqref{eq:2ndOrdSecondLine}, the second integral after the last equality of both \eqref{eq:2ndOrdFourthLine} and \eqref{eq:2ndOrdFourthLine1}, and the second integral of \eqref{eq:2ndOrdFifthLine6} to get  

\begin{equation}
\label{eq:2ndOrdProofFinal2}
\begin{split}
&-\frac{1}{\omega_1\omega_2}\frac{1}{\pi}\hat{\mathcal{P}}^{(+)}_{\mathcal{K}^*_{\omega}}\hat{\mathcal{P}}^{(\Gamma_1^+)}_{12}i\int d\varepsilon\,\rho_0(\varepsilon)\text{tr}(\mathcal{O}_{i_0}(G^r_{\omega_1+\omega_2}-G^r_{\omega_2})J_{i_1}G^r_{\omega_2}J_{i_2}-\mathcal{O}_{i_0}(G^r_{\omega_1}-G^r)J_{i_1}G^rJ_{i_2}
\\
&\qquad\qquad\qquad\qquad+\frac12 J_{i_1}(G^a_{-\omega_1}-G^a)\mathcal{O}_{i_0}(G^r_{\omega_2}-G^r)J_{i_2})G^{r-a})
\\
&=-\frac{C^r_{\mathcal{O}_{i_0}J_{i_1}J_{i_2}}(\omega_1,\omega_2)-C^r_{\mathcal{O}_{i_0}J_{i_1}J_{i_2}}(\omega_1,0)-C^r_{\mathcal{O}_{i_0}J_{i_1}J_{i_2}}(0,\omega_2)+C^r_{\mathcal{O}_{i_0}J_{i_1}J_{i_2}}(0,0)}{\omega_1\omega_2}.
\end{split}
\end{equation}

The only remaining terms are the second integral after the last equality of \eqref{eq:2ndOrdThirdLine} and the third integral of \eqref{eq:2ndOrdFifthLine6} which yield

\begin{equation}
\label{eq:2ndOrdProofFinal3}
\begin{split}
&\frac{e^2}{4\pi}\hat{\mathcal{P}}^{(+)}_{\mathcal{K}^*_{\omega}}i\int d\varepsilon\,\rho_0(\varepsilon)\text{tr}\left(\mathcal{O}_{i_0}G^r_{\omega_1+\omega_2}(G^r_{\omega_1}-G^r_{\omega_2})([x_{i_1},\mathcal{A}_{i_2}]-[x_{i_2},\mathcal{A}_{i_1}]-[\mathcal{A}_{i_1},\mathcal{A}_{i_2}])G^{r-a}\right)
\\
&=-\frac{e^2}{4\pi}\hat{\mathcal{P}}^{(+)}_{\mathcal{K}^*_{\omega}}i\int d\varepsilon\,\rho_0(\varepsilon)\text{tr}\left(\mathcal{O}_{i_0}G^r_{\omega_1+\omega_2}(G^r_{\omega_1}-G^r_{\omega_2})[r_{i_1},r_{i_2}]G^{r-a}\right).
\end{split}
\end{equation}

We have thus reduced \eqref{eq:secondOrderStart} to the expressions \eqref{eq:2ndOrdProofFinal1}, \eqref{eq:2ndOrdProofFinal2} and \eqref{eq:2ndOrdProofFinal3} containing the curved current operator which constitute the second order response formula \eqref{eq:secondOrderFinal}, as claimed.

\section{Proof of formula \eqref{eq:secondOrderGreenFinal}}
\label{secondOrderGreenProof}

We shall make ample use of the identities discussed in Appendix \ref{GreenIdentities}.\\
We would first like to combine the first and third lines of \eqref{eq:secondOrderFinal} and start by manipulating the latter

\begin{equation}
\label{eq:2ndOrdGreenProof1}
\begin{split}
&-\frac{e^2}{4\pi}\hat{\mathcal{P}}^{(+)}_{\mathcal{K}^*_{\omega}}i\int d\varepsilon\,\rho_0(\varepsilon)\text{tr}\left(\mathcal{O}_{i_0}G^r_{\omega_1+\omega_2}(G^r_{\omega_1}-G^r_{\omega_2})[r_{i_1},r_{i_2}]G^{r-a}\right)
\\
&=-\frac{e^2}{4\pi}\hat{\mathcal{P}}^{(+)}_{\mathcal{K}^*_{\omega}}i\int d\varepsilon\,\rho_0(\varepsilon)\text{tr}\left(\mathcal{O}_{i_0}G^r_{\omega_1+\omega_2}(G^r_{\omega_1}-G^r_{\omega_2})G^rG^{-1}[r_{i_1},r_{i_2}]G^{r-a}\right)
\\
&=\frac{e^2}{4\pi}\hat{\mathcal{P}}^{(+)}_{\mathcal{K}^*_{\omega}}i\int d\varepsilon\,\rho_0(\varepsilon)\text{tr}\left(\mathcal{O}_{i_0}G^r_{\omega_1+\omega_2}(G^r_{\omega_1}-G^r_{\omega_2})G^r[[r_{i_1},r_{i_2}],G^{-1}]G^{r-a}\right)
\\
&=-\frac{\hbar^2}{\pi}\hat{\mathcal{P}}^{(+)}_{\mathcal{K}^*_{\omega}}i\int d\varepsilon\,\rho_0(\varepsilon)\frac{ie}{4\hbar}\text{tr}\left(\mathcal{O}_{i_0}G^r_{\omega_1+\omega_2}(G^r_{\omega_1}-G^r_{\omega_2})G^r([r_{i_1},J_{i_2}]-[r_{i_2},J_{i_1}])G^{r-a}\right),
\end{split}
\end{equation}

where, for the final equality, we used the Jacobi identity and applied the definition of the curved current operator. Moving on to the first line of \eqref{eq:secondOrderFinal}, we see that it is expressed with Green's operators after the first equality of \eqref{eq:2ndOrdProofFinal1}. Applying \eqref{eq:greenId3Point2} to the latter and combining with the result of \eqref{eq:2ndOrdGreenProof1} we have

\begin{equation}
\begin{split}
&\frac{\hbar^2}{\pi}\hat{\mathcal{P}}^{(+)}_{\mathcal{K}^*_{\omega}}i\int d\varepsilon\,\rho_0(\varepsilon)\frac{ie}{4\hbar}\text{tr}\left(\mathcal{O}_{i_0}G^r_{\omega_1+\omega_2}(G^r_{\omega_1}+G^r_{\omega_2})G^r([r_{i_1},J_{i_2}]+[r_{i_2},J_{i_1}])G^{r-a}\right)
\\
&-\frac{\hbar^2}{\pi}\hat{\mathcal{P}}^{(+)}_{\mathcal{K}^*_{\omega}}i\int d\varepsilon\,\rho_0(\varepsilon)\frac{ie}{4\hbar}\text{tr}\left(\mathcal{O}_{i_0}G^r_{\omega_1+\omega_2}(G^r_{\omega_1}-G^r_{\omega_2})G^r([r_{i_1},J_{i_2}]-[r_{i_2},J_{i_1}])G^{r-a}\right)
\\
&=\frac{ie\hbar}{2\pi}\hat{\mathcal{P}}^{(-)}_{\mathcal{K}^*_{\omega}}i\int d\varepsilon\,\rho_0(\varepsilon)\text{tr}\left(\mathcal{O}_{i_0}G^r_{\omega_1+\omega_2}G^r(G^r_{\omega_1}[r_{i_2},J_{i_1}]+G^r_{\omega_2}[r_{i_1},J_{i_2}])G^{r-a}\right).
\end{split}
\end{equation}

With $[r_{i},J_{k}]=i\hbar[\mathcal{D}_i,J_k]$, this becomes the first integral of \eqref{eq:secondOrderGreenFinal}.\\

Next we look at the second line of \eqref{eq:secondOrderFinal}. This is expressed with Green's operators on the left hand side of \eqref{eq:2ndOrdProofFinal2}. Consider the following manipulation of the first two terms in the trace of the latter

\begin{equation}
\begin{split}
&\mathcal{O}_{i_0}(G^r_{\omega_1+\omega_2}-G^r_{\omega_2})J_{i_1}G^r_{\omega_2}J_{i_2}-\mathcal{O}_{i_0}(G^r_{\omega_1}-G^r)J_{i_1}G^rJ_{i_2}
\\
&=\mathcal{O}_{i_0}(G^r_{\omega_1+\omega_2}-G^r_{\omega_2})J_{i_1}G^rJ_{i_2}-\hbar\omega_2\mathcal{O}_{i_0}(G^r_{\omega_1+\omega_2}-G^r_{\omega_2})J_{i_1}G^r_{\omega_2}G^rJ_{i_2}-\mathcal{O}_{i_0}(G^r_{\omega_1}-G^r)J_{i_1}G^rJ_{i_2}
\\
&=\mathcal{O}_{i_0}(G^r_{\omega_1+\omega_2}-G^r_{\omega_2}-G^r_{\omega_1}+G^r)J_{i_1}G^rJ_{i_2}-\hbar\omega_2\mathcal{O}_{i_0}(G^r_{\omega_1+\omega_2}-G^r_{\omega_2})J_{i_1}G^r_{\omega_2}G^rJ_{i_2}
\\
&=\hbar^2\omega_1\omega_2(\mathcal{O}_{i_0}G^r_{\omega_1+\omega_2}(G^r_{\omega_1}+G^r_{\omega_2})G^rJ_{i_1}G^rJ_{i_2}+\mathcal{O}_{i_0}G^r_{\omega_1+\omega_2}G^r_{\omega_2}J_{i_1}G^r_{\omega_2}G^rJ_{i_2}),
\end{split}
\end{equation}
where we used $G^r_{\omega_2}=G^r-\hbar\omega_2G^r_{\omega_2}G^r$ for the first equality and \eqref{eq:greenId2Point}, \eqref{eq:greenId3Point2} for the final equality. Similarly, we can use \eqref{eq:greenId2Point} for the third term in the trace on the left hand side of \eqref{eq:2ndOrdProofFinal2} and altogether get

\begin{equation}
\begin{split}
&-\frac{\hbar^2}{\pi}\hat{\mathcal{P}}^{(+)}_{\mathcal{K}^*_{\omega}}\hat{\mathcal{P}}^{(\Gamma_1^+)}_{12}i\int d\varepsilon\,\rho_0(\varepsilon)\text{tr}((\mathcal{O}_{i_0}G^r_{\omega_1+\omega_2}(G^r_{\omega_1}+G^r_{\omega_2})G^rJ_{i_1}G^rJ_{i_2}+\mathcal{O}_{i_0}G^r_{\omega_1+\omega_2}G^r_{\omega_2}J_{i_1}G^r_{\omega_2}G^rJ_{i_2}
\\
&\qquad\qquad\qquad\qquad-\frac12 J_{i_1}G^a_{-\omega_1}G^a\mathcal{O}_{i_0}G^r_{\omega_2}G^rJ_{i_2})G^{r-a}),
\end{split}
\end{equation}

which is the second integral of \eqref{eq:secondOrderGreenFinal}.
\end{widetext}

\nocite{*}

\bibliography{2ndord2}

\end{document}